\documentclass[twocolumn,showkeys,showpacs,prb]{revtex4}
\usepackage{amsfonts, amsmath, amssymb, latexsym}
\usepackage[dvips]{graphicx}

\newtheorem{Pr1}{Proposition}
\newtheorem{Pr2}[Pr1]{Proposition}
\newtheorem{1}{Theorem}
\newtheorem{2}[1]{Theorem}
\newtheorem{3}[1]{Theorem}
\newtheorem{4}[1]{Theorem}
\newtheorem{6}{Corollary}
\newtheorem{C1}[Pr1]{Proposition}
\newtheorem{C2}[Pr1]{Proposition}
\newtheorem{C3}[Pr1]{Proposition}
\newtheorem{C4}[Pr1]{Proposition}
\newtheorem{C5}[Pr1]{Proposition}
\newtheorem{C6}[Pr1]{Proposition}
\newtheorem{C7}[1]{Theorem}
\newtheorem{C8}[1]{Theorem}
\newtheorem{D1}[1]{Theorem}
\newtheorem{D2}[Pr1]{Proposition}
\newtheorem{Def1}{Definition}

\begin{document}

\title{Reweighted techniques: definition and asymptotic convergence}

\author{Cristian Predescu}

\affiliation{
Department of Chemistry, Brown University, Providence, Rhode Island 02912
}
\date{\today}
\begin{abstract}
I  define and characterize the reweighted methods, which are techniques used in conjunction with the random series implementation 
of the Feynman-Ka\c{c} formula. I prove several convergence results valid for 
all series representations and then I specialize the results  for the L\'evy-Ciesielski and Wiener-Fourier series. As opposed to the partial averaging method on which they are based, the reweighted techniques do not involve any modification of the physical potential. Rather, the underlying idea is to develop some specialized constructions of the Brownian bridge that enters the Feynman-Ka\c{c} formula, so as to simulate the  partial averaging effect. For the L\'evy-Ciesielski series representation, I develop a reweighted technique which has $o(1/n^2)$ convergence for potentials having first order Sobolev derivatives. It is suggested that the asymptotic convergence may reach $O(1/n^3)$ for potentials having second order Sobolev derivatives.  The method preserves the favorable $\propto \log(n)$ scaling for the time necessary to compute a path at a given discretization point.  For the Wiener-Fourier series representation, the particular reweighted method designed in the present article is shown to have  ${O}(1/n^3)$ convergence if the potential has second order Sobolev derivatives. The convergence constant has superior dependence with the inverse temperature as compared to the partial averaging method for the same series. Because the expression of the convergence constant does not actually involve the second order derivatives of the potential, it is conjectured that the $O(1/n^3)$ convergence extends to the potentials having first order Sobolev derivatives only.  
\end{abstract}
\maketitle

\section{Introduction}
Numerical recipes based on the discretization of the Feynman-Ka\c{c} formula\cite{Fey65, Sim79} are the most successful ways to account for the quantum contributions in equilibrium statistical simulations for many-body systems. The Feynman-Ka\c{c} formula represents the density matrix and the partition function of a canonical quantum
system as an infinite dimensional stochastic integral of a Brownian
bridge functional.\cite{Sim79, Pre02}  This stochastic integral is then
approximated by a sequence of finite dimensional integrals, which in turn
are evaluated by Monte Carlo techniques.  The random series implementations,\cite{Dol90,Pre02} which lead to some of the most effective discretization techniques, 
utilize the Ito-Nisio theorem\cite{Kwa92} to construct the Brownian
bridge in the Feynman-Ka\c{c} formula. 

There are two discretization techniques of practical importance that can be derived from a given random series representation: the partial averaging\cite{Pre02, Dol85} and the reweighted methods.\cite{Pre02, Pre02b} As a direct method, the partial averaging (PA) technique features many desirable properties, as for instance (i) good behavior at low temperature (avoiding the so-called ``classical collapse''),\cite{Pre02d}   (ii) convergence for a wide class of potentials, including those that have negative coulombic singularities,\cite{Pre02c} and (iii) favorable asymptotic convergence of up to ${O}(1/n^3)$ for smooth enough potentials.\cite{Pre02a}

However, a major concern in the practical applications of the Feynman-Ka\c{c} formula is the asymptotic rate of convergence as measured against the number of variables  used to
parameterize the paths $n$ as well as the computational time necessary
to evaluate the various quantities involved. In this respect, the main disadvantage of the partial averaging method is that it involves the Gaussian transform of the potential. This makes the PA technique hard to implement if the Gaussian transform is not analytically known. 

It has been recently recognized that there is an indirect application of the partial averaging method.\cite{Pre02} More precisely, since the  properties of the partial averaging method are mathematically well understood, one may try and devise methods that (asymptotically) simulate the effect of partial averaging, with the hope of capturing at least some of its desirable features.  As such, the reweighted (RW) random series techniques, which will be rigorously  defined in the following section, are designed  to achieve superior asymptotic convergence  by simulating the  partial averaging approach. By construction, the reweighted techniques do not involve  any modifications of the physical potential. Rather, they are based on some specialized constructions of the Brownian bridge that enters the Feynman-Ka\c{c} formula. This makes the RW techniques straightforward to implement yet, for  any given series representation, there are reweighted techniques that achieve an asymptotic convergence better than that of the corresponding partial averaging method. 

As opposed to the partial averaging method, the reweighted techniques are not uniquely defined by the random series representation. In fact, we talk about a family of methods called reweighted techniques that are derived from a given random series representation. They share the property that their asymptotic convergence is characterized by the same convergence theorems. Then, these theorems can be used to design reweighted methods having improved asymptotic behavior. In this article, I use the aforementioned convergence theorems to design two special reweighted techniques based on the L\'evy-Ciesielski and Wiener-Fourier series, respectively. The technique based on the L\'evy-Ciesielski representation will be shown to have $o(1/n^2)$ asymptotic convergence for Kato-class potentials that have first order Sobolev derivatives.\cite{Footnote} I also conjecture that the asymptotic rate of convergence may reach $O(1/n^3)$ for the potentials that have second order Sobolev derivatives. Quite important, the method preserves the favorable $\propto \log(n)$ scaling for the time necessary to compute a path at a given discretization point.  The technique based on the Wiener-Fourier series is proven to have $O(1/n^3)$ asymptotic convergence for potentials that have second order Sobolev derivatives. However, the convergence constant has more favorable dependence with the inverse temperature as compared to the partial averaging convergence constant. Moreover, the convergence constant does not depend upon the second order derivatives and I speculate that the cubic asymptotic convergence extends to the class of potentials that have first order Sobolev derivatives only.  The special treatment received by the Wiener-Fourier series is motivated by the fact that it is the optimal series representation of the Brownian bridge with regard to the minimization of the number of variables used to parameterize the paths.\cite{Pre02}

In this paper, our primary interest is the development of the mathematical ideas lying behind the reweighted techniques. As such, the two L\'evy-Ciesielski and Wiener-Fourier reweighted techniques developed in the paper may still not be the optimal ones. Rather, they serve as examples demonstrating that there is a large class of methods which have cubic convergence for sufficiently smooth potentials and which are direct methods in the sense that they do not use ``effective potentials'' for their implementation.

The content of  the article is organized
as follows. In the next section, I give a short review of the random series and the partial
averaging methods with the purpose of establishing the notation. Next, I define the reweighted methods and justify the mathematical relevance of the definition. 
In the first part of Section~III, I provide a description of the main class of potentials for which the reweighted method is expected to converge to the correct Feynman-Ka\c{c} result. In the second part of the same section, I state some important
convergence theorems, which are valid for all series representations.  In Section~IV, I utilize  the results of Section~III to design two specific reweighted methods based on
 the L\'evy-Ciesielski and Wiener-Fourier series, respectively. Their convergence properties are reinforced by numerical studies of a simple harmonic oscillator.  I conclude the article with Section~V, where I summarize and discuss the main results of the paper.

\section{Definitions of the main random series techniques}
\newcommand{\ud}{\mathrm{d}}
\subsection{Random series technique and partial averaging}
In this section, I shall give a short review of
the random series technique and of the partial averaging method  with the  purpose of establishing the
notation. For a more complete discussion, the reader should consult
Refs.~(\onlinecite{Pre02}) and~(\onlinecite{Pre02a}). The starting point
is the Feynman-Ka\c{c} formula
\begin{equation}
%EQUATION 1
\label{eq:1}
\frac{\rho(x,x';\beta)}{\rho_{fp}(x,x';\beta)}=\mathbb{E}\exp\left\{-\beta\int_{0}^{1}\! 
\!  V\Big[x_r(u)+\sigma B_u^0 \Big]\ud u\right\},
\end{equation}
where $\rho(x,x';\beta)$ is the density matrix for a monodimensional 
canonical system characterized by the inverse temperature 
$\beta=1/(k_B T)$ and made up of identical particles of mass $m_0$ 
moving in the potential $V(x)$.
The stochastic element that appears in Eq.~(\ref{eq:1}), $\{B_u^0,\, u\geq
0\}$, is a so-called standard Brownian bridge defined as follows: if 
$\{B_u,\, u\geq
0\}$ is a standard Brownian motion starting at zero, then the Brownian
bridge is the stochastic process~$\{B_u |\,B_1=0,\, 0 \leq u \leq 1\}$
i.e., a Brownian motion conditioned on~$B_1=0$. In this
paper, we shall reserve the symbol~$\mathbb{E}$ to denote the expected
value (average value) of  a certain random variable against the
underlying probability measure of the Brownian bridge~$B_u^0$. To
complete the description of Eq~(\ref{eq:1}), we set $x_r(u)=x+(x'-x)u$
(called the reference path), $\sigma= (\hbar^2\beta  /m_0)^{1/2}$, and
let $\rho_{fp}(x,x';\beta)$ denote the density matrix for a similar free
particle.

The generalization of the Eq.~(\ref{eq:1}) to a $d$-dimensional
system is straightforward. The symbol $B_u^0$ now denotes a
$d$-dimensional standard Brownian bridge, which is a vector $(B_{u,1}^0,
B_{u,2}^0,\ldots, B_{u, d}^0)$ with the components being independent
standard Brownian bridges. The symbol $\sigma$ stands for the vector
$(\sigma_1, \sigma_2, \ldots, \sigma_d)$ with components defined by
$\sigma_i^2=\hbar^2\beta/m_{0,i}$, while the product $\sigma B_u^0$ is
interpreted as the $d$-dimensional vector of components $\sigma_i B_{u,i}^0$.
Finally, $x$ and $x'$ are points in the configuration space
$\mathbb{R}^d$ while $x_r(u)=x+(x'-x)u$ is a line in $\mathbb{R}^d$
connecting the points $x$ and $x'$. In this paper, we shall conduct the
proofs for monodimensional systems and only state the
easily obtainable $d$-dimensional results. 

	The most general series representation of the Brownian bridge is given
by the Ito-Nisio theorem,\cite{Kwa92} the statement of which is
reproduced below.
Assume given $\{\lambda_k(\tau)\}_{k \geq 1}$
a system of functions on the interval $[0,1]$, which, together
with the constant function
  $\lambda_0(\tau)=1$, makes up an orthonormal basis in $L^2[0,1]$.
  Let 
\[ \Lambda_k(t)=\int_0^t \lambda_k(u)\ud u.\]
If $\Omega$
is the space of infinite sequences $\bar{a}\equiv(a_1,a_2,\ldots)$ and
\begin{equation}
\label{eq:2}
P[\bar{a}]=\prod_{k=1}^{\infty}\mu(a_k)
\end{equation}
  is the probability measure on $\Omega$ such that the coordinate maps 
$\bar{a}\rightarrow a_k$ are independent identically distributed 
(i.i.d.) variables with distribution probability
\begin{equation}
\label{eq:3}
\mu(a_k\in A)= \frac{1}{\sqrt{2\pi}}\int_A e^{-z^2/2}\,\ud z,
\end{equation}
then
\begin{equation}
\label{eq:4}
B_u^0(\bar{a}) = \sum_{k=1}^{\infty}a_k\Lambda_{k}(u),\; 0\leq u\leq1
\end{equation}
is equal in distribution to a standard Brownian bridge. Moreover,
the convergence of the above series is almost surely uniform on the
interval $[0,1]$.

Using the Ito-Nisio representation of the Brownian bridge,
the Feynman-Ka\c{c} formula (\ref{eq:1}) takes the form
\begin{eqnarray}
\label{eq:5}
  \frac{\rho(x, x' ;\beta)}{\rho_{fp}(x, x' 
;\beta)}&=&\int_{\Omega}\ud P[\bar{a}]\nonumber  \exp\bigg\{-\beta 
\int_{0}^{1}\! \!  V\Big[x_r(u) \\& +& \sigma \sum_{k=1}^{\infty}a_k 
\Lambda_k(u) \Big]\ud u\bigg\}.
\end{eqnarray}
The independence of the coordinates $a_k$, which physically
amounts to choosing those representations in which the kinetic energy
operator is diagonal, is the key to the use of the partial averaging
method. Denoting by $\mathbb{E}_{n}$ the average over the coefficients
beyond the rank~$n$, the partial averaging formula reads
\begin{eqnarray}
\label{eq:6}
  &&\frac{\rho_n^{{PA}}(x, x' ;\beta)}{\rho_{fp}(x, x' 
;\beta)}=\int_{\mathbb{R}}\ud \mu(a_1)\ldots \int_{\mathbb{R}}\ud 
\mu(a_n)\nonumber \\ &&\times \exp\bigg\{-\beta \; 
\mathbb{E}_n\int_{0}^{1}\! \!
V\Big[x_r(u)+\sigma \sum_{k=1}^{\infty}a_k
\Lambda_k(u) \Big]\ud u\bigg\}. \qquad
\end{eqnarray}

To make more sense of the above formula, it is convenient to use the
notation introduced in Ref.~(\onlinecite{Pre02})
\[S_u^n(\bar{a})=\sum_{k=1}^n a_k \Lambda_k(u) \quad \text{and} \quad 
B_u^n(\bar{a})=\sum_{k=n+1}^\infty a_k \Lambda_k(u), \]
which denote the  $n$th order partial sum and and $n$th order tail
series, respectively.
By construction, the random variables $S_u^n(\bar{a})$ and 
$B^n_u(\bar{a})$ are independent and
\[B_u^0(\bar{a})=\sum_{k=1}^\infty a_k 
\Lambda_k(u)=S_u^n(\bar{a})+B_u^n(\bar{a}).\]
Moreover, the variables $B_u^n$ and $B_\tau^n$ have a joint
Gaussian distribution of covariances
\begin{eqnarray}\nonumber
\label{eq:7}&&
\mathbb{E}_n(B_u^n)^2=\sum_{k=n+1}^\infty \Lambda_k(u)^2, \quad 
\mathbb{E}_n(B_\tau^n)^2=\sum_{k=n+1}^\infty 
\Lambda_k(\tau)^2,\nonumber \\&&\text{and} \quad 
\gamma_n(u,\tau)=\mathbb{E}_n(B_u^nB_\tau^n)=\sum_{k=n+1}^\infty 
\Lambda_k(u)\Lambda_k(\tau). \qquad \end{eqnarray}
Equivalently, by using the fact that $S_u^n(\bar{a})$ and 
$B^n_u(\bar{a})$ are independent and their sum is  $B^0_u(\bar{a})$, 
one may evaluate the above series to be
\begin{eqnarray*}
\mathbb{E}_n(B_u^n)^2=\mathbb{E}(B_u^0)^2-\sum_{k=1}^n 
\Lambda_k(u)^2=u(1-u)-\sum_{k=1}^n \Lambda_k(u)^2,\\ 
\mathbb{E}_n(B_\tau^n)^2=\mathbb{E}(B_\tau^0)^2-\sum_{k=1}^n 
\Lambda_k(\tau)^2=\tau(1-\tau)-\sum_{k=1}^n \Lambda_k(\tau)^2,
\end{eqnarray*}
and
\begin{eqnarray*}&&
\gamma_n(u,\tau)=\mathbb{E}(B_u^0B_\tau^0)-\sum_{k=1}^n 
\Lambda_k(u)\Lambda_k(\tau)\\&&=\min(u,\tau)-u\tau-\sum_{k=1}^n 
\Lambda_k(u)\Lambda_k(\tau) .\end{eqnarray*}
A straightforward  computation of $\mathbb{E}(B_u^0)^2$, $\mathbb{E}(B_\tau^0)^2$,
and $\mathbb{E}(B_u^0B_\tau^0)$ is presented in  Appendix~A of Ref.~(\onlinecite{Pre02a}).

Going back to Eq.~(\ref{eq:6}),  one inverts the order of
integration in the exponent and computes
\begin{eqnarray}
\label{eq:8}&&
\mathbb{E}_n\!\int_0^1\! \ud u V[x_r(u)+\sigma 
B_u^0(\bar{a})]\nonumber \\&& = \int_0^1\! \ud u \mathbb{E}_n 
V[x_r(u)+\sigma S_u^n(\bar{a})+\sigma B_u^n(\bar{a})] 
\\&&=\int_0^1\!  \overline{V}_{u,n}[x_r(u)+\sigma S_u^n(\bar{a})]\ud 
u, \nonumber
\end{eqnarray}
where
\begin{equation}
\label{eq:9}
\overline{V}_{u,n}(y)=\int_{\mathbb{R}}\frac{1}{\sqrt{2\pi\Gamma_{n}^2(u)}} 
\exp\left[-\frac{z^2}{2\Gamma_{n}^2(u)}\right]V(y+z) \ud z,
\end{equation}
with~$\Gamma_n^2(u)$ defined by
\begin{equation}
\label{eq:10}
\Gamma_{n}^2(u)=\sigma^2 \mathbb{E}_n(B_u^n)^2=
\sigma^2\left[u(1-u)-
\sum_{k=1}^{n}\Lambda_k(u)^2\right].
\end{equation}
In deducing Eq.~(\ref{eq:8}), one uses the fact that the
variable $B_u^n$ has a Gaussian distribution of variance
$\mathbb{E}_n(B_u^n)^2$. To summarize, we define the $n$th order partial
averaging approximation to the diagonal density matrix by the formula
\begin{eqnarray}
\label{eq:11}&&
\frac{\rho^{PA}_n(x, x' ;\beta)}{\rho_{fp}(x, x' 
;\beta)}=\int_{\mathbb{R}}\ud \mu(a_1)\ldots \int_{\mathbb{R}}\ud 
\mu(a_n)\nonumber  \\&& \times \exp\bigg\{-\beta \; \int_{0}^{1}\! \!
\overline{V}_{u,n}\Big[x_r(u)+\sigma
\sum_{k=1}^{n}a_k \Lambda_k(u) \Big]\ud u\bigg\}.\qquad
\end{eqnarray}

In the present paper, the cosine-Fourier basis $ \{\lambda_k(\tau) =\sqrt{2} \cos(k 
\pi \tau);\; k \geq 1\}$ will receive a special treatment. The corresponding series representation for 
the Brownian bridge is  the Wiener-Fourier construction
\[
B_u^0(\bar{a})\stackrel{d}{=} 
\sqrt{\frac{2}{\pi^2}}\sum_{k=1}^{\infty}a_k\frac{\sin({k\pi 
u})}{k},\; 0\leq u\leq1.
\]
This series has superior convergence properties as demonstrated in Refs.~(\onlinecite{Pre02}) and~(\onlinecite{Pre02a}). The corresponding methods are denoted by the acronym WFPI. 

\subsection{Definition of the reweighted methods}

The reweighted method was first introduced  by Predescu and Doll on an ``intuitive'' basis as a method to account for the  effects of the tail series in a way that does not
involve any modification of the associated potential.\cite{Pre02} As opposed to the partial averaging method, the reweighted method is not uniquely defined by the random series representation on which it is based. Rather, for each series representation, we talk about a family of methods called reweighted methods, that are constructed according to certain guidelines.  The original idea was to replace the collection of random variables $\{B_u^n(\bar{a}); 0\leq u \leq 1\}$  by another collection, $\{R_u^n(b_1,\cdots,b_n); 0 \leq u \leq 1\}$, which
is supported by an $n$-dimensional  underlying probability space and is chosen such that:
\begin{enumerate}
\item{The variance at the point~$u$ of $R_u^n(b_1,\cdots,b_n)$, 
denoted by ${\Gamma'}^2_n(u)$, be as close as possible to $\Gamma_n^2(u)$. }
\item{The variables $S_u^n(a_1,\cdots, a_n)$ and $R^n_u(b_1,\cdots,b_n)$ be 
independent and their sum have a joint distribution as close to a Brownian 
bridge as possible. }  
\end{enumerate}

It is clear that these requirements are rather lax and they involve a lot of experimentation in order to construct a good reweighted method. However,  the recent availability of sharp theorems on the convergence of the partial averaging method\cite{Pre02a} allows us  to provide a clearer definition of the reweighted methods. We begin our analysis by defining the reweighted methods and then, in the remainder of the section, we justify this definition.  

\begin{Def1}
\label{Def:1}
A reweighted method constructed from the random series $\sum_{k=1}^\infty a_k \Lambda_k(u)$ is any sequence of approximations to the density matrix of the form
\begin{eqnarray}
\label{eq:12}
\frac{\rho^{RW}_n(x, x' ;\beta)}{\rho_{fp}(x, x' 
;\beta)}=\int_{\mathbb{R}}\ud \mu(a_1)\ldots \int_{\mathbb{R}}\ud 
\mu(a_{qn+p})\nonumber  \\ \times \exp\bigg\{-\beta \; \int_{0}^{1}\! \!
V\Big[x_r(u)+\sigma
\sum_{k=1}^{n}a_k \Lambda_k(u) \\ \nonumber  + \sigma \sum_{k=n+1}^{qn+p}a_k \tilde{\Lambda}_{n,k}(u)\Big]\ud u\bigg\},\qquad
\end{eqnarray}
where $q$ and $p$ are some fixed integers and where the functions $\tilde{\Lambda}_{n,k}(u)$ satisfy the relation
\begin{equation}
\label{eq:13}
\sum_{k=n+1}^{qn+p}\tilde{\Lambda}_{n,k}(u)^2=\sum_{k=n+1}^{\infty}\Lambda_{k}(u)^2=\mathbb{E}\left(B^n_u\right)^2.
\end{equation}
\end{Def1}

In contrast to the original definition, here the equality $\Gamma_n^2(t)={\Gamma'}^2_n(t)$ is enforced exactly. As such, the reweighted Fourier path integral method initially utilized by Predescu and Doll can be no longer categorized as a reweighted technique. However, the reweighted technique introduced by the same authors in Ref.~(\onlinecite{Pre02b}) for the L\'evy-Ciesielski series representation remains a veritable reweighted method. 

To justify the relevance of  Definition~\ref{Def:1}, we first have to introduce some additional notation. We define 
\begin{equation}
\label{eq:14}
\tilde{B}^{n}_u(\bar{a})= \sum_{k=n+1}^{qn+p}a_k \tilde{\Lambda}_{n,k}(u)
\end{equation}
and
\begin{equation}
\label{eq:15}
\tilde{\gamma}_n(u,\tau)= \mathbb{E}\left(\tilde{B}^{n}_u \tilde{B}^{n}_\tau\right)=\sum_{k=n+1}^{qn+p}\tilde{\Lambda}_{n,k}(u) \tilde{\Lambda}_{n,k}(\tau).
\end{equation}
Then, Eq.~(\ref{eq:13}) says that 
\begin{equation}
\label{eq:16}
\tilde{\gamma}_n(u,u)=\mathbb{E}\left(\tilde{B}^{n}_u\right)^2= \mathbb{E}\left(B^{n}_u\right)^2= \gamma_n(u,u).
\end{equation}
Next, we define
\[
\tilde{U}_n(x,x',\beta; \bar{a})=\int_0^1V[x_r(u)+\sigma 
S^n_u(\bar{a})+\sigma \tilde{B}^n_u(\bar{a})] \ud u
\]
and
\[
\tilde{X}_n(x,x',\beta;\bar{a})=\rho_{fp}(x,x';\beta)\exp[-\beta \, 
\tilde{U}_n(x,x',\beta;\bar{a})].
\]
This functions are  analogous to the  variables $U_n(x,x',\beta;\bar{a})$, $U_\infty(x,x',\beta;\bar{a})$, $X_n(x,x',\beta;\bar{a})$, and $X_\infty(x,x',\beta;\bar{a})$ introduced in the beginning of Section~III of Ref.~(\onlinecite{Pre02b}). To save typographical space, I shall not  redefine the latter variables here, but I refer the reader to the cited paper instead. In terms of the new functions, the $n$-th order reweighted technique approximation to the density matrix given by Eq.~(\ref{eq:12}) is expressed as
\begin{equation}
\label{eq:17} 
\rho_{n}^{\text{RW}}(x,x';\beta)=\mathbb{E}\,[\tilde{X}_n(x,x',\beta;\bar{a})].
\end{equation}

We now make a crucial observation which shall eventually justify Definition~\ref{Def:1}. With the new notation, the relation (\ref{eq:8}) reads 
\begin{equation}
\label{eq:18}
U_n(x,x',\beta;\bar{a})=\mathbb{E}_n 
\left[U_{\infty}(x,x',\beta;\bar{a})\right].
\end{equation}
However, because of Eq.~(\ref{eq:16}), we also have
\begin{equation}
\label{eq:19}
U_n(x,x',\beta;\bar{a})=\mathbb{E}_n 
\left[\tilde{U}_{n}(x,x',\beta;\bar{a})\right]
\end{equation}
and this last equality implies that 
\begin{Pr1}
\label{Pr:1}
The partial averaging method derived from any reweighted technique is identical to the partial averaging method derived from the original random series representation.
\end{Pr1}
The reader should stop and ponder that Proposition~\ref{Pr:1} is not really a theorem. Rather, the underlying idea is to define the reweighted method such that Proposition~\ref{Pr:1} be true. The value of this idea is that almost all  identities relating $\rho_{n}(x,x';\beta)$ and $\rho_{n}^{\text{PA}}(x,x';\beta)$ are also true of $\rho_{n}^{\text{RW}}(x,x';\beta)$ and $\rho_{n}^{\text{PA}}(x,x';\beta)$. For instance, an application of Jensen's inequality shows that 
\[\mathbb{E}_n e^{-\beta \, 
\tilde{U}_n(x,x',\beta;\bar{a})} \geq e^{-\beta \, 
\mathbb{E}_n \tilde{U}_n(x,x',\beta;\bar{a})}= e^{-\beta \, 
U_n(x,x',\beta;\bar{a})},\]
from which, by taking the total expectation and then multiplying with $\rho_{fp}(x,x';\beta)$, we learn that 
\begin{equation}
\label{eq:20}
\rho_{n}^{\text{RW}}(x,x';\beta)\geq \rho_{n}^{\text{PA}}(x,x';\beta).
\end{equation}
This relation is analogous to the well-known inequality
\[
\rho(x,x';\beta)\geq \rho_{n}^{\text{PA}}(x,x';\beta).
\]
More generally, any proofs that are based solely on the common property of the random variables $B_u^n(\bar{a})$ and  $\tilde{B}_u^n(\bar{a})$ of being Gaussian distributed variables are likely to extend to the reweighted technique, too. Therefore, it should not be surprising that the asymptotic behavior of the difference 
\begin{equation}
\label{eq:21}
\rho_{n}^{\text{RW}}(x,x';\beta)-\rho_{n}^{\text{PA}}(x,x';\beta)
\end{equation}
is described by theorems similar to those that characterize the difference
\begin{equation}
\label{eq:22}
\rho(x,x';\beta)-\rho_{n}^{\text{PA}}(x,x';\beta).
\end{equation}

Such theorems are presented in Section~III and they can be employed for the design of reweighted methods that have specific asymptotic properties. More precisely, let us assume that the asymptotic convergence of the partial averaging method is 
\[
\rho(x,x';\beta)-\rho_{n}^{\text{PA}}(x,x';\beta)\approx \frac{C_{\text{PA}}(x,x';\beta)}{n^s}.
\]
Let us also assume that we are able to devise a reweighted technique for which the difference~(\ref{eq:21}) has the asymptotic behavior
\[
\rho_{n}^{\text{RW}}(x,x';\beta)-\rho_{n}^{\text{PA}}(x,x';\beta)\approx \frac{\tilde{C}_{\text{RW}}(x,x';\beta)}{n^s}.
\]
Then the asymptotic behavior of the reweighted method is 
\begin{eqnarray} 
\label{eq:23}\nonumber &&
\rho(x,x';\beta)-\rho_{n}^{\text{RW}}(x,x';\beta)=\Big[\rho(x,x';\beta)\\&& -\rho_{n}^{\text{PA}}(x,x';\beta)\Big]  -\left[\rho_{n}^{\text{RW}}(x,x';\beta)-\rho_{n}^{\text{PA}}(x,x';\beta)\right] \qquad \\&& \approx \frac{C_{\text{PA}}(x,x';\beta)-\tilde{C}_{\text{RW}}(x,x';\beta)}{n^s}, \nonumber
\end{eqnarray}
whenever $C_{\text{PA}}(x,x';\beta)\neq \tilde{C}_{\text{RW}}(x,x';\beta)$. Actually, we would like the difference 
\begin{equation}
\label{eq:24}
C_{\text{RW}}(x,x';\beta)=C_{\text{PA}}(x,x';\beta)- \tilde{C}_{\text{RW}}(x,x';\beta)
\end{equation} 
 to be as close to zero as possible. In  the ideal situation that the difference is zero, the convergence of the reweighted technique becomes $o(1/n^s)$ and so, even if we do not know its asymptotic behavior, we know that the new method is faster than  the corresponding partial averaging method. For the cases where a perfect cancellation does not happen, one may still exploit the possibility that the constant $C_{\text{RW}}(x,x';\beta)$ might have better properties than its components, as for instance more favorable dependence upon temperature or upon the derivatives of the potential.  
 
To conclude the justification of the reweighted method, we  notice that the $n$th order approximation utilizes in fact $qn+p$ Gaussian variables $a_k$. As such,  the asymptotic convergence of the reweighted method as measured with respect to the actual number of coefficients used to parameterize the paths is 
\begin{equation}
\label{eq:25}
\frac{q^s{C}_{\text{RW}}(x,x';\beta)}{(qn+p)^s}
\end{equation}
i.e., the convergence order is maintained but the convergence constant becomes $q^s$-times larger as modulus. This observation also explains why we forced the maximum number of additional parameters  to scale at most linearly with $n$. Finally, Eq.~(\ref{eq:25}) indicates that, as far as the total computational cost of the method is concerned, we should try and  minimize the number of additional coefficients, if this does not affect the asymptotic behavior of the resulting method or some desirable features of the convergence constant $C_{\text{RW}}(x,x';\beta)$.

\section{The asymptotic convergence of the reweighted technique}
\subsection{The class of potentials for which the reweighted techniques are expected to converge}
We begin this section by considering the class of potentials for which the reweighted techniques are expected to converge to the correct Feynman-Ka\c{c} result. As discussed in Refs.~(\onlinecite{Pre02c}) and (\onlinecite{Pre02a}), 
the Feynman-Ka\c{c} formula  is known to represent the Green's function of the corresponding Bloch equation for a large class of potentials called the Kato class. 
A Kato-class potential is defined as follows.\cite{Aiz82} Let
\[
g(y)=\left\{\begin{array}{cc} |y| &  d=1, \\ \ln({\|y\|^{-1}})& d=2, 
\\ \|y\|^{2-d}& d\geq 3 \end{array}\right.
\]
and define the Kato class $K_d$ as the set of all measurable 
functions $f: \mathbb{R}^d \to \mathbb{R}$ such that
\begin{equation*}
\lim_{\alpha \downarrow 0} \sup_{x\in \mathbb{R}^d} \int_{\|x-y\|\leq 
\alpha}|f(y)g(x-y)|\ud y =0.
\end{equation*}
We say that $f$ is in $K_{d}^{\text{loc}}$ if $1_{D}f \in K_{d}$ 
for all bounded domains $D \subset \mathbb{R}^d$. Clearly, $K_d \subset K_d^{\text{loc}}$. We say that  $V(x)$ 
is a Kato-class potential or that it is Kato-decomposable if its negative part $V_{-}=\max\{0, -V\}$ 
is in $K_d$ while its positive part $V_{+}=\max\{0,V\}$ is in 
$K_d^{\text{loc}}$.

Since the Kato class contains almost all physically relevant potentials, we shall restrict the analysis and the definition of the reweighted techniques to this class of potentials.  It has been shown that if the potential has also finite Gaussian transform,  the partial averaging method is convergent to the Feynman-Ka\c{c} formula.\cite{Pre02c} For definitions and further discussions, we refer the reader to the cited bibliography.  The condition that the potentials have finite Gaussian transform is not necessary for the convergence of the reweighted techniques, which are expected to converge even for systems that are confined to a certain region of space (for example, a particle in a box). 
However, as opposed to the partial averaging method,  the non-averaged methods are known not to converge for potentials having sufficiently ``bad'' negative singularities, as for instance negative coulombic singularities.\cite{Law69, Kle95, Mus97, Kol01} Such non-averaged methods include the trapezoidal Trotter discrete path integral methods and the primitive random series methods. Unfortunately, they also include the reweighted techniques. 

To identify the problems that might appear if the potential has negative singularities, it is useful to work out an analogy with the proof of the convergence of the partial averaging method. Theorem~1 of Ref.~(\onlinecite{Pre02c}) says that if the potential $V(x)$ is a Kato-class potential that has finite Gaussian transform, then 
\begin{equation}\
\label{eq:26}
\lim_{n \to \infty} U_n(x,x',\beta;\bar{a})=U_\infty(x,x',\beta;\bar{a})
\end{equation}
$P$-almost everywhere (probabilists say $P$-almost surely because $P$ is a probability measure). Of course, this immediately implies 
\begin{equation}
\label{eq:27}
\lim_{n \to \infty} X_n(x,x',\beta;\bar{a})=X_\infty(x,x',\beta;\bar{a})
\end{equation}
$P$-almost surely. In analogy with these equations, for the reweighted techniques the relations
\begin{equation}
\label{eq:28}
\lim_{n \to \infty} \tilde{U}_n(x,x',\beta;\bar{a})=U_\infty(x,x',\beta;\bar{a})
\end{equation}
and 
\begin{equation}
\label{eq:29}
\lim_{n \to \infty} \tilde{X}_n(x,x',\beta;\bar{a})=X_\infty(x,x',\beta;\bar{a})
\end{equation}
are expected to hold $P$-almost surely. However, in this case the requirement that $V(x)$ have finite Gaussian transform is no longer necessary because the functions $\tilde{U}_n(x,x',\beta;\bar{a})$ and $\tilde{X}_n(x,x',\beta;\bar{a})$ no longer involve the Gaussian transform of the potential. In this paper, we shall admit without proof that Eqs.~(\ref{eq:28}) and (\ref{eq:29}) are true for all Kato-class potentials (functions).   

However, Eq.~(\ref{eq:29}) does not necessarily imply 
\begin{equation}
\label{eq:30}
\lim_{n \to \infty} \int_{\Omega} \tilde{X}_n(x,x',\beta;\bar{a})\ud P(\bar{a})= \int_{\Omega} X_\infty(x,x',\beta;\bar{a})\ud P(\bar{a}).
\end{equation}
To give a simple example, the sequence of functions $f_n(x)=1/(n |x|)$ converges to zero for any $x \neq 0$, thus Lebesgue almost everywhere. However, 
\[ \int_{\mathbb{R}} \frac{1}{n |x|}\ud x = \infty \not \to 0 = \int_{\mathbb{R}} 0 \,\ud x. 
\]
A similar situation  happens for Eq.~(\ref{eq:30}) whenever the potential has coulombic negative singularities. The left-hand term of Eq.~(\ref{eq:30}) is $+\infty$ for all $n \geq 1$, yet the right-hand term is finite [according to Theorem~5 of Ref.~(\onlinecite{Pre02c})]. 
In the case of the partial averaging method, it happens that the sequence  $X_n(x,x',\beta;\bar{a})$ is dominated from above by the sequence $\mathbb{E}_n X_\infty(x,x',\beta;\bar{a})$, as follows from the Jensen's inequality. The latter sequence has the properties 
\begin{equation}
\label{eq:31}
 \lim_{n \to \infty} \mathbb{E}_n X_\infty (x,x',\beta;\bar{a})=X_\infty(x,x',\beta;\bar{a})
\end{equation}
and
\begin{equation}
\label{eq:32}
\mathbb{E}\left[\mathbb{E}_n X_\infty (x,x',\beta;\bar{a})\right]=\mathbb{E} X_\infty (x,x',\beta;\bar{a})=\rho(x,x';\beta) < \infty. 
\end{equation}
In this conditions, a standard theorem from measure theory, namely the dominated convergence theorem,\cite{Fol99} guaranties that 
\begin{equation*}
\lim_{n \to \infty} \int_{\Omega} X_n(x,x',\beta;\bar{a})\ud P(\bar{a})= \int_{\Omega} X_\infty(x,x',\beta;\bar{a})\ud P(\bar{a}).
\end{equation*}
By analogy, it follows that in order to enforce Eq.~(\ref{eq:30}), we need to restrict the potential $V(x)$ to those potentials for which the sequence $\tilde{X}_n(x,x',\beta;\bar{a})$ is dominated by an appropriate sequence of random variables $\tilde{Y}_n(x,x',\beta;\bar{a})$ satisfying the relations
\[
\lim_{n \to \infty} \tilde{Y}_n(x,x',\beta;\bar{a}) =\tilde{Y}_\infty(x,x',\beta;\bar{a})
\]
$P$-almost surely and 
\[
\lim_{n \to \infty} \mathbb{E} \left[ \tilde{Y}_n(x,x',\beta;\bar{a})\right] =\mathbb{E}\left[\tilde{Y}_\infty(x,x',\beta;\bar{a}) \right]
\]
for some integrable $\tilde{Y}_\infty(x,x',\beta;\bar{a})$. This is the strategy we adopt in proving the following proposition. 
\begin{Pr2}
\label{Pr:2}
Let  $V(x)$ be a Kato-class potential such that   $e^{-\beta V(x)}$ is a $K_d^{\text{loc}}$ function that has finite Gaussian transform  for all $\beta > 0$. Then,
\begin{equation}
\label{eq:33}
\lim_{n \to \infty} \rho_n^{\text{RW}}(x,x';\beta)= \rho(x,x';\beta) 
\end{equation}
for all $(x,x')\in \mathbb{R}^d\times \mathbb{R}^d$ and $\beta >0$.
If in addition 
\[
Z_{\text{cl}}(\beta)=  \left(\prod_{i=1}^d\frac{1}{\sqrt{2\pi \sigma_i^2}}\right)\int_{\mathbb{R}^d} e^{-\beta V(x)} \ud x < \infty, 
\] 
then we also have
\begin{equation}
\label{eq:34}
\lim_{n \to \infty}Z_n^{\text{RW}}(\beta)=Z(\beta)\leq Z_{cl}(\beta).
\end{equation}
\end{Pr2}

\emph{Proof.} Let us define the random variables 
\begin{eqnarray*}
\tilde{Y}_n(x,x',\beta;\bar{a})= \rho_{fp}(x,x';\beta)\int_0^1 \exp\Big\{-\beta V[x_r(u)\\ +\sigma 
S^n_u(\bar{a})+\sigma \tilde{B}^n_u(\bar{a})]\Big\}\ud u
\end{eqnarray*}
and
\begin{eqnarray*}
\tilde{Y}_\infty(x,x',\beta;\bar{a})= \rho_{fp}(x,x';\beta)\int_0^1 \exp\Big\{-\beta V[x_r(u)\\+ \sigma B^0_u(\bar{a})]\Big\}\ud u.
\end{eqnarray*}
Then Jensen's inequality guaranties that
\[
\tilde{X}_n(x,x',\beta;\bar{a})\leq \tilde{Y}_n(x,x',\beta;\bar{a})
\]
and
\[
\tilde{X}_\infty(x,x',\beta;\bar{a})\leq \tilde{Y}_\infty(x,x',\beta;\bar{a}),
\]
respectively. Because $e^{-\beta V(x)}$ is a positive $K_d^{\text{loc}}$ function, it is Kato-decomposable and replacing $V(x)$ with  $e^{-\beta V(x)}$ in  Eq.~(\ref{eq:26}), we learn that
\[\lim_{n \to \infty}\tilde{Y}_n(x,x',\beta;\bar{a}) = \tilde{Y}_\infty(x,x',\beta;\bar{a}) \] 
$P$-almost surely. 
Moreover, using Eq.~(\ref{eq:16}), one readily computes 
\begin{eqnarray*}&&
\mathbb{E}\left[\tilde{Y}_n(x,x',\beta;\bar{a})\right]= \mathbb{E}\left[ \tilde{Y}_\infty(x,x',\beta;\bar{a})\right]= \rho_{fp}(x,x';\beta)\\&& \times \int_0^1 \ud u (2\pi)^{-d/2}\int_{\mathbb{R}^d}\ud z  e^{-\|z\|^2/2} e^{-\beta V[x_r (u)+\sigma \Gamma_0(u)z] } < \infty.
\end{eqnarray*}
The fact that the last value is finite is guarantied by Theorem~4 of Ref.~(\onlinecite{Pre02c}) and the fact that $e^{-\beta V(x)}$ is a Kato-class function of finite Gaussian transform [again, replace $V(x)$ with $e^{-\beta V(x)}$ in Eq.~(A2) of the cited reference]. 

In these conditions, Eq.~(\ref{eq:29}) and the dominated convergence theorem imply Eqs.~(\ref{eq:30}) and (\ref{eq:33}). Moreover, we have
\[
Z_n^{\text{RW}}(\beta) \leq \int_{\mathbb{R}^d}  \mathbb{E} \left[\tilde{Y}_n(x,x,\beta;\bar{a})\right]\ud x= Z_{cl}(\beta)
\]
and
\[
Z(\beta) \leq \int_{\mathbb{R}^d}  \mathbb{E}\left[\tilde{Y}_\infty(x,x,\beta;\bar{a})\right] \ud x= Z_{cl}(\beta),
\]
respectively. As such, if $Z_{cl}(\beta)$ is finite, Eq.~(\ref{eq:33}) and again the dominated convergence theorem imply Eq.~(\ref{eq:34}). The proof is concluded. \hspace{\stretch{1}}$\Box$

The reader should realize that the hypothesis of Proposition~\ref{Pr:2} is likely to be satisfied for many of the  physical systems of interest for the chemical physicist. It is trivial to show that if a Kato-class potential is bounded from below, then $e^{-\beta V(x)}$ has finite Gaussian transform and lies in $K_d^{\text{loc}}$ [use the fact that $e^{-\beta V(x)}$ is bounded from above by a constant and that the constant functions are  $K_d$-functions]. For such potentials, Proposition~\ref{Pr:2} guaranties the pointwise convergence of the density matrices, as shown by Eq.~(\ref{eq:33}). Moreover, the quantum partition function is recovered as the limit
\[
Z(\beta)=\lim_{n \to \infty}Z_n^{\text{RW}}(\beta),
\] whenever the classical partition function is finite. 
 
\subsection{Asymptotic convergence of the reweighted techniques}

For the remainder of this paper, we shall conduct the proofs for  monodimensional systems and only state the multidimensional analogues. Also, we shall assume that the potential $V(x)$ is a Kato-class potential that satisfies the hypothesis of Proposition~\ref{Pr:2}. However, for the purpose of studying the asymptotic convergence, we also demand that the  potential $V(x)$ have finite Gaussian transform, so that the partial averaging method is convergent too. This requirement is motivated by the fact that we do not  study  directly the asymptotic behavior of the quantity
\[
\rho(x,x';\beta)-\rho_{n}^{\text{RW}}(x,x';\beta)\]
but instead, as argued in Section~II.B, we study the difference
\begin{eqnarray*}&&
\left[\rho(x,x';\beta)-\rho_{n}^{\text{PA}}(x,x';\beta)\right] \\ &&  -\left[\rho_{n}^{\text{RW}}(x,x';\beta)-\rho_{n}^{\text{PA}}(x,x';\beta)\right]. 
\end{eqnarray*}
The convergence of the first term in the expression above was extensively characterized in Ref.~(\onlinecite{Pre02a}). Therefore, the main focus in the present article is on the asymptotic behavior of the difference 
\begin{equation}
\label{eq:35}
\rho_{n}^{\text{RW}}(x,x';\beta)-\rho_{n}^{\text{PA}}(x,x';\beta).
\end{equation}

As emphasized in Section~II.B, the techniques utilized for proving the asymptotic convergence of the partial averaging method can be straightforwardly applied for the difference (\ref{eq:35})  as long as they are solely based on the common property of the random variables $B_u^n(\bar{a})$ and  $\tilde{B}_u^n(\bar{a})$ of being Gaussian distributed. A little algebra shows that
\begin{eqnarray*}
\mathbb{E}_n \tilde{X}_n(x,x',\beta;\bar{a})-X_n(x,x',\beta;\bar{a})=X_n(x,x',\beta;\bar{a})\\ 
\times \,\mathbb{E}_n\left\{e^{-\beta 
[\tilde{U}_n(x,x',\beta;\bar{a})-U_n(x,x',\beta;\bar{a})]}-1\right\}.
\end{eqnarray*}
It is clear that 
\begin{eqnarray*}
\lim_{n \to \infty} [\tilde{U}_n(x,x',\beta;\bar{a})-U_n(x,x',\beta;\bar{a})]=0,
\end{eqnarray*}
because both terms of the above difference converge to the common value $U_\infty(x,x',\beta;\bar{a})$, as shown by Eqs.~(\ref{eq:26}) and (\ref{eq:28}). Then, for large $n$, it makes sense to expand the exponential in a Taylor 
series and retain the first non-vanishing positive term in the 
series, which is also the one that controls the asymptotic 
convergence. In analogy with Eq.~(26) of Ref.~(\onlinecite{Pre02a}), we have
\begin{eqnarray}
\label{eq:36} \nonumber
\mathbb{E}_n 
\tilde{X}_n(x,x',\beta;\bar{a})-X_n(x,x',\beta;\bar{a})\approx 
X_n(x,x',\beta;\bar{a})\\ \times \frac{\beta^2}{2}\mathbb{E}_n 
\left[\tilde{U}_{n}(x,x',\beta;\bar{a})-U_n(x,x',\beta;\bar{a})\right]^2,\quad
\end{eqnarray}
where we used the fact that the first term of the Taylor expansion cancels exactly because of the identity~(\ref{eq:19}). The mathematical significance of the symbol $\approx$  is that the ratio between the left-hand and the right-hand sides of Eq.~(\ref{eq:36}) converges to $1$ in the limit $n \to \infty$. As discussed in Ref.~(\onlinecite{Pre02a}), Eq.~(\ref{eq:36}) is expected to
be true for all potentials $V(x)\in \cap_\alpha 
L^2_\alpha(\mathbb{R})$ which are not constant. The space $V(x)\in \cap_\alpha 
L^2_\alpha(\mathbb{R})$ is the space of all potentials $V(x)$ whose square has finite Gaussian transform. For additional information, the reader is referred to the cited bibliography. 
Taking the total expectation $\mathbb{E}$ in Eq.~(\ref{eq:36}) and remembering that $X_n(x,x',\beta;\bar{a})$ does not depend upon the coefficients beyond the rank $n$, we obtain
\begin{eqnarray}
\label{eq:37} \nonumber
\rho_n^{\text{RW}}(x,x';\beta)-\rho_n^{\text{PA}}(x,x';\beta)\approx\frac{\beta^2}{2}\mathbb{E}\,\bigg\{ 
X_n(x,x',\beta;\bar{a})\\ \times 
\left[\tilde{U}_{n}(x,x',\beta;\bar{a})-U_n(x,x',\beta;\bar{a})\right]^2 \bigg\}.\quad
\end{eqnarray}

Sharper descriptions of the asymptotic behavior of the difference 
\[
\rho_n^{\text{RW}}(x,x';\beta)-\rho_n^{\text{PA}}(x,x';\beta)
\]
can be obtained for the potentials that lie in the classes $\cap_\alpha W^{1,2}_\alpha (\mathbb{R})$ and $\cap_\alpha W^{2,2}_\alpha (\mathbb{R})$. As discussed in Ref.~(\onlinecite{Pre02a}), the former class comprises the potentials $V(x)\in \cap_\alpha L^2_\alpha(\mathbb{R})$ for which  the  first order Sobolev derivatives are also $\cap_\alpha L^2_\alpha(\mathbb{R})$ functions. For the potentials in the latter class, the second order Sobolev derivatives are $\cap_\alpha L^2_\alpha(\mathbb{R})$ functions, too. Because their proofs were based solely on the common property of the random variables $B_u^n(\bar{a})$ and  $\tilde{B}_u^n(\bar{a})$ of being Gaussian distributed, Theorems~1 and~2 of Ref.~(\onlinecite{Pre02a}) extend to our problem in a straightforward fashion. Remembering that the potentials considered in the present paper are Kato decomposable and satisfy the hypothesis of Proposition~\ref{Pr:2}, we have

\begin{1}
\label{th:1}
Assume $V(x) \in \cap_\alpha 
W^{1,2}_\alpha (\mathbb{R})$. Then,
\begin{eqnarray}
\label{eq:38} \nonumber &&
\rho_n^{\text{RW}}(x,x';\beta)-\rho_n^{\text{PA}}(x,x';\beta)\\&& \lessapprox 
\frac{\beta^2}{2}  \int_0^1 \! \ud u \! \int_0^1\! \ud \tau 
\tilde{\gamma}_n(u,\tau)K_{x,x'}^{\beta,n}(u,\tau),
\end{eqnarray}
where
\begin{eqnarray}
\label{eq:39}\nonumber &&
K_{x,x'}^{\beta,n}(u,\tau)=\sigma^2\mathbb{E}\Big\{ 
\tilde{X}_{n}(x,x',\beta;\bar{a})\\ && \times  V'[x_r(u)+\sigma S_{u}^n(\bar{a})+ \sigma \tilde{B}_{u}^n(\bar{a})]  \\ && \times 
V'[x_r(\tau)+\sigma S_{\tau}^n(\bar{a})+ \sigma \tilde{B}_{\tau}^n(\bar{a})]\Big\}. \nonumber
\end{eqnarray}

If in addition $\tilde{\gamma}_n(u,\tau)\geq 0$, then the following stronger 
result holds
\begin{eqnarray}
\label{eq:40} \nonumber 
\rho_n^{\text{RW}}(x,x';\beta)-\rho_n^{\text{PA}}(x,x';\beta) \approx 
\frac{\beta^2}{2} \\ \times \int_0^1 \! \ud u \! 
\int_0^1\! \ud \tau  \tilde{\gamma}_n(u,\tau)K_{x,x'}^{\beta,n}(u,\tau).
\end{eqnarray}
\end{1}
\begin{2}
\label{th:2}
Assume $V(x) \in \cap_\alpha 
W^{2,2}_\alpha (\mathbb{R})$. Then,
\begin{eqnarray}
\label{eq:41} \nonumber 
\rho_n^{\text{RW}}(x,x';\beta)-\rho_n^{\text{PA}}(x,x';\beta) \approx 
\frac{\beta^2}{2}\bigg[  \int_0^1 \! \ud u \! \int_0^1\! \ud \tau 
\tilde{\gamma}_n(u,\tau)\\ \times K_{x,x'}^{\beta,n}(u,\tau)-\frac{1}{2} \int_0^1 \! \ud u \! 
\int_0^1\! \ud \tau 
\tilde{\gamma}_n^2(u,\tau)Q_{x,x'}^{\beta,n}(u,\tau)\bigg],\nonumber
\end{eqnarray}
where
\begin{eqnarray}\nonumber
\label{eq:42} &&
Q_{x,x'}^{\beta,n}(u,\tau)=
{\sigma^4} \mathbb{E}\Big\{ 
\tilde{X}_{n}(x,x',\beta;\bar{a})\\ && \times  V''[x_r(u)+\sigma S_{u}^n(\bar{a})+ \sigma \tilde{B}_{u}^n(\bar{a})]  \\ && \times 
V''[x_r(\tau)+\sigma S_{\tau}^n(\bar{a})+ \sigma \tilde{B}_{\tau}^n(\bar{a})]\Big\}. \nonumber
\end{eqnarray}
\end{2}

There is one important observation we make about the  functions $K_{x,x'}^{\beta,n}(u,\tau)$ and $Q_{x,x'}^{\beta,n}(u,\tau)$.  One may replace the $\tilde{X}_n(x,x',\beta;\bar{a})$ functions appearing in Eqs.~(\ref{eq:39}) and (\ref{eq:41}) with  $X_n(x,x',\beta;\bar{a})$ or $X_\infty(x,x',\beta;\bar{a})$. It makes no difference as to the validity of the theorems.  This observation can be  justified by using  the fact that $\tilde{\gamma}_n(u,\tau)$ and $\tilde{\gamma}_n(u,\tau)^2$ are positive definite integral kernels. In this respect,  the reader may look up the argument  utilized  to demonstrate the equivalence between Eqs.~(32) and (33)  of Ref.~(\onlinecite{Pre02a}). 

As discussed in Section~II.B, there are an infinite number of random series techniques associated with a given random series representation of the Brownian bridge. Some of them have better asymptotic convergence, others are worse. However, they all share the property that their asymptotic convergence is described by the theorems presented in this section. Then, we may use these theorems to find those reweighted techniques that have superior asymptotic convergence. As Theorems~\ref{th:1} and \ref{th:2} show, the asymptotic convergence is controlled by the two-point expectation $\tilde{\gamma}_n(u,\tau)$. As a rule of thumb, the better methods are those for which $\tilde{\gamma}_n(u,\tau)$ is as close as possible to ${\gamma}_n(u,\tau)$,  the two-point expectation of the Brownian bridge.

\section{Applications to the design of optimal reweighted techniques}

\subsection{A reweighted L\'evy-Ciesielski path integral construction}

In Ref.~(\onlinecite{Pre02b}), the authors analyzed the asymptotic convergence of the random series technique based on the L\'evy-Ciesielski series representation of the Brownian bridge. They found that the asymptotic convergence of the corresponding partial averaging method is $O(1/n^2)$ and they also proposed a reweighted technique which has $O(1/n^2)$ convergence. In this section, I shall use the theory developed so far to design a reweighted technique having $o(1/n^2)$ convergence on the class of potentials $V(x) \in \cap_\alpha 
W^{1,2}_\alpha (\mathbb{R})$.  The trick consists in finding a particular reweighted technique for which the constants $C_{\text{PA}}(x,x';\beta)$ and $ \tilde{C}_{\text{RW}}(x,x';\beta)$ appearing in Eq.~(\ref{eq:24}) are equal and which uses a minimal number of additional variables. 

\begin{figure}[!tbp] 
   \includegraphics[angle=270,width=8.5cm,clip=t]{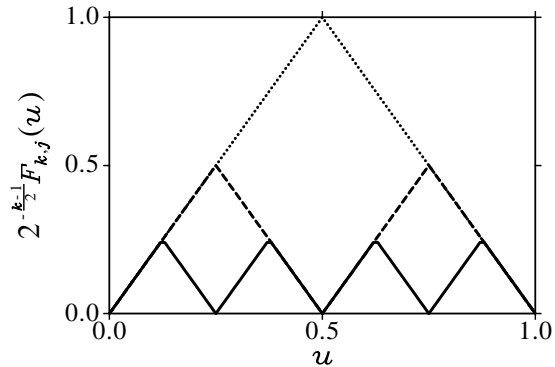} 
%    \BoxedEPSF{Potential.eps scaled 300}
 \caption[sqr]
{\label{Fig:2}
A plot of the renormalized Schauder functions for the layers $k=1,2,\,\text{and}\,3$ showing the pyramidal structure.
}
\end{figure}

The L\'evy-Ciesielski representation of the Brownian bridge is based on the so-called Haar basis which is defined as follows.  For $k=1,2,\ldots$ and $j=1,2,\ldots,2^{k-1}$, the Haar function $f_{k,j}$  is defined by
\begin{equation}
\label{eq:LC1}
f_{k,j}(t)=\left\{\begin{array}{cc} 2^{(k-1)/2},& t \in [(l-1)/2^k, l/2^k]\\ - 2^{(k-1)/2},& t \in [l/2^k, (l+1)/2^k]\\ 0, &\text{elsewhere,} \end{array}\right.
\end{equation}
where $l=2j-1$.
Together with $f_0\equiv 1$, these functions make up a complete orthonormal basis in $L^2([0,1])$. Their primitives 
\begin{widetext}
\begin{equation}
\label{eq:LC2}
F_{k,j}(t)=\left\{\begin{array}{cc} 2^{(k-1)/2}[t-(l-1)/2^k],& t \in [(l-1)/2^k, l/2^k]\\ 2^{(k-1)/2}[(l+1)/2^k-t],& t \in [l/2^k, (l+1)/2^k]\\ 0, &\text{elsewhere} \end{array}\right.
\end{equation}
\end{widetext}
are called the \emph{Schauder functions}. The Schauder functions resemble some ``little tents'' that can be obtained one from the other by dilatations and translations. 
More precisely, we have
\begin{equation}
\label{eq:LC3}
F_{k,1}(u)=2^{-(k-1)/2}F_{1,1}(2^{k-1}u)
\end{equation}
for $k \geq 1$ and 
\begin{equation}
\label{eq:LC4}
F_{k,j}(u)=F_{k,1}\left(u-\frac{j-1}{2^{k-1}}\right)
\end{equation}
for $k\geq 1$ and $ 1\leq j \leq 2^{k-1}$. The scaling relations above have to do with the fact that the original Haar wavelet basis makes up a multiresolution analysis of $L^2([0,1])$  organized in ``layers'' indexed by $k$. If we disregard the factor $2^{(k-1)/2}$, the Schauder functions make up a pyramidal structure as shown in Fig.~\ref{Fig:2}.

Let $\{a_{k,j}; k=1,2,\ldots; j=1,2,\ldots,2^{k-1}\}$ be i.i.d. standard normal variables and define $Y_0(u,\bar{a})\equiv 0$ and
\[Y_k(u,\bar{a})=\sum_{j=1}^{2^{k-1}}a_{k,j}F_{k,j}(u).\] Then by the Ito-Nisio theorem, 
\begin{equation}
\label{eq:LC5}
B^0_u(\bar{a})=\sum_{k=1}^{\infty}Y_k(u,\bar{a})
\end{equation} is equal in distribution to a standard Brownian bridge and the convergence of the right hand side random series is uniform almost surely. The introduction of the intermediate random variables $Y_k(u,\bar{a})$ summing all variables corresponding to the layer $k$ is quite natural and we shall only consider the $n=2^k-1$ subsequences of the various methods analyzed. In these conditions,  we have
\begin{equation}
\label{eq:LC6}
S^n_u(\bar{a})=\sum_{l=1}^{k}Y_l(u,\bar{a}) 
\end{equation}
and 
\begin{equation}
\label{eq:LC7}
 B^n_u(\bar{a})=\sum_{l=k+1}^{\infty}Y_l(u,\bar{a}) .
\end{equation} 
To define the partial averaging method, besides the sum (\ref{eq:LC6}), we need to evaluate
 \[\Gamma_n^2(u)=\sigma^2 \mathbb{E}\, [B^n_u(\bar{a})^2]=\sigma^2\sum_{l=k+1}^\infty \sum_{j=1}^{2^{l-1}} F_{l,j}(u)^2, \]   quantity that enters the expression of the   ``effective'' potential
\[
\overline{V}_{u,n}(x)=\int_{\mathbb{R}}\frac{1}{\sqrt{2\pi\Gamma_{n}^2(u)}} \exp\left[-\frac{z^2}{2\Gamma_{n}^2(u)}\right]V(x+z) \ud z.
\]  
As shown in Ref.~(\onlinecite{Pre02b}), 
\begin{equation}
\label{eq:LC8}
\Gamma^2_n(u)=\sigma^2\sum_{p=1}^{2^{k}}g_{n,p}(u),		
\end{equation} 
where
 \begin{equation}
\label{eq:LC9}
g_{n,1}(u)=\left\{\begin{array}{l l} u(1-2^k u), & 0\leq u <2^{-k}, \\ 0, & \text{otherwise} \end{array} \right.
 \end{equation}
 and 
 \begin{equation}
 \label{eq:LC10}
g_{n,p}(u)= g_{n,1}[u-(p-1)/2^{k}].
\end{equation}
Clearly, for a given $n$, the functions $\{g_{n,p}(u),\; 1\leq p \leq 2^k\}$ have \emph{disjoint} support. Moreover, if we extend the function 
\[\gamma_0(u,u)=\mathbb{E}\left(B_u^0\right)^2=u(1-u)\] outside the interval $[0,1]$ by setting it to zero, we have 
\begin{equation}
\label{eq:LC11}
g_{n,1}(u)=2^{-k}\gamma_0(2^k u, 2^k u). 
\end{equation}

Again in Ref.~(\onlinecite{Pre02b}), the partial averaging  method was shown to have the asymptotic behavior
\[
\rho(x,x';\beta)- \rho_n^{\text{PA}}(x,x';\beta)\approx \frac{C_{\text{PA}}(x,x';\beta)}{(n+1)^2},
\]
where
\begin{eqnarray}
\label{eq:LC12} \nonumber &&
C_{\text{PA}}(x,x';\beta)=
\frac{\hbar^2 \beta^3}{24 m_0} \\ && \times \int_{0}^1\left\langle x\left|e^{-\theta \beta H}\|\nabla V\|^2e^{-(1-\theta)\beta H}\right|x'\right\rangle \ud \theta. 
\end{eqnarray}

We now focus on the construction of our special reweighted technique. We begin by altering the shape of the Schauder functions for the next two layers beyond the layer $k$:
\begin{eqnarray}\nonumber
\tilde{B}^n_u(\bar{a})= \sum_{p=1}^{2^k} \Big[ a_{k+1,p}C_{n,p}(u)+ a_{k+2,2p-1}L_{n,p}(u)\\ +a_{k+2,2p}R_{n,p}(u)\Big].
\end{eqnarray}
However, these additional $3\cdot 2^k= 3n+3$ functions are not chosen independently, but we assume that there are some nonnegative functions $C_0(u)$, $L_0(u)$, and $R_0(u)$ of support $[0,1]$, $[0,1/2]$, and $[1/2,1]$ respectively, such that 
\begin{eqnarray}
\label{eq:LC13} \nonumber
C_{n,1}(u)= 2^{-k}C_{0}\left( 2^k u \right), \quad L_{n,1}(u)=2^{-k} L_{0}\left(2^k u \right), \\ R_{n,1}(u)=2^{-k} R_{0}\left(2^k u \right), \qquad \qquad \qquad
\end{eqnarray}
as well as
\begin{eqnarray}
\label{eq:LC14} \nonumber
C_{n,p}(u)= C_{n,1}\left[u-(p-1)/2^{k}\right],\\ L_{n,p}(u)= L_{n,1}\left[u-(p-1)/2^{k}\right], \\ R_{n,p}(u)= R_{n,1}\left[u-(p-1)/2^{k}\right]. \nonumber  
\end{eqnarray}

Let me pause a moment and explain why this special construction is desirable. One of the major advantages of the L\'evy-Ciesielski series representation is the fact that in order to compute the series $S_u^n(\bar{a})$ at a point $u$, one needs a number of operations proportional to  $\log(n+1)$ only. This is due to the fact that for a given $k$, the Schauder functions $\left\{F_{k,j}(u), \; 1\leq j \leq 2^k \right\}$ have disjoint support. By the way they were constructed, for a given $n$, the set of  functions $\left\{C_{n,p}(u), 1\leq p \leq 2^k\right\}$ have disjoint support too and the same is true of the family $\left\{L_{n,p}(u), R_{n,p}(u); \; 1\leq p \leq 2^k\right\}$. Therefore, in order to evaluate the series $S_u^n(\bar{a})+\tilde{B}^n_u(\bar{a})$ at a point $u$, one needs to perform a number of operations proportional to $2+\log(n+1)$. This is a desirable feature which,  not surprisingly, is also compatible with Eqs.~(\ref{eq:LC10}) and (\ref{eq:LC11}). In fact, using these last equations, one easily shows that the definitional equation (\ref{eq:13}) is satisfied if and only if  
\begin{equation}
\label{eq:LC15}
\gamma_0(u,u)=u(1-u)=C_{0}(u)^2+L_{0}(u)^2+R_{0}(u)^2.
\end{equation}

In these conditions, the function $\tilde{\gamma}_n(u,\tau)$ defined by Eq.~(\ref{eq:15}) becomes
\begin{eqnarray*}
\tilde{\gamma}_n(u,\tau)=\sum_{p=1}^{2^k} \Big[ C_{n,p}(u)C_{n,p}(\tau)+L_{n,p}(u)L_{n,p}(\tau)\\ +R_{n,p}(u)R_{n,p}(\tau) \Big].
\end{eqnarray*}
I leave it for the reader to use the fact that the functions $C_{n,p}(u)$, $L_{n,p}(u)$, and $R_{n,p}(u)$ have compact support shrinking to zero and prove
\[
\lim_{n \to \infty} \frac{\tilde{\gamma}_n(u,\tau)}{\int_0^1\int_0^1 \tilde{\gamma}_n(u,\tau) \ud u \ud \tau} \Rightarrow \delta(u-\tau) 
\]
in the sense of distributions.
Moreover, one computes 
\begin{eqnarray*} \nonumber 
\int_0^1 \int_0^1 \tilde{\gamma}_n(u,\tau) \ud u \ud \tau = 2^k \Bigg\{ \left[\int_0^{1/2^k} C_{n,1}(u)\ud u\right]^2 \\ + \left[\int_0^{1/2^k} L_{n,1}(u)\ud u\right]^2+ \left[\int_0^{1/2^k} R_{n,1}(u)\ud u\right]^2 \Bigg\} \nonumber  \\ = 2^{-2k} \Bigg\{ \left[\int_0^{1} C_{0}(u)\ud u\right]^2  + \left[\int_0^{1} L_{0}(u)\ud u\right]^2\\ + \left[\int_0^{1} R_{0}(u)\ud u\right]^2 \Bigg\}. \nonumber
\end{eqnarray*}
Since $\tilde{\gamma}_n(u,\tau)\geq 0$ and $2^k=n+1$, an application of the second part of Theorem~\ref{th:1} shows that
\[
\nonumber 
\rho_n^{\text{RW}}(x,x';\beta)-\rho_n^{\text{PA}}(x,x';\beta) \approx \frac{\tilde{C}_{\text{RW}}(x,x';\beta)} {(n+1)^2},
\]
where
\begin{eqnarray*}
\tilde{C}_{\text{RW}}(x,x';\beta)=\frac{\beta^2\sigma^2}{2}   \Bigg\{ \left[\int_0^{1} C_{0}(u)\ud u\right]^2 \\ + \left[\int_0^{1} L_{0}(u)\ud u\right]^2  + \left[\int_0^{1} R_{0}(u)\ud u\right]^2 \Bigg\}\\ \times \int_{0}^1\left\langle x\left|e^{-\theta \beta H}\|\nabla V\|^2e^{-(1-\theta)\beta H}\right|x'\right\rangle \ud \theta.
\end{eqnarray*}
We now compare the above equation with Eq.~(\ref{eq:LC12}) and see that the equality
\[C_{\text{RW}}(x,x';\beta)= \tilde{C}_{\text{RW}}(x,x';\beta)\]
holds if and only if 
\begin{eqnarray}
\label{eq:LC16} \nonumber
\left[\int_0^{1} C_{0}(u)\ud u\right]^2  + \left[\int_0^{1} L_{0}(u)\ud u\right]^2 \\ + \left[\int_0^{1} R_{0}(u)\ud u\right]^2 =\frac{1}{12}. 
\end{eqnarray} 
Then, by virtue of Eq.~(\ref{eq:24}), the following result holds. 
\begin{4}
\label{th:LC}
Assume $V(x)\in \cap_{\alpha}W^{1,2}_\alpha(\mathbb{R})$ is a Kato-class potential that satisfies the hypothesis  of Proposition~(\ref{Pr:2}). Then, a L\'evy-Ciesielski reweighted method specified by the nonnegative functions $C_0(u)$, $L_0(u)$, and $R_0(u)$ has $o(1/n^2)$ convergence if and only if Eqs.~(\ref{eq:LC15}) and (\ref{eq:LC16}) hold true. 
\end{4}

\begin{figure}[!tbp]
   \includegraphics[angle=270,width=8.5cm,clip=t]{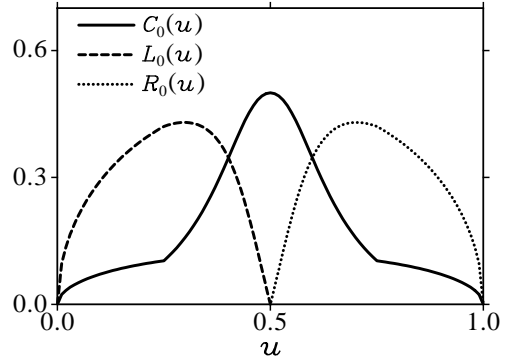}
%    \BoxedEPSF{Potential.eps scaled 300}
  \caption[sqr]
{\label{Fig:LC1}
Shapes of the functions $C_0(u)$, $L_0(u)$, and $R_0(u)$ utilized for the the construction of the reweighted technique. }
\end{figure}

However, to finish the proof of the above theorem, one has to construct at least on example of functions $C_0(u)$, $L_0(u)$, and $R_0(u)$ that satisfy Eqs.~(\ref{eq:LC15}) and (\ref{eq:LC16}). 
We start with the approximations
\begin{eqnarray*} \nonumber
C_{0}(u)\approx w_0 F_{1,1}(u),\quad L_{0}(u)\approx w_1 F_{2,1}(u), \\
\text{and}\ R_{0}(u)\approx w_1 F_{2,2}(u). \qquad \qquad 
\end{eqnarray*}
where the nonnegative coefficients $w_0$ and $w_1$ are yet to be determined. First, we approximately enforce the equality~(\ref{eq:LC15}) through  integral values
\begin{eqnarray*}
\int_0^{1} u(1-u)\ud u=w_0^2 \int_{0}^{1} F_{1,1}(u)^2 \ud u \\ + w_1^2\int_{0}^{1} \left[ F_{2,1}(u)^2+F_{2,2}(u)^2\right]  \ud u.
\end{eqnarray*}
The weights of the functions $L_{0}(u)$ and $R_{0}(u)$ were taken equal for symmetry reasons.
Computing the integrals, one obtains the equation 
\[
2w_0^2+w_1^2=4,
\]
which implies $w\equiv w_0\in \left[0,\sqrt{2}\right]$ and $w_1= \sqrt{4-2w^2}$.
Next, we define the function
\[
r_w(u)= \left\{\frac{u(1-u)}{w^2F_{1,1}(u)^2+\left(4-2w^2\right)\left[F_{2,1}(u)^2+F_{2,2}(u)^2\right]}\right\}^{1/2}
\]
and set
\[C^{(w)}_{0}(u)= w r_w(u) F_{1,1}(u),\]
\[L^{(w)}_{0}(u)= \sqrt{4-2w^2} r_w(u) F_{2,1}(u),\]
and
\[R^{(w)}_{0}(u)= \sqrt{4-2w^2} r_w(u) F_{2,2}(u).\]
It is clear that  the above defined functions satisfy Eq.~(\ref{eq:LC15}) for all $w\in \left[0,\sqrt{2}\right]$. However, Eq.~(\ref{eq:LC16}) is  satisfied only for the roots of the function
\begin{eqnarray}
\label{eq:LC17} \nonumber
h(w)=\left[\int_0^{1} C^{(w)}_{0}(u)\ud u\right]^2  + \left[\int_0^{1} L^{(w)}_{0}(u)\ud u\right]^2 \\ + \left[\int_0^{1} R^{(w)}_{0}(u)\ud u\right]^2 -\frac{1}{12}, 
\end{eqnarray} 
if there are any. Numerical analysis shows that the equation $h(w)=0$ has precisely one solution on the interval $\left[0,\sqrt{2}\right]$. The first few digits of the solution are
\begin{equation}
\label{eq:LC18}
w_r=0.62258\ldots
\end{equation}
The functions  $C_0(u)=C^{(w_r)}_0(u)$, $L_0(u)=L^{(w_r)}_0(u)$, and $ R_0(u)=R^{(w_r)}_0(u)$ are plotted in Fig.~\ref{Fig:LC1}. From them, one generates the functions $C_{n,p}(u)$, $L_{n,p}(u)$, and $R_{n,p}(u)$ that make up the additional layers in the reweighted technique by dilatations (actually contractions) and translations, as shown by Eqs.~(\ref{eq:LC13}) and (\ref{eq:LC14}). 

I conclude this section by making a quite interesting observation. Using arguments similar to the ones employed in the proof of Theorem~2 of Ref.~(\onlinecite{Pre02b}), it is not difficult to prove that if $n=2^k-1$, then
\begin{eqnarray}
\label{eq:LC19}
\rho_n^{\text{RW}}(x,x';\beta)=\int_{\mathbb{R}}\ud x_1 \ldots \int_{\mathbb{R}}\ud x_n\; \rho_0^{\text{RW}}\left(x,x_1;\frac{\beta}{n+1}\right)\nonumber \\ \ldots \rho_0^{\text{RW}}\left(x_n,x';\frac{\beta}{n+1}\right),\qquad
\end{eqnarray}
where 
\begin{eqnarray}\nonumber
\label{eq:LC20}
\frac{\rho_0^{\text{RW}}(x,x';\beta)}{\rho_{fp}(x,x';\beta)}=\frac{1}{\left(2\pi\right)^{3/2}}\int_{\mathbb{R}}\int_{\mathbb{R}}\int_{\mathbb{R}} e^{-\left(a_1^2+a_2^2+a_3^2\right)/2}\\\times \exp\bigg\{-\beta \int_0^1 V[x+(x'-x)u+a_1\sigma C_0(u)\\ +a_2 \sigma L_0(u)+ a_3 \sigma R_0(u)] \ud u\bigg\} \ud a_1 \ud a_2 \ud a_3. \nonumber
\end{eqnarray}
However, for arbitrary $n$, the right-hand side of Eq.~(\ref{eq:LC19}) defines a new quantity which we shall denote by $\rho^{\text{DPI}}_n(x,x';\beta)$. The acronym DPI stands for discrete path integral methods and it is commonly used for methods obtained by Trotter composing some short-time approximation [in our case, the one given by Eq.~(\ref{eq:LC20})]. The results in the present section show that the subsequence $n=2^k-1$ of $\rho^{\text{DPI}}_n(x,x';\beta)$ has $o(1/n^2)$ asymptotic convergence for sufficiently smooth potentials. However, it is not farfetched to assume that $\rho^{\text{DPI}}_n(x,x';\beta)$ has in general $o(1/n^2)$ convergence. If true, this would be a quite surprising result because the conventional wisdom is that one cannot obtain an asymptotic convergence better than $O(1/n^2)$ for DPI methods without employing short-time approximations that depend upon the derivatives of the potential.

\begin{figure}[!tbp]
   \includegraphics[angle=270,width=8.5cm,clip=t]{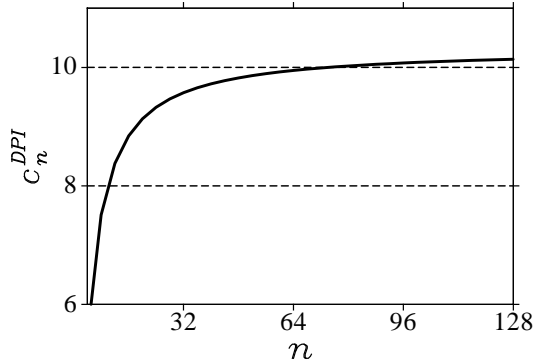}
%    \BoxedEPSF{Potential.eps scaled 300}
  \caption[sqr]
{\label{Fig:LC2}
The constant $c_n^{\text{DPI}}$ is seen to increase to a certain finite value, demonstrating the $O(1/n^3)$ convergence of the reweighed technique for the harmonic oscillator.}
\end{figure} 

We can test the validity of this assumption by studying the asymptotic convergence of the DPI method (and implicitly, that of the reweighted method) for the simple example of an harmonic oscillator. Let us see that Theorem~\ref{th:LC} predicts that the convergence of the reweighted technique is strictly better than $O(1/n^2)$ but it does not say how much better. The experience with the asymptotic convergence of the partial averaging suggests that if the potential has also second order Sobolev derivatives, then we might gain an order of magnitude in the asymptotic convergence. As such, we expect that the asymptotic convergence of the DPI method for the harmonic oscillator is $O(1/n^3)$. We study the convergence of the quantity
\[
c_n^{\text{DPI}} = n^3 \frac{Z_n^{\text{DPI}}(\beta)-Z(\beta)}{Z(\beta)}
\]
where $Z(\beta)$ is the exact partition function of the harmonic oscillator and $Z_n^{\text{DPI}}(\beta)$ is the $n$-th order DPI approximation to the density matrix. For the harmonic oscillator, the short-time approximation given by Eq.~(\ref{eq:LC20}) can be computed analytically because it is the exponential of a quadratic expression. Then, the Trotter composition shown by Eq.~(\ref{eq:LC19}) can be performed by the numerical matrix multiplication method.\cite{Kle73, Thi83}  The constants $c_n^{\text{DPI}}$ plotted in Fig.~\ref{Fig:LC2} were computed in atomic units  for an harmonic oscillator of mass $m_0=1$ and frequency $\omega=1$, at the inverse temperature $\beta=10$. The plot strongly suggests that the sequence $c_n^{\text{DPI}}$  converges to a finite positive constant, demonstrating the $O(1/n^3)$ convergence for our special DPI and reweighted techniques.

\subsection{A reweighted Wiener-Fourier path integral construction}

As argued in Ref.~(\onlinecite{Pre02}),  the Wiener-Fourier series representation of the Brownian bridge is expected to be  the optimal series representation as far as the asymptotic convergence of the primitive, partial averaging, and reweighted methods is concerned. This is clearly demonstrated by the asymptotic convergence of the partial averaging method for potentials having second order Sobolev derivatives. As shown by Theorem~4 of Ref.~(\onlinecite{Pre02a}), this convergence is 
\begin{eqnarray}
\label{eq:PAFPI}
\nonumber &&
\lim_{n \to \infty} n^3 
\left[\rho(x,x';\beta)-\rho_n^{\text{PA}}(x,x';\beta)\right]
\\ && = \frac{\hbar^2\beta^3}{3\pi^4 m_0}
\rho(x,x';\beta)\left[V'(x)^2+V'(x')^2\right]\\&&+\frac{\hbar^4\beta^4}{12\pi^4m_0^2} 
\int_0^1\left\langle x\Big|e^{-\beta \theta H} V''^2 e^{-\beta 
(1-\theta) H}\Big|x'\right\rangle \ud \theta.\nonumber
\end{eqnarray}
Our experience  with the L\'evy-Ciesielski series representation in the preceding section suggests that it might be possible to devise a Wiener-Fourier reweighted method that has $o(1/n^3)$ asymptotic convergence. However, my attempts of finding a reweighted method satisfying the relation
\[C_{\text{PA}}(x,x';\beta)=\tilde{C}_{\text{PA}}(x,x';\beta) \] were unsuccessful. Though it remains an open question, I believe that such a method does not exist.

In this paper,  I shall present a special Wiener-Fourier reweighted method (RW-WFPI) for which the term in Eq.~(\ref{eq:PAFPI}) that depends upon the second order derivative eventually disappears. Clearly, this is the best thing to enforce short of perfect cancellation. The resulting method will have more favorable dependence of the convergence constant with the inverse temperature ($\beta^3$ instead of $\beta^4$). Moreover, the convergence constant will depend only on the first order derivatives of the potential and I speculate that the $O(1/n^3)$ convergence extends to the $\cap_\alpha W^{1,2}_\alpha(\mathbb{R})$ class of potentials. 

 The reweighted method analyzed in this section is constructed as follows. First, remember that the Wiener-Fourier series is defined by the primitives
\begin{equation}
\label{eq:FPI}
\Lambda_k(u)= \sqrt{\frac{2}{\pi^2}} \frac{\sin(k\pi u)} {k},\quad k \geq 1. 
\end{equation} 
Next, let $q=4$ and $p=0$ in Eq.~(\ref{eq:12}) and define
\begin{equation}
\label{eq:42a}
r_n(u)=\sqrt{\frac{\gamma_n(u,u)}{{\gamma}^\circ_n(u,u)}}, 
\end{equation}
where
\begin{equation}
\label{eq:43}
{\gamma}_n^\circ(u,\tau)=\frac{2\alpha_n}{3n}\sum_{k=n+1}^{4n} \sin(k\pi u)\sin(k\pi \tau)
\end{equation}
and
\begin{equation}
\label{eq:44}
\alpha_n= \int_0^1 \gamma_n(u,u)\ud u=\frac{1}{\pi^2}\sum_{k=n+1}^\infty \frac{1}{k^2}=\frac{1}{6}-\frac{1}{\pi^2}\sum_{k=1}^n \frac{1}{k^2}. 
\end{equation}
Next, we let
\begin{equation}
\label{eq:45}
\tilde{\Lambda}_{n,k}(u)=\sqrt{\frac{2\alpha_n}{3n}}r_n(u)\sin(k\pi u),\quad n\geq 1, \; n<k \leq 4n.
\end{equation}
Then,
\begin{eqnarray*}
\tilde{\gamma}_n(u,\tau)= \sum_{k=n+1}^{4n}\tilde{\Lambda}_{n,k}(u)\tilde{\Lambda}_{n,k}(\tau) = \frac{2\alpha_n}{3n}r_n(u)r_n(\tau) \\ \times \sum_{k=n+1}^{4n}\sin(k\pi u) \sin(k\pi \tau)
\end{eqnarray*}
satisfies the equality
\[
\tilde{\gamma}_n(u,u)={\gamma}_n(u,u)\quad \forall\, u \in [0,1].
\]
Therefore, the method defined by Eqs.~(\ref{eq:12}), (\ref{eq:FPI}), and (\ref{eq:45}) is  a reweighted technique derived from the Wiener-Fourier series representation. While not the only  one, the particular RW-WFPI technique that we study has superior asymptotic convergence in a sense that will become clear from the expression of the convergence constant to be derived. 

We begin the convergence analysis by first considering the class of potentials $ \cap_\alpha W^{2,2}_\alpha (\mathbb{R})$, for which we anticipate an asymptotic convergence of the type $\mathit{O}(1/n^3)$. As such, the scope of this section is to compute the limit 
\[
\lim_{n \to \infty} n^3 \left[\rho_n^{\text{RW}}(x,x';\beta)-\rho_n^{\text{PA}}(x,x';\beta)\right]
\] with the help of  Theorem~\ref{th:2}. Unfortunately, this theorem is not given in a particularly useful form because the computation of the limit 
\begin{eqnarray}
\label{eq:46}
\lim_{n \to \infty} n^3 \frac{\beta^2}{2} \int_0^1 \! \ud u \! \int_0^1\! \ud \tau 
\tilde{\gamma}_n(u,\tau) K_{x,x'}^{\beta,n}(u,\tau)
\end{eqnarray}
is quite involved. As discussed in the preceding section, we can replace the function $\tilde{X}_n(x,x',\beta;\bar{a})$ appearing in the definition of $K_{x,x'}^{\beta,n}(u,\tau)$ [see Eq.~(\ref{eq:39})] with ${X}_n(x,x',\beta;\bar{a})$ without changing the asymptotic behavior of 
\[
\int_0^1 \! \ud u \! \int_0^1\! \ud \tau \tilde{\gamma}_n(u,\tau) K_{x,x'}^{\beta,n}(u,\tau).
\]
Doing so and noticing that [see Appendix~C of Ref.~(\onlinecite{Pre02a})] 
\begin{widetext}
\begin{eqnarray}\nonumber
\label{eq:47}
\mathbb{E}\left\{ 
{X}_{n}(x,x',\beta;\bar{a}) \mathbb{E}_n \left\{V'[x_r(u)+\sigma S_{u}^n(\bar{a})+ \sigma \tilde{B}_{u}^n(\bar{a})] 
V'[x_r(\tau)+\sigma S_{\tau}^n(\bar{a})+ \sigma \tilde{B}_{\tau}^n(\bar{a})]\right\}\right\}\\ = 
 \sum_{k=0}^\infty \frac{\sigma^{2k}}{k!}\tilde{\gamma}_n(u,\tau)^k\mathbb{E}\left\{ 
{X}_{n}(x,x',\beta;\bar{a})  \overline{V}^{(k+1)}_{n,u}[x_r(u)+\sigma S_{u}^n(\bar{a})]\overline{V}^{(k+1)}_{n,\tau}[x_r(\tau)+\sigma S_{\tau}^n(\bar{a})]\right\},
\end{eqnarray}
\end{widetext} 
we obtain
\begin{eqnarray}
\nonumber &&
\label{eq:48}
\int_0^1 \! \ud u \! \int_0^1\! \ud \tau \tilde{\gamma}_n(u,\tau) K_{x,x'}^{\beta,n}(u,\tau)   \approx 
\sigma^2\int_0^1  \int_0^1 \sum_{k=0}^\infty \frac{\sigma^{2k}}{k!} 
\\ && \nonumber \times \tilde{\gamma}_n(u,\tau)^{k+1} \mathbb{E}\Big\{ 
{X}_{n}(x,x',\beta;\bar{a})  \overline{V}^{(k+1)}_{n,u}[x_r(u)+\sigma S_{u}^n(\bar{a})]
\\ && \times \overline{V}^{(k+1)}_{n,\tau}[x_r(\tau)+\sigma S_{\tau}^n(\bar{a})]\Big\}\ud u \ud \tau.
\end{eqnarray}
Because the kernels $\tilde{\gamma}_n(u,\tau)^{k}$ are positive definite for all $k \geq 1$, the terms of the series appearing in Eq.~(\ref{eq:48}) are positive. We now argue that we can truncate the series to the first two terms without changing the asymptotic behavior, at least as far as the $\mathit{O}(1/n^3)$ convergence is concerned. That is, we want to prove that
\begin{eqnarray}
\label{eq:49}\nonumber &&
\int_0^1 \! \ud u \! \int_0^1\! \ud \tau \tilde{\gamma}_n(u,\tau) K_{x,x'}^{\beta,n}(u,\tau) \approx 
\sigma^2\int_0^1  \int_0^1 \sum_{k=0}^1 \frac{\sigma^{2k}}{k!}
 \\ && \nonumber \times \tilde{\gamma}_n(u,\tau)^{k+1} \mathbb{E}\Big\{ 
{X}_{n}(x,x',\beta;\bar{a})  \overline{V}^{(k+1)}_{n,u}[x_r(u)+\sigma S_{u}^n(\bar{a})]\\ && \times \overline{V}^{(k+1)}_{n,\tau}[x_r(\tau)+\sigma S_{\tau}^n(\bar{a})]\Big\}\ud u \ud \tau+o\left(1/n^3\right).
\end{eqnarray}
 
We have the estimate 
\begin{widetext}
\begin{eqnarray}
\label{eq:50}\nonumber &&
\sigma^2\sum_{k=2}^\infty \frac{\sigma^{2k}}{k!} \int_0^1 \! \ud u \! \int_0^1\! \ud \tau \tilde{\gamma}_n(u,\tau)^{k+1}\mathbb{E} \left\{X_n(x,x',\beta;\bar{a})
  \overline{V}^{(k+1)}_{n,u}[x_r(u)+\sigma S_{u}^n(\bar{a})]\overline{V}^{(k+1)}_{n,\tau}[x_r(\tau)+\sigma S_{\tau}^n(\bar{a})]\right\}
  \\ && \leq
\sigma^4  \sum_{k=1}^\infty \frac{\sigma^{2k}}{k!} \int_0^1 \! \ud u \! \int_0^1\! \ud \tau \tilde{\gamma}_n(u,\tau)^{k+2} \mathbb{E} \left\{X_n(x,x',\beta;\bar{a})
  \overline{V}^{(k+2)}_{n,u}[x_r(u)+\sigma S_{u}^n(\bar{a})]\overline{V}^{(k+2)}_{n,\tau}[x_r(\tau)+\sigma S_{\tau}^n(\bar{a})]\right\} 
 \\ && = \nonumber
  \sigma^2 \int_0^1 \! \ud u \! \int_0^1\! \ud \tau \tilde{\gamma}_n(u,\tau)^{2} \Xi_n(u,\tau),
\end{eqnarray}
\end{widetext}
where the function 
\begin{widetext}
\begin{eqnarray*}\nonumber
 \Xi_n(u,\tau)=\mathbb{E}\bigg[ X_n(x,x',\beta;\bar{a})\bigg(\mathbb{E}_n \left\{
V''[x_r(u)+\sigma S_{u}^n(\bar{a})+ \sigma \tilde{B}_{u}^n(\bar{a})] V''[x_r(\tau)+\sigma S_{\tau}^n(\bar{a})+ \sigma \tilde{B}_{\tau}^n(\bar{a})] \right\} \\ -\left\{\mathbb{E}_n 
V''[x_r(u)+\sigma S_{u}^n(\bar{a})+ \sigma \tilde{B}_{u}^n(\bar{a})]\right\} \left\{\mathbb{E}_n V''[x_r(\tau)+\sigma S_{\tau}^n(\bar{a})+ \sigma \tilde{B}_{\tau}^n(\bar{a})]\right\}\bigg)\bigg]
\end{eqnarray*}
\end{widetext}
has the property 
\[
\lim_{n \to \infty} \Xi_n(u,\tau) =0. 
\]
Using this relation together with Eq.~(\ref{eq:D1}), we deduce that the last term of Eq.~(\ref{eq:50}) has the property
\begin{eqnarray*}
\lim_{n \to \infty} n^3 \sigma^2 \int_0^1 \! \ud u \! \int_0^1\! \ud \tau \tilde{\gamma}_n(u,\tau)^{2} \Xi_n(u,\tau) \\ = \frac{\sigma^2}{3\pi^4}\lim_{n \to \infty} \int_0^1\Xi_n(u,u)\ud u=0
\end{eqnarray*}
and therefore, Eq.~(\ref{eq:49}) is proved. 

From Eq.~(\ref{eq:49}) and again Eq.~(\ref{eq:D1}), we learn that 
\begin{eqnarray}
\label{eq:51}\nonumber &&
\lim_{n \to \infty} n^3 \int_0^1 \! \ud u \! \int_0^1\! \ud \tau \tilde{\gamma}_n(u,\tau) K_{x,x'}^{\beta,n}(u,\tau)=I(x,x';\beta) 
\\ && \nonumber +\frac{\sigma^4}{3\pi^4}\int_0^1 \mathbb{E}\Big\{ 
{X}_{\infty}(x,x',\beta;\bar{a})  {V}''[x_r(u) +\sigma B_{u}^0(\bar{a})]^2\Big\} \ud u \\ &&
 = I(x,x';\beta)+ \frac{\sigma^4}{3\pi^4} \int_0^1 \rho(x,y;u\beta)
 \\ && \qquad \qquad \qquad \times \rho[y,x'; (1-u)\beta]V''(y)^2 \ud u, \nonumber
\end{eqnarray}
where
\begin{eqnarray*} &&
I(x,x';\beta)=\lim_{n \to \infty}  n^3
\sigma^2\int_0^1  \int_0^1 \tilde{\gamma}_n(u,\tau)
\\ && \times \mathbb{E}\Big\{ 
{X}_{n}(x,x',\beta;\bar{a})  \overline{V}'_{n,u}[x_r(u)+\sigma S_{u}^n(\bar{a})]
\\ && \times \overline{V}'_{n,\tau}[x_r(\tau)+\sigma S_{\tau}^n(\bar{a})]\Big\}\ud u \ud \tau.
\end{eqnarray*} 

Now, let us consider the function 
\begin{eqnarray}
\label{eq:52}\nonumber
K_{x,x'}^\beta(u,\tau)=\sigma^2\mathbb{E}\Big\{ 
X_{\infty}(x,x',\beta;\bar{a})V'[x_r(u)+\sigma B_{u}^0(\bar{a})]\\ 
\quad \times V'[x_r(\tau)+\sigma B_{\tau}^0(\bar{a})]\Big\}.\quad
\end{eqnarray}
[this is Eq.~(35) of  Ref.~(\onlinecite{Pre02a})]. Proceeding in a way similar to  the case of the related functions $K_{x,x'}^{\beta,n} (u,\tau)$, one may replace $X_{\infty} (x,x',\beta;\bar{a})$ with $X_{n} (x,x',\beta;\bar{a})$, use the fact that $\tilde{\gamma}_n(u,\tau)\gamma_n(u,\tau)^k$ are positive definite kernels for all $k \geq 0$, and employ Eq.~(\ref{eq:D2}) of Appendix~B to justify the equality
\begin{eqnarray}
\label{eq:53}\nonumber &&
\int_0^1 \! \ud u \! \int_0^1\! \ud \tau \tilde{\gamma}_n(u,\tau) K_{x,x'}^{\beta}(u,\tau) \approx 
\sigma^2\int_0^1  \int_0^1 \sum_{k=0}^1 \frac{\sigma^{2k}}{k!}
 \\ &&\times \tilde{\gamma}_n(u,\tau){\gamma}_n(u,\tau)^{k} \mathbb{E}\Big\{ 
{X}_{n}(x,x',\beta;\bar{a})  \overline{V}^{(k+1)}_{n,u}[x_r(u) \\ && +\sigma S_{u}^n(\bar{a})]\overline{V}^{(k+1)}_{n,\tau}[x_r(\tau)+\sigma S_{\tau}^n(\bar{a})]\Big\}\ud u \ud \tau+o\left(1/n^3\right). \nonumber \qquad
\end{eqnarray}
From Eqs.~(\ref{eq:53}) and (\ref{eq:D2}), one obtains
\begin{eqnarray}
\label{eq:54} \nonumber 
\lim_{n \to \infty} n^3 \int_0^1 \! \ud u \! \int_0^1\! \ud \tau \tilde{\gamma}_n(u,\tau) K_{x,x'}^{\beta}(u,\tau) = I(x,x';\beta) \\ + \frac{\sigma^4}{4\pi^4} \int_0^1 \rho(x,y;u\beta)\rho[y,x'; (1-u)\beta]V''(y)^2. \qquad
\end{eqnarray}
We eliminate the unknown functions $I(x,x';\beta)$ from Eqs.~(\ref{eq:51}) and (\ref{eq:54}) to obtain 
\begin{eqnarray}
\label{eq:55}\nonumber &&
\lim_{n \to \infty} n^3 \int_0^1 \! \ud u \! \int_0^1\! \ud \tau \tilde{\gamma}_n(u,\tau) K_{x,x'}^{\beta,n}(u,\tau)
 \\ && =\lim_{n \to \infty} n^3 \int_0^1 \! \ud u \! \int_0^1\! \ud \tau \tilde{\gamma}_n(u,\tau) K_{x,x'}^{\beta}(u,\tau) 
 \\ && \nonumber + \frac{\sigma^4}{12\pi^4} \int_0^1 \rho(x,y;u\beta)\rho[y,x'; (1-u)\beta]V''(y)^2.
\end{eqnarray}
With the help of the this last equation, Theorem~\ref{th:2}, and Eq.~(\ref{eq:D1}), we learn that
\begin{eqnarray}&&
\label{eq:56}\nonumber
\lim_{n \to \infty} n^3 \left[\rho_n^{\text{RW}}(x,x';\beta)-\rho_n^{\text{PA}}(x,x';\beta)\right]
\\ &&= \frac{\beta^2}{2}\lim_{n \to \infty} n^3 \int_0^1 \! \ud u \! \int_0^1\! \ud \tau \tilde{\gamma}_n(u,\tau) K_{x,x'}^{\beta}(u,\tau)
\\ && \nonumber - \frac{\beta^2 \sigma^4}{24 \pi^4}\int_0^1 \rho(x,y;u\beta)\rho[y,x'; (1-u)\beta]V''(y)^2.
\end{eqnarray}

The remainder of this section deals with the limit 
\begin{equation}
\label{eq:57}
\lim_{n \to \infty} n^3 \int_0^1 \! \ud u \! \int_0^1\! \ud \tau \tilde{\gamma}_n(u,\tau) K_{x,x'}^{\beta}(u,\tau).
\end{equation} 
As discussed in Section~IV.B of Ref.~(\onlinecite{Pre02a}), the integral appearing in this limit can be written as 
\[
\int_0^1 \! \ud u \! \int_0^u\! \ud \tau \tilde{\gamma}_n(u,\tau) K(u,\tau),
\] 
where 
\[K(u,\tau)=K_{x,x'}^{\beta}(u,\tau)+K_{x',x}^{\beta}(u,\tau).\]
The limit in Eq.~(\ref{eq:57}) becomes
\begin{eqnarray*} 
\lim_{n \to \infty} \frac{2\alpha_n n^2}{3}\sum_{k=n+1}^{4n} \int_0^1 \! \ud u \! \int_0^u\! \ud \tau K(u,\tau)r_n(u)r_n(\tau)\\ \times \sin(k \pi u)  \sin(k  \pi \tau)= \lim_{n \to \infty} \frac{2 n}{3\pi^2} \sum_{k=n+1}^{4n} \int_0^1 \! \ud u \! \int_0^u\! \ud \tau \\ \times K(u,\tau) f_{n,k}(u)f_{n,k}(\tau),
\end{eqnarray*}
where we  used the fact that  the product $n\alpha_n$ converges to $1/\pi^2$ [see Eq.~(\ref{eq:C7}) of Appendix~A] and introduced the functions 
\[
f_{n,k}(u)= r_n(u)\sin(k\pi u)
\][see Eq.~(\ref{eq:C1}) of Appendix~A]. 

In these conditions, Theorem~\ref{th:C8} of Appendix~A shows that the limit in Eq.~(\ref{eq:57}) is
\begin{eqnarray*}&&
\nonumber
\frac{1}{3\pi^4}\left\{\int_{1/4}^1 [1+H(a)]^2 \ud a\right\} [K(0,0)+K(1,1)]\\&& -\frac{1}{8\pi^4}[K(1,1)-K(0,0)]+\frac{1}{4\pi^4} \int_0^1 \frac{\partial}{\partial \tau} K(u,\tau)\Big|_{\tau=u} \ud u,
\end{eqnarray*}
where the function $H(a)$ is defined by Eq.~(\ref{eq:C56}). 
Now, Eq.~(68) of Ref.~(\onlinecite{Pre02a}) says
\[
K(0,0)=K(1,1)=\sigma^2\rho(x,x';\beta)\left[V'(x)^2+V'(x')^2\right],
\]
while Eq.~(72) of the same reference shows that
\begin{eqnarray*}
&&
\int_0^1 \frac{\partial}{\partial \tau}K(u,\tau)\bigg|_{\tau=u} \ud u 
= 
\frac{\hbar^2\sigma^2\beta}{m_0}\\&& \times \nonumber 
\int_0^1\int_{\mathbb{R}}\rho(x,y;u\beta)\rho[y,x';(1-u)\beta] 
V''(y)^2 \ud y \ud u.
\end{eqnarray*}
The limit in Eq.~(\ref{eq:57}) becomes
\begin{eqnarray*}
\nonumber
\frac{2\sigma^2}{3\pi^4}\left\{\int_{1/4}^1 [1+H(a)]^2 \ud a\right\} \rho(x,x';\beta)\left[V'(x)^2+V'(x')^2\right]\\+\frac{\hbar^2 \sigma^2 \beta}{4\pi^4 m_0} \int_0^1\int_{\mathbb{R}}\rho(x,y;u\beta)\rho[y,x';(1-u)\beta] 
V''(y)^2 \ud y \ud u.
\end{eqnarray*}

Replacing the last expression in Eq.~(\ref{eq:56}) and performing some straightforward simplifications, we conclude
\begin{eqnarray*}
\nonumber &&
\lim_{n \to \infty} n^3 
\left[\rho_n^{\text{RW}}(x,x';\beta)-\rho_n^{\text{PA}}(x,x';\beta)\right]
 = \frac{\hbar^2\beta^3}{3\pi^4 m_0}\\&& \times  \left\{\int_{1/4}^1 [1+H(a)]^2 \ud a\right\}
\rho(x,x';\beta)\left[V'(x)^2+V'(x')^2\right]\nonumber \\&&+\frac{\hbar^4\beta^4}{12\pi^4m_0^2} 
\int_0^1\left\langle x\Big|e^{-\beta \theta H} V''^2 e^{-\beta 
(1-\theta) H}\Big|x'\right\rangle \ud \theta .
\end{eqnarray*}
Combining this result with Eq.~(\ref{eq:PAFPI}), we end up with the following theorem. 
\begin{3}
\label{th:3}
Assume $V(x)\in \cap_{\alpha}W^{2,2}_\alpha(\mathbb{R})$ is a Kato-class potentials satisfying the hypothesis of Proposition~\ref{Pr:2}. Then,
\begin{eqnarray}
\label{eq:58}\nonumber 
\lim_{n \to \infty} n^3 
\left[\rho(x,x';\beta)-\rho_n^{\text{RW}}(x,x';\beta)\right]
 = \frac{\hbar^2\beta^3}{N_{0}\pi^4 m_0}\\ \times  
\rho(x,x';\beta)\left[V'(x)^2+V'(x')^2\right], \qquad
\end{eqnarray}
where 
\begin{equation}
\label{eq:59}
N_{0}=3 \left/ \left\{1-\int_{1/4}^1 [1+H(a)]^2 \ud a\right\}\right. .
\end{equation}
\end{3}
The integral in Eq.~(\ref{eq:59}) has been computed numerically, the resulting value of the constant  $N_{0}$ being
\begin{equation}
\label{eq:60}
N_{0}\approx 15.0045.
\end{equation}

In the case of a $d$-dimensional system, the statement of Theorem~\ref{th:3} becomes 
\begin{eqnarray}
\label{eq:61}\nonumber 
\lim_{n \to \infty} n^3 
\left[\rho(x,x';\beta)-\rho_n^{\text{RW}}(x,x';\beta)\right]
 = \frac{\hbar^2\beta^3}{N_{0}\pi^4}\\ \times  
\rho(x,x';\beta)\left\{\sum_{i=1}^d \frac{[\partial_i V(x)]^2+[\partial_i V(x')]^2}{m_{0,i}}\right\}, 
\end{eqnarray}
where $\partial_i V(x)$ denotes the first order partial derivative $\partial V(x)/ \partial x_i$. For a better understanding of this formula, the reader is advised to compare it with Eq.~(76) of Ref.~(\onlinecite{Pre02a}). 

We conclude this section by numerically verifying the findings of 
Theorem~\ref{th:3} for the simple case of an harmonic oscillator. In fact, we 
shall verify the following corollary, which is 
a direct consequence of Theorem~\ref{th:3}.
\begin{6}
\[
\lim_{n \to \infty} n^3 
\frac{Z(\beta)-Z^{\text{RW}}_n(\beta)}{Z(\beta)}=\frac{\int_{\mathbb{R}} 
\rho(x;\beta)\Delta Z(x;\beta)\ud x}{\int_{\mathbb{R}} 
\rho(x;\beta)\ud x},
\]
where
\[\Delta Z(x;\beta)=\frac{2\hbar^2\beta^3}{N_0\pi^4 m_0} 
V'(x)^2.\]
\end{6}
 Corollary~1 gives an estimate for the relative error of the 
partition function as an average of a suitable estimating function. 
In practice, such an average can be evaluated by Monte Carlo 
integration if so desired. 

The harmonic oscillator we study is described by the quadratic  potential $V(x)=m_0\omega^2x^2/2$. 
The computations are performed in atomic units for a particle of mass 
$m_0=1$ and for the frequency $\omega=1$. The inverse temperature is 
set to $\beta=10$. Since the density matrix for an harmonic oscillator is analytically 
known,\cite{Fey94} the convergence constant for the relative error of 
the partition function can be evaluated directly from Corollary~1 to 
be
\[
c_{\text{RW}}=\frac{\int_{\mathbb{R}} \rho(x;\beta)\Delta Z(x;\beta)\ud 
x}{\int_{\mathbb{R}} \rho(x;\beta)\ud x}=0.684.
\]
I remind the reader that, according to Eq.~(\ref{eq:25}), the actual convergence constant as measured against the number of variables used to parameterize the paths is $4^3 \cdot 0.684\approx 43.8$. 

We can also evaluate the convergence constant $c_{\text{RW}}$ by studying the limit of the sequence
\[
c^{\text{RW}}_n=n^3 \frac{Z(\beta)-Z^{\text{RW}}_n(\beta)}{Z(\beta)}.
\]
The computation of the terms $Z^{\text{RW}}_n(\beta)$ is discussed in Appendix~C.
Because the even and the odd sequences $c^{\text{RW}}_{2n}$ and 
$c^{\text{RW}}_{2n+1}$ have a slightly different asymptotic behavior that 
generates an oscillatory pattern in plots, we  plot them 
separately.
\begin{figure}[!tbp]
   \includegraphics[angle=270,width=8.5cm,clip=t]{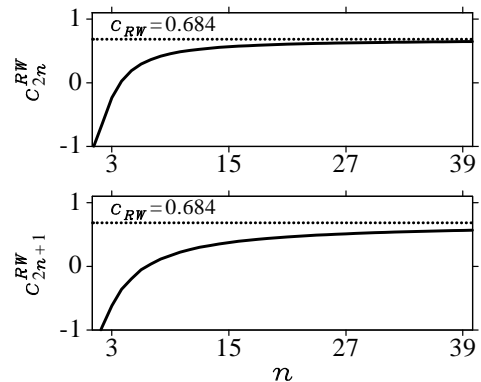}
%    \BoxedEPSF{Potential.eps scaled 300}
  \caption[sqr]
{\label{Fig:1}
Both the even sequence  $c^{\text{RW}}_{2n}$ and the odd sequence 
$c^{\text{RW}}_{2n+1}$ are seen  to converge to  $c_{\text{RW}}=0.684$, which is 
the value predicted by Corollary~1.
}
\end{figure}
In these conditions, Corollary~1 says that
\[
c_{\text{RW}}= \lim_{n \to \infty} c^{\text{RW}}_n
\]
and the prediction is well verified for the harmonic oscillator, as 
Fig~\ref{Fig:1} shows. I regard this numerical example as strong 
evidence that the analysis performed in the present paper is correct. 
It is hard to believe that the exact agreement between theory 
and numerical analysis for the harmonic oscillator is accidental, especially in the view of the complicated definitional expression for the constant $N_0$ [see Eq.~(\ref{eq:59})]. 

\section{Summary and Discussion}

In this paper, I have defined and characterized the reweighted methods, which are  computational techniques used in conjunction with the random series implementation of path integral simulations.  I have also analyzed the class of potentials for which the methods are expected to converge and I stated some general convergence theorems of use for arbitrary basis. Using the developed theory, I have constructed two special reweighted techniques starting with the L\'evy-Ciesielski and Wiener-Fourier basis, respectively. The methods were shown to have superior asymptotic behavior as follows. The reweighted L\'evy-Ciesielski method has $o(1/n^2)$ asymptotic convergence for potentials having first order Sobolev derivatives and it preserves the favorable $\propto \log(n)$ scaling for the time necessary to evaluate a path at a given discretization point. For potentials having second order Sobolev derivatives, I conjecture that the asymptotic convergence of the reweighted L\'evy-Ciesielski method is $O(1/n^3)$.  The reweighted Wiener-Fourier method was shown to have $O(1/n^3)$ convergence for potentials having second order Sobolev derivatives. The convergence constant has more favorable dependence with the temperature and with the derivatives of the potentials as compared to the constant for the corresponding partial averaging method. Since the expression of the convergence constant does not actually depend  on the second order derivatives of the potentials, it is suggested that the $O(1/n^3)$ convergence extends to the class of potentials having first order derivatives only. 

The examples analyzed in the present paper suggest that in general for any basis one may construct a  reweighted technique which has an asymptotic convergence equal to or faster than that of the partial averaging method. More important, this fast asymptotic convergence is achieved with the help of some special constructions of the Brownian bridge that enters the Feynman-Ka\c{c} formula and not by modifying the potentials. This make the numerical algorithms fast and straightforward  to implement. The main conclusion of this paper is that there are path integral techniques capable of  reaching  $O(1/n^3)$ asymptotic convergence without any recourse to effective potentials.

\begin{acknowledgments}
The author acknowledges support from the National Science Foundation through 
awards CHE-0095053 and CHE-0131114.  He  would also like to
thank Professor J. D. Doll for continuing discussions concerning the present developments.
\end{acknowledgments}

\appendix
\section{}
In Section IV, we introduced the  functions
\begin{equation}
\label{eq:C1}
f_{n,k}(u)=r_n(u)\sin(k\pi u),\quad n\geq 1, \; n<k \leq 4n,
\end{equation}
where 
\begin{equation}
\label{eq:C2}
r_n(u)=\sqrt{\frac{\gamma_n(u,u)}{{\gamma}_n^\circ(u,u)}}.
\end{equation}
The functions $\gamma_n(u,\tau)$ and ${\gamma}_n^\circ(u,\tau)$ are defined as follows:
\begin{eqnarray}
\label{eq:C3}&&
\gamma_n(u,\tau)=\frac{2}{\pi^2}\sum_{k=n+1}^{\infty}\frac{ \sin(k\pi u)\sin(k\pi \tau)}{k^2}\nonumber \\&&=
\min(u,\tau)-u\tau - \frac{2}{\pi^2}\sum_{k=1}^{n}\frac{ \sin(k\pi u)\sin(k\pi \tau)}{k^2}\quad
\end{eqnarray}
and
\begin{equation}
\label{eq:C4}
{\gamma}^\circ_n(u,\tau)=\frac{2\alpha_n}{3n}\sum_{k=n+1}^{4n} \sin(k\pi u)\sin(k\pi \tau),
\end{equation}
respectively. The sequence $\alpha_n$ appearing in the previous formula is 
\begin{equation}
\label{eq:C5}
\alpha_n= \int_0^1 \gamma_n(u,u)\ud u=\frac{1}{\pi^2}\sum_{k=n+1}^\infty \frac{1}{k^2}=\frac{1}{6}-\frac{1}{\pi^2}\sum_{k=1}^n \frac{1}{k^2}. 
\end{equation}

The purpose of this appendix is to study the asymptotic properties of the functions $f_{n,k}(u)$, which are used in the construction of the reweighted method.  In particular, we are interested in establishing an integration by parts formula. Before giving the main results of the section, it is convenient to rearrange the expression for the functions $r_n(u)$ as follows. We write
\begin{eqnarray}\nonumber 
\label{eq:C5a}
\gamma_{n}(u,u)=\frac{2}{\pi^2}\sum_{k=n+1}^{\infty}\frac{ \sin(k\pi u)^2}{k^2}=
\frac{1}{\pi^2}\sum_{k=n+1}^{\infty}\frac{ 1}{k^2}\\-\frac{1}{\pi^2}\sum_{k=n+1}^{\infty}\frac{ \cos(2k\pi u)}{k^2}=\alpha_n[1-g_n(u)],
\end{eqnarray}
where
\[g_n(u)=\frac{1}{\pi^2\alpha_n}\sum_{k=n+1}^{\infty}\frac{ \cos(2k\pi u)}{k^2}.\]
Similarly, we have
\begin{eqnarray}
\label{eq:C5b}
{\gamma}_{n}^\circ(u,u)=\frac{2\alpha_n}{3n}\sum_{k=n+1}^{4n} \sin(k\pi u)^2=\alpha_n[1-{g}^\circ_n(u)],
\end{eqnarray}
where
\[{g}^\circ_n(u)=\frac{1}{3n}\sum_{k=n+1}^{4n}{ \cos(2k\pi u)}.\]
Therefore, 
\begin{equation}
\label{eq:C6}
r_n(u)=\sqrt{\frac{1-g_n(u)}{1-{g}^\circ_n(u)}}.
\end{equation}

Using Eq.~(\ref{eq:C6}), one can prove the following proposition.
\begin{C1}
 $r_n(u) \to 1$  on $(0,1)$.
\end{C1}

\emph{Proof.} By letting $n\to \infty$ in the sequence of inequalities
\begin{eqnarray*}\frac{n}{n+1}=n\int_{n+1}^\infty x^{-2}\ud x \leq n\sum_{k=n+1}^\infty \frac{1}{k^2}\\ \leq n\int_{n+1}^\infty x^{-2}\ud x =1,\end{eqnarray*}
one readily proves
\begin{equation}
\label{eq:C7}
\lim_{n \to \infty}n \pi^2 \alpha_n =1 \ \text{and}\ \frac{1}{\pi^2(n+1)}\leq \alpha_n \leq \frac{1}{\pi^2 n}.
\end{equation}
Now, 
\begin{eqnarray*}
\sin(\pi u) \sum_{k=n+1}^{\infty}\frac{ \cos(2k\pi u)}{k^2}=\frac{1}{2} \sum_{k=n+1}^{\infty}\frac{\sin[(2k+1)\pi u]}{k^2} \nonumber \\- \frac{1}{2}\sum_{k=n+1}^{\infty}\frac{\sin[(2k-1)\pi u]}{k^2}=-\frac{1}{2}\frac{\sin[(2n+1)\pi u]}{(n+1)^2} \quad  \\ \nonumber +\frac{1}{2}\sum_{k=n+1}^{\infty}\left[\frac{1}{k^2}-\frac{1}{(k+1)^2} \right]\sin[(2k+1)\pi u]
\end{eqnarray*}
and therefore, 
\begin{eqnarray}
\label{eq:C8}\nonumber
\left|\sin(\pi u) \sum_{k=n+1}^{\infty}\frac{ \cos(2k\pi u)}{k^2}\right|\leq \frac{1}{2}\frac{1}{(n+1)^2} \quad  \\  +\frac{1}{2}\sum_{k=n+1}^{\infty}\left[\frac{1}{k^2}-\frac{1}{(k+1)^2} \right]= \frac{1}{(n+1)^2}. 
\end{eqnarray}
From Eqs.~(\ref{eq:C7}) and (\ref{eq:C8}) we learn that 
\begin{equation}
\label{eq:C9}
|g_n(u)|\leq \frac{1}{\pi^2 \alpha_n (n+1)^2}\frac{1}{\sin(\pi u)}\leq \frac{1}{n}\frac{1}{\sin( \pi u)}.
\end{equation}
On the other hand, we have
\begin{eqnarray*}&&
\sin(\pi u) \sum_{k=n+1}^{4n}{ \cos(2k\pi u)}\\&&= \frac{1}{2}\sum_{k=n+1}^{4n}\Big\{ \sin[(2k+1)\pi u]-\sin[(2k-1)\pi u]\Big\}\\&&=\frac{1}{2}\left\{ \sin[(8n+1)\pi u]-\sin[(2n+1)\pi u]\right\}
\end{eqnarray*}
and therefore,
\begin{eqnarray*}
\left|\sum_{k=n+1}^{4n}{ \cos(2k\pi u)}\right|\leq  \frac{1}{\sin(\pi u)}.
\end{eqnarray*}
The last inequality  implies 
\begin{equation}
\label{eq:C10}
|{g}^\circ_n(u)|\leq  \frac{1}{3n}\frac{1}{\sin(\pi u)}
\end{equation}
for all $u\in(0,1)$. 
Equations (\ref{eq:C9}) and (\ref{eq:C10}) readily imply  
\begin{equation}
\label{eq:C11}
\lim_{n \to \infty} r_n(u)= 1, \quad \forall\, u\in (0,1) 
\end{equation}
and the proof of the proposition is concluded. \hspace{\stretch{1}} $\Box$

\emph{Observation.}
The technique utilized in establishing the bounds given by Eqs.~(\ref{eq:C9}) and (\ref{eq:C10}) can be used to establish a similar bound for the series
\[
\sum_{k=n+1}^\infty \frac{\sin(2 k \pi u)}{k},
\] 
which will be needed later. More precisely, we have
\begin{eqnarray*}
\sin(\pi u) \sum_{k=n+1}^\infty \frac{\sin(2 k \pi u)}{k}= \frac{1}{2}\sum_{k=n+1}^\infty \frac{\cos[(2k-1)\pi u]}{k}\\ -\frac{1}{2}\sum_{k=n+1}^\infty \frac{\cos[(2k+1)\pi u]}{k}=\frac{1}{2}\frac{\cos[(2n+1)\pi u]}{n+1}\\-\frac{1}{2}\sum_{k=n+1}^\infty\left[\frac{1}{k}-\frac{1}{k+1}\right] \cos[(2k+1)\pi u].
\end{eqnarray*}
This equality implies 
\begin{eqnarray*}
\sin(\pi u)\left| \sum_{k=n+1}^\infty \frac{\sin(2 k \pi u)}{k}\right|\leq  \frac{1}{2}\frac{1}{n+1}\\+\frac{1}{2}\sum_{k=n+1}^\infty\left[\frac{1}{k}-\frac{1}{k+1}\right]= \frac{1}{n+1},
\end{eqnarray*}
from which 
\begin{equation}
\label{eq:C12}
\left| \sum_{k=n+1}^\infty \frac{\sin(2 k \pi u)}{k}\right|\leq \frac{1}{n+1}\frac{1}{\sin(\pi u)}.
\end{equation}

The convergence of  $r_n(u)$ to $1$ is not uniform. In fact,  the functions $r_n(u)$ present singularities of the form $u^{-1/2}$ at the end points $0$ and $1$. This can be proved as follows. From Eq.~(\ref{eq:C3}) we obtain for $u$ small
\begin{eqnarray*}
\gamma_n(u,u)=u-u^2 - \frac{2}{\pi^2}\sum_{k=1}^{n}\frac{ \sin(k\pi u)^2}{k^2} \approx u.
\end{eqnarray*}
On the other hand, from Eq.~(\ref{eq:C4}), we have
\begin{eqnarray*}
{\gamma}_n^\circ(u,u)=\frac{2\alpha_n}{3n}\sum_{k=n+1}^{4n} \sin(k\pi u)^2  \approx \frac{2\alpha_n}{3n}\pi^2\left(\sum_{k=n+1}^{4n} k^2 \right)u^2
\end{eqnarray*}
and from Eq.~(\ref{eq:C2}), we deduce
\[r_n(u)\approx \left[{2\alpha_n \pi^2 \left(\sum_{k=n+1}^{4n} k^2 \right)}\Big/(3n)\right]^{-1/2} u^{-1/2}\]
for $u$ small enough. The situation is similar at the other end point and follows by symmetry [trivially, $r_n(u)=r_n(1-u)$]. However, the functions $f_{n,k}(u)$ themselves have no discontinuities because they clearly behave like 
$u^{1/2}$ near the end point $0$ and like $(1-u)^{1/2}$ near the end point $1$. 

The next few propositions in this appendix help establish Proposition~\ref{Pr:C6} which characterizes  the asymptotic behavior of the primitives
\begin{equation}
\label{eq:C13}
G_{n,k}(t)=\int_0^t [r_n(u)-1]\sin(k\pi u)\ud u. 
\end{equation}
It is also helpful to introduce the functions
\[\phi_{n,k}(u)=[g_n(u)-{g}^\circ_n(u)]\sin(k\pi u)\]
and
\[\psi_{n,k}(u)=\left\{r_n(u)-1+\frac{1}{2}\left[g_n(u)-{g}^\circ_n(u)\right]\right\}\sin(k\pi u).\]
 The primitives of these two functions will be denoted by 
\[ \Phi_{n,k}(t)=\int_0^t \phi_{n,k}(u)\ud u \ \text{and} \ \Psi_{n,k}(t)=\int_0^t \psi_{n,k}(u)\ud u,\]
respectively. Therefore, we have 
\begin{equation}
\label{eq:C14}
G_{n,k}(t)= -\frac{1}{2}\Phi_{n,k}(t)+\Psi_{n,k}(t).
\end{equation}
Finally, let us notice that 
\begin{equation}
\label{eq:C14a}
 f_{n,k}(u)=\sin(k\pi u) -\frac{1}{2} \phi_{n,k}(u)+\psi_{n,k}(u)
\end{equation}
and that the primitives of $\sin(k\pi u)$ are known analytically.
 
 Symmetry arguments which shall be invoked at the end of the appendix show that it is enough to establish the asymptotic behavior of these primitives on the interval $(0,1/2]$ and therefore, we shall turn our attention to this problem. We begin by analyzing the behavior of the primitives $\Phi_{n,k}(t)$ on the interval $(0,1/2]$ and for this purpose, we first need to prove the following proposition.
\begin{C2}
\label{Pr:C2}
 Let $q_i\to \infty$ and $p_i\to \infty$ be two sequences of natural numbers such that $q_i/p_i \to a$. Assume $a \neq 1$ and let $t\in (0,1/2]$.   Then
\begin{equation}
\label{eq:C15}
\lim_{i \to \infty} \int_0^t \frac{\sin(q_i \pi u)\sin(p_i \pi u)}{\sin(\pi u)}\ud u =\frac{1}{2\pi} \ln\left(\left|\frac{1+a}{1-a}\right|\right).
\end{equation}
We also have
\begin{eqnarray}
\label{eq:C16} \nonumber
\sup_{t\in(0,1/2]}\left|\int_0^t \frac{\sin(q \pi u)\sin(p \pi u)}{\sin(\pi u)}\ud u\right|\\ \leq \frac{1}{2}\frac{\pi+1}{\pi} \ln\left(\left|\frac{1+q/p}{1-q/p}\right|\right)
\end{eqnarray}
and 
\begin{eqnarray}
\label{eq:C16a} \nonumber
\sup_{t\in(0,1/2]}\left|\int_0^t \frac{\sin[(2q+1) \pi u]^2}{\sin(\pi u)}\ud u\right|\\ \leq \frac{1}{\pi}\left[1+\frac{1}{2}\ln(2q+1)\right].
\end{eqnarray}
\end{C2}

\emph{Proof.} First, let us notice that by the symmetry of the function
\[\int_0^t \frac{\sin(q_i \pi u)\sin(p_i \pi u)}{\sin(\pi u)}\ud u\]
 in the variables $q_i$ and $p_i$, we may assume without loss of generality that $a<1$. Moreover, since Eq.~(\ref{eq:C16}) is trivial for $q_i=p_i$, we may also assume that $q_i < p_i$. In these conditions, $p_i-q_i \to \infty$ as $i \to \infty$.   Then 
\begin{eqnarray}&&
\label{eq:C17}\nonumber
 \int_{0}^{t} \frac{\sin(q_i \pi u)\sin(p_i \pi u)}{\sin(\pi u)}\ud u \\&&=\frac{1}{2}  \int_{0}^{t} \frac{\cos[(p_i-q_i)\pi u]-\cos[(p_i+q_i)\pi u]}{\sin(\pi u)}\ud u \nonumber \\&& = \frac{1}{2} \int_{0}^{t} \int_{1-q_i/p_i}^{1+q_i/p_i} p_i \sin(p_i \pi u v)f(u)\ud v\ud u, \end{eqnarray}
where
\[f(u)=\frac{\pi u}{\sin(\pi u)}.\]
The function $f(u)$ is well defined and in fact it has continuous derivatives of any order on the set $(0, 1/2]$. The function $f(u)$ and its first order derivative  can be shown to be positive and increasing on the interval $(0, 1/2]$. Using this observation, the reader may prove that
\begin{equation}
\label{eq:C18}
\sup_{u\in(0,1/2]} |f(u)|= \pi/2 \; \text{and}\; \sup_{u\in(0,1/2]} |f'(u)|= \pi.
\end{equation}

Inverting the order of integration in Eq.~(\ref{eq:C17}) and then integrating by parts against the variable $u$, we obtain
\begin{eqnarray}&&
\label{eq:C19}\nonumber
 \frac{1}{2} \int_{1-q_i/p_i}^{1+q_i/p_i}\int_{0}^{t}  p_i \sin(p_i \pi u v)f(u)\ud u\ud v\\&& =\frac{1}{2} \int_{1-q_i/p_i}^{1+q_i/p_i}\frac{1}{\pi v} [1-\cos(p_i \pi t v)f(t)] \ud v\\ && +  \frac{1}{2} \int_{1-q_i/p_i}^{1+q_i/p_i}\int_{0}^{t} \frac{1}{\pi v} \cos(p_i \pi u v)f'(u)\ud u \ud v. \nonumber
\end{eqnarray}
From Eqs.~(\ref{eq:C17}), (\ref{eq:C18}), and (\ref{eq:C19}) we learn that 
\begin{eqnarray}&&
\label{eq:C20}\nonumber
 \left| \int_{0}^{t} \frac{\sin(q_i \pi u)\sin(p_i \pi u)}{\sin(\pi u)}\ud u- \frac{1}{2\pi}\ln\left(\frac{1+q_i/p_i}{1-q_i/p_i}\right)\right|\\ && \leq  \frac{1}{4} \left|\int_{1-q_i/p_i}^{1+q_i/p_i}\frac{1}{ v} \cos(p_i \pi t v) \ud v\right|\\ && +  \frac{1}{2}\int_{0}^{t}\left| \int_{1-q_i/p_i}^{1+q_i/p_i} \frac{1}{ v} \cos(p_i \pi u v)\ud v\right| \ud u.  \nonumber
\end{eqnarray}
Integrating by parts the first term of the right-hand side of Eq.~(\ref{eq:C20}), we obtain
\begin{equation}
\label{eq:C21}
\left|\int_{1-q_i/p_i}^{1+q_i/p_i}\frac{1}{ v} \cos(p_i \pi t v) \ud v\right| \leq \int_{1-q_i/p_i}^{1+q_i/p_i}\frac{1}{ v} \ud v=\ln\left(\frac{1+q_i/p_i}{1-q_i/p_i}\right)
\end{equation}
and so, 
\begin{eqnarray}
\label{eq:C22}\nonumber
 \left| \int_{0}^{t} \frac{\sin(q_i \pi u)\sin(p_i \pi u)}{\sin(\pi u)}\ud u- \frac{1}{2\pi}\ln\left(\frac{1+q_i/p_i}{1-q_i/p_i}\right)\right| \quad \\ \leq   \left(\frac{1}{4}+\frac{1}{2}t\right)\ln\left(\frac{1+q_i/p_i}{1-q_i/p_i}\right)\leq   \frac{1}{2}\ln\left(\frac{1+q_i/p_i}{1-q_i/p_i}\right). \quad
\end{eqnarray}
Then, Eq.~(\ref{eq:C16}) follows trivially from Eq.~(\ref{eq:C22}). 

On the other hand, by integration by parts, we obtain 
\begin{eqnarray}
\label{eq:C23}\nonumber &&
\left|\int_{1-q_i/p_i}^{1+q_i/p_i}\frac{1}{ v} \cos(p_i \pi t v) \ud v\right| =
\Bigg| \frac{\sin[\pi(p_i+q_i)t]}{\pi(p_i+q_i)t}\\ &&-\frac{\sin[\pi(p_i-q_i)t]}{\pi(p_i-q_i)t}  +\frac{1}{\pi t p_i}\int_{1-q_i/p_i}^{1+q_i/p_i}\frac{1}{ v^2} \sin(p_i \pi t v) \ud v\Bigg| \nonumber \\ &&\leq \frac{1}{\pi t } \left( \frac{1}{p_i+q_i}+\frac{1}{p_i-q_i} \right)\quad + \frac{1}{\pi t p_i} \\ && \bigg(\frac{1}{1-q_i/p_i} -\frac{1}{1+q_i/p_i}\bigg)\nonumber = \frac{2}{\pi t(p_i-q_i) }. 
\end{eqnarray}
Thus, 
\begin{eqnarray}
\label{eq:C24}\lim_{i \to \infty}
\left|\int_{1-q_i/p_i}^{1+q_i/p_i}\frac{1}{ v} \cos(p_i \pi t v) \ud v\right| =0
\end{eqnarray}
and by the dominated convergence theorem [use also Eq.~(\ref{eq:C21})], 
\begin{eqnarray}
\label{eq:C25}\lim_{i \to \infty}\int_0^t
\left|\int_{1-q_i/p_i}^{1+q_i/p_i}\frac{1}{ v} \cos(p_i \pi u v) \ud v\right|\ud u =0.
\end{eqnarray}
Eq.~(\ref{eq:C15}) then follows easily from Eqs.~(\ref{eq:C20}), (\ref{eq:C24}), and (\ref{eq:C25}) by letting $i \to \infty$. 

To prove Eq.~(\ref{eq:C16a}), we first notice that 
\begin{eqnarray*}&&
2\sin(\pi u) \sum_{k=1}^q \cos(2k\pi u)= \sum_{k=1}^q \sin[(2k+1)\pi u] \\&& -\sum_{k=1}^q \sin[(2k-1)\pi u]= \sin[(2q+1)\pi u] - \sin(\pi u) 
\end{eqnarray*}
and so, 
\[ \frac{\sin[(2q+1)\pi u]}{\sin(\pi u)}= 1+ 2 \sum_{k=1}^q \cos(2k \pi u). \]
Then, by the positivity of the integrand,
\begin{eqnarray*}&&
\sup_{t\in(0,1/2]}\left|\int_0^t \frac{\sin[(2q+1) \pi u]^2}{\sin(\pi u)}\ud u\right|\\&& =\int_0^{1/2} \frac{\sin[(2q+1) \pi u]^2}{\sin(\pi u)}\ud u = \int_0^{1/2} \sin[(2q+1)\pi u] \ud u \\&& +2 \sum_{k=1}^q \int_0^{1/2} \cos(2k\pi u) \sin[(2q+1)\pi u] \ud u.
\end{eqnarray*}
Using the trigonometric identity 
\[
2 \sin(\alpha)\cos(\beta)= \sin(\alpha-\beta)+\sin(\alpha +\beta),
\] 
the above quantity becomes
\begin{eqnarray*}
\int_0^{1/2} \sin[(2q+1)\pi u] \ud u + \sum_{k=1}^q \int_0^{1/2}\sin[(2q-2k+1)\pi u] \ud u \\ + \sum_{k=1}^q \int_0^{1/2}\sin[(2q+2k+1)\pi u] \ud u = \frac{1}{\pi} \frac{1}{2q+1}\\ +\frac{1}{\pi}\sum_{k=1}^q \frac{1}{2q-2k+1}+\frac{1}{\pi}\sum_{k=1}^q \frac{1}{2q+2k+1}= \frac{1}{\pi} \sum_{k=0}^{2q}\frac{1}{2k+1}.
\end{eqnarray*}
We write the last sum as 
\begin{eqnarray*}
\frac{1}{\pi}\left(1+\frac{1}{2} \sum_{k=1}^{2q} \frac{1}{k+1/2}\right)
\end{eqnarray*}
and observe that 
\[
\int_{k}^{k+1} \frac{1}{x} \ud x \geq \left(\int_{k}^{k+1} x\ud x\right)^{-1}= \frac{1}{k+1/2},
\]
by the convexity of the function $1/x$ and by Jensen's inequality.
Therefore, 
\begin{eqnarray*}&&
\sup_{t\in(0,1/2]}\left|\int_0^t \frac{\sin[(2q+1) \pi u]^2}{\sin(\pi u)}\ud u\right|\\&& =\frac{1}{\pi}\left(1+\frac{1}{2} \sum_{k=1}^{2q} \frac{1}{k+1/2}\right) \leq \frac{1}{\pi}\left(1+\frac{1}{2} \sum_{k=1}^{2q} \int_{k}^{k+1}\frac{1}{x}\ud x\right)\\&& = \frac{1}{\pi} \left(1+\frac{1}{2}\int_1^{2q+1}\frac{1}{x} \ud x \right)= \frac{1}{\pi}\left[1+\frac{1}{2}\ln(2q+1)\right].
\end{eqnarray*}
The proof of Eq.~(\ref{eq:C16a}) and of the proposition is concluded. \hspace{\stretch{1}}$\Box$

\begin{C3}
\label{Pr:C3}
Let $k_i\to \infty$ and $n_i\to \infty$ be two sequences such that $n_i/k_i \to a$. Assume $a \neq 1/2$  and $t \in (0,1/2]$. Then
\begin{eqnarray}
\label{eq:C26}\nonumber
\lim_{i \to \infty} \left[k_i\pi  \Phi_{n_i,k_i}(t)\right]=1-a \ln\left(\left|\frac{1+2a}{1-2a}\right|\right)\\ + \frac{1}{12a} \left[\ln\left(\left|\frac{1+2a}{1-2a}\right|\right)-\ln\left(\left|\frac{1+8a}{1-8a}\right|\right)\right].
\end{eqnarray} 
Moreover, there are positive constants $c'_1$, $c'_2$, $c'_3$, and $c'_4$ such that 
\begin{equation}
\label{eq:C27}
 \sup_{t\in (0,1/2]}\left|k\pi  \Phi_{n,k}(t)\right|\leq c'_1-c'_2\ln\left(\left|1-\frac{2n+1}{k}\right|\right), 
\end{equation}
if $k \neq 2n+1$ and 
\begin{equation}
\label{eq:C27a}
 \sup_{t\in (0,1/2]}\left|k\pi  \Phi_{n,k}(t)\right|\leq c'_3+c'_4\ln\left(2n+1\right), 
\end{equation}
if $k=2n+1$.
\end{C3}

\emph{Proof.}
We begin by analyzing the term 
\[
\int_0^t g_n(u)\sin(k\pi u)\ud u.
\]
Integrating by parts twice, we obtain
\begin{widetext}
\begin{eqnarray}\nonumber \label{eq:C28}
&&
k\pi \int_0^t g_n(u)\sin(k\pi u)\ud u=\frac{k  }{\pi \alpha_n} \sum_{j=n+1}^\infty \int_0^t\frac{\cos(2j\pi u)}{j^2}\sin(k \pi u)\ud u
=\frac{k  }{\pi \alpha_n}\Bigg\{\frac{1}{k\pi} \Bigg(\sum_{j=n+1}^\infty\frac{1}{j^2}\Bigg) 
\\ &&
-\frac{\cos(k\pi t)}{k\pi}\Bigg[\sum_{j=n+1}^\infty \frac{\cos(2j\pi t)}{j^2}\Bigg]\Bigg\}  -
\frac{2}{\pi \alpha_n}\sum_{j=n+1}^\infty \int_0^t\frac{\sin(2j\pi u)}{j}\cos(k \pi u) \ud u  =
1- \frac{\cos(k\pi t)}{\pi^2 \alpha_n} 
\\ &&
 \times\Bigg[\sum_{j=n+1}^\infty \frac{\cos(2j\pi u)}{j^2}\Bigg]- \frac{2 \sin(k\pi t)}{\pi^2 \alpha_n k}\Bigg[\sum_{j=n+1}^\infty \frac{\sin(2j\pi t)}{j}\Bigg] \nonumber
+ \frac{4}{\pi \alpha_n k} \sum_{j=n+1}^\infty \int_0^t{\sin(2j\pi u)}\cos(k \pi u)\ud u.
\end{eqnarray}
\end{widetext}
The integration term by term performed in the above equation is justified by the fact that the series \[\sum_{j=n+1}^\infty \frac{\cos(2j\pi u)}{j^2}\sin(k \pi u)\] is  uniformly convergent. The last term in Eq.~(\ref{eq:C28}) can be evaluated as follows
\begin{widetext}
\begin{eqnarray}
\nonumber \label{eq:C29} &&
\sum_{j=n+1}^\infty \int_0^t{\sin(2j\pi u)}\cos(k \pi u)\ud u= \lim_{N \to \infty} \int_0^t\sum_{j=n+1}^N {\sin(2j\pi u)} \cos(k \pi u)\ud u \\ &&  = -\frac{1}{2}\int_0^t \frac{\sin[(2n+1)\pi u] \sin(k\pi u)}{\sin(\pi u)}\ud u+ \frac{1}{2}\lim_{N \to \infty} \int_0^t \frac{\sin[(2N+1)\pi u] \sin(k\pi u)}{\sin(\pi u)}\ud u.
\end{eqnarray}
\end{widetext}
Noticing that ${\sin(k\pi u)}/{\sin(\pi u)}$ is  integrable on $[0,1/2]$ and employing  the Riemann-Lebesgue lemma, we learn that 
\[\lim_{N \to \infty} \int_0^t \frac{\sin[(2N+1)\pi u] \sin(k\pi u)}{\sin(\pi u)}\ud u=0.\]
Therefore,
\begin{eqnarray}
\nonumber \label{eq:C30} &&
\sum_{j=n+1}^\infty \int_0^t{\sin(2j\pi u)}\cos(k \pi u)\ud u \\ && =  -\frac{1}{2}\int_0^t \frac{\sin[(2n+1)\pi u] \sin(k\pi u)}{\sin(\pi u)}\ud u. 
\end{eqnarray}

A little thought and the inequalities~(\ref{eq:C9}) and (\ref{eq:C12}) show that
\begin{eqnarray*}\nonumber&&
\lim_{i \to \infty}k_i\pi \int_0^t g_{n_i}(u)\sin(k_i\pi u)\ud u= 1-\\&& \lim_{i\to \infty}\frac{2}{\pi \alpha_{n_i} k_i}  \int_0^t \frac{\sin[(2n_i+1)\pi u] \sin(k_i\pi u)}{\sin(\pi u)}\ud u. 
\end{eqnarray*}
The last limit can be evaluated with the help of Eqs.~(\ref{eq:C7}) and (\ref{eq:C15}). The result is
\begin{equation} \label{eq:C31}
\lim_{i \to \infty}k_i\pi \int_0^t g_{n_i}(u)\sin(k_i\pi u)\ud u= 1-a \ln\left(\left|\frac{1+2a}{1-2a}\right|\right). 
\end{equation}

The magnitude of the terms of the last sum in Eq.~(\ref{eq:C28}) can be established as follows.
Clearly, 
\begin{equation}\label{eq:C32}
\left|\frac{\cos(k\pi t)}{\pi^2 \alpha_n}\sum_{j=n+1}^\infty \frac{\cos(2j\pi u)}{j^2}\right|\leq \frac{1}{\pi^2 \alpha_n}\sum_{j=n+1}^\infty \frac{1}{j^2}=1.
\end{equation}
Then, with the help of Eq.~(\ref{eq:C12}), we have
\begin{eqnarray}\nonumber \label{eq:C33}
\left|\frac{2 \sin(k\pi t)}{\pi^2 \alpha_n k}\sum_{j=n+1}^\infty \frac{\sin(2j\pi t)}{j}\right|\leq 
\frac{2 }{\pi^2 \alpha_n (n+1)} \\ \times \left|\frac{\sin(k\pi t)}{k \pi t}\right|\frac{ \pi t}{\sin(\pi t)} \leq 2 \frac{ \pi t}{\sin(\pi t)} \leq \pi. 
\end{eqnarray}
For the last two inequalities, we used the fact that $|\sin(x)/x|\leq 1$ as well as the fact that $\pi t / \sin(\pi t)$ is positive and increasing on the interval $(0,1/2]$ and therefore it is bounded by $\pi/2$. 
Form Eqs.~(\ref{eq:C28}), (\ref{eq:C32}), and (\ref{eq:C33}) we learn that 
\begin{widetext}
\begin{eqnarray}\nonumber
\label{eq:C34}
\sup_{t\in (0,1/2]}\left|k\pi  \int_0^t g_{n}(u)\sin(k\pi u)\ud u\right|\leq 2+\pi  + \frac{2}{\pi \alpha_{n} k} \sup_{t\in (0,1/2]}  \left|\int_0^t \frac{\sin[(2n+1)\pi u] \sin(k\pi u)}{\sin(\pi u)}\ud u\right| \\ \leq
2+\pi + 2\pi \sup_{t\in (0,1/2]} \left|\int_0^t \frac{\sin[(2n+1)\pi u] \sin(k\pi u)}{\sin(\pi u)}\ud u\right|, 
\end{eqnarray}
\end{widetext}
where for the last inequality we used  the relations $\pi^2 \alpha_{n}(n+1) \geq 1$ [see Eq.~(\ref{eq:C7})] and $k\geq n+1$. Now, if $k \neq 2n+1$, Eqs.~(\ref{eq:C16}) and (\ref{eq:C34}) produce 
\begin{widetext}
\begin{eqnarray}
\label{eq:C34a}\nonumber 
\sup_{t\in (0,1/2]}\left|k\pi  \int_0^t g_{n}(u)\sin(k\pi u)\ud u\right| \leq 2+\pi + (1+\pi)\ln\left(\left|\frac{1+(2n+1)/k}{1-(2n+1)/k}\right|\right)\\ = 2+\pi + (1+\pi)\ln\left[1+(2n+1)/k\right] -(1+\pi) \ln\left[\left|1-(2n+1)/k\right|\right]\\ \leq 
2+\pi +(1+\pi) \ln(3)-(1+\pi) \ln\left(\left|1-\frac{2n+1}{k}\right|\right), \nonumber
\end{eqnarray} 
\end{widetext}
where for the last inequality we utilized the relation $k\geq n+1 \geq n+1/2$ to conclude 
\begin{equation}
\label{eq:C34aa}
\ln\left[1+(2n+1)/k\right]\leq \ln(3).
\end{equation}
If $k=2n+1$, then Eqs.~(\ref{eq:C16a}) and (\ref{eq:C34}) produce 
\begin{equation}
\label{eq:C34b}
\sup_{t\in (0,1/2]}\left|k\pi  \int_0^t g_{n}(u)\sin(k\pi u)\ud u\right| \leq 4+\pi +  \ln\left(2n+1\right). 
\end{equation} 

The analysis of the term 
\[
k \pi \int_0^t g^\circ_n(u)\sin(k\pi u)\ud u
\]
is more straightforward. We start with the identity 
\begin{eqnarray}\nonumber \label{eq:C35}&&
k \pi \int_0^t g^\circ_n(u)\sin(k\pi u)\ud u  \\ &&= \frac{k \pi}{6 n} \int_0^t \frac{\sin[(8n+1)\pi u] \sin(k\pi u)}{\sin(\pi u)}\ud u \\&& - \frac{k \pi}{6 n}\int_0^t \frac{\sin[(2n+1)\pi u] \sin(k\pi u)}{\sin(\pi u)}\ud u. \nonumber
\end{eqnarray}
Then, from Proposition~\ref{Pr:C2} we readily learn that 
\begin{eqnarray}\nonumber \label{eq:C36}
\lim_{i \to \infty}k_i\pi \int_0^t g^\circ_{n_i}(u)\sin(k_i\pi u)\ud u= \frac{1}{12 a}\\ \times  \left[\ln\left(\left|\frac{1+8a}{1-8a}\right|\right)-\ln\left(\left|\frac{1+2a}{1-2a}\right|\right)\right]. 
\end{eqnarray}
Now, using the inequality $k\leq 4n$ together Eq.~(\ref{eq:C35}), we conclude 
\begin{eqnarray*}\nonumber && 
 \sup_{t\in (0,1/2]}\left|\int_0^t g^\circ_{n_i}(u)\sin(k_i\pi u)\ud u\right|  \leq \frac{1+\pi}{3} \\ && \times  \left[\ln\left(\left|\frac{1+(8n+1)/k}{1-(8n+1)/k}\right|\right)+\ln\left(\left|\frac{1+(2n+1)/k}{1-(2n+1)/k}\right|\right)\right]. \nonumber \\&& 
\end{eqnarray*}
Using Eq.~(\ref{eq:C34aa}), the fact that the function
$
\ln\left[\left|(1+x)/(1-x)\right|\right]
$
is decreasing on the interval $(1,\infty)$, and the fact  that $(8n+1)/k \geq 2$, we obtain
\begin{eqnarray}\nonumber  \label{eq:C37}
 \sup_{t\in (0,1/2]}\left|\int_0^t g^\circ_{n_i}(u)\sin(k_i\pi u)\ud u\right|  \leq \frac{1+\pi}{3} \\  \times  \left[2\ln\left(3\right)-\ln\left(\left|1-\frac{2n+1}{k}\right|\right)\right]. 
\end{eqnarray}
Proceeding in an analogous manner but also using Eq.~(\ref{eq:C16a}), one proves 
\begin{eqnarray}\nonumber  \label{eq:C37a}
 \sup_{t\in (0,1/2]}\left|\int_0^t g^\circ_{n_i}(u)\sin(k_i\pi u)\ud u\right|  \leq \frac{1+\pi}{3} \\ \times \ln\left(3\right) +\frac{2}{3}+\frac{1}{3}\ln\left(2n+1\right). 
\end{eqnarray}

The statements of the  proposition follow from Eqs.~(\ref{eq:C31}), (\ref{eq:C34a}), (\ref{eq:C34b}),  (\ref{eq:C36}), (\ref{eq:C37}),  (\ref{eq:C37a}), and the relation 
\[
\Phi_{n,k}(t)=\int_0^t \left[g_n(u)-g^\circ_n(u)\right]\sin(k\pi u)\ud u.
\]
The proof is concluded. \hspace{\stretch{1}} $\Box$

We now focus out attention on the asymptotic behavior of the  primitives $\Psi_{n,k}(t)$ on the interval $(0,1/2]$. We begin by establishing the following bound.
\begin{C4}
\label{Pr:C4}
Let $R: (0, \infty)\to (0, \infty)$ be the continuous, positive, and integrable function
\begin{equation}
\label{eq:C38}
R(u)= \left\{\begin{array}{ l l} u^{-1/2}, & u< 1, \\ u^{-2}, & u\geq1.\end{array}\right.
\end{equation}
Then there is $M_0>0$ a positive constant such that 
\begin{equation}
\label{eq:C39}
\left|r_n(u)-1+\frac{1}{2}\left[g_n(u)-g^\circ_n(u)\right]\right|\leq M_0 R(nu) 
\end{equation} for all $u \in (0,1/2]$ and $n\geq 1$. 
\end{C4}

\emph{Proof.}
According to Eq.~(\ref{eq:C3}), we have 
\begin{equation}
\label{eq:C40}
\gamma_n(u,u)\leq u. 
\end{equation}
Next, let us notice that the function $\sin(\pi u)$ is positive and concave (convex down) on the interval $(0, 1/2]$ and so, Jensen's inequality says that 
\begin{eqnarray}
\label{eq:C41}&&
\sin(\pi u) = \sin[ (1-2u)0+ 2u (\pi /2)]\nonumber \\ && \geq (1-2u)\sin(0)+ 2u \sin(\pi/2)= 2u 
\end{eqnarray}
for all $u \in (0,1/2]$. It follows that if  $k\leq 2n$, we have 
\[\sin(k \pi u)^2\geq 4k^2u^2\] for all $u<1/(4n)$. Using this observation, one shows
\begin{eqnarray*}
\gamma^\circ_n(u,u)\geq \frac{2\alpha_n}{3n}\sum_{k=n+1}^{2n}\sin(k\pi u)^2 \geq \frac{2\alpha_n}{3n}\sum_{k=n+1}^{2n}4k^2u^2 \nonumber \\ \geq \frac{8\alpha_n u^2}{3n}\int_{n}^{2n}x^2 \ud x= \frac{56\alpha_n u^2 n^2}{9} \qquad 
\end{eqnarray*}
for all $u< 1/(4n)$. On the other hand, if $u\in[ 1/(4n), 1/2]$, then Eqs.~(\ref{eq:C10}) and (\ref{eq:C41}) show that 
\begin{eqnarray*}&&
\gamma^\circ_n(u,u)=\alpha_n[1-g^\circ_n(u)]\geq \alpha_n[1-|g^\circ_n(u)|] \\&& \geq \alpha_n\left(1-\frac{1}{6nu}\right)\geq \alpha_n\left(1-\frac{1}{6n/4n}\right)= \frac{\alpha_n}{3}.
\end{eqnarray*}
As such, for all $u\in (0,1/2]$, the following inequality holds
\[\gamma^\circ_n(u,u) \geq \frac{\alpha_n}{3}\min \left\{ 1, \frac{56 u^2 n^2}{3}\right\}.
\]
This inequality implies that
\begin{equation}
\label{eq:C42}
\gamma^\circ_n(u,u)\geq \frac{\alpha_n}{3}n^2 u^2
\end{equation}
for all $u< 1/n$. Indeed, if $u<1/n$, then clearly $n^2u^2< 1$ and $n^2 u^2 \leq 56u^2 n^2/3$. Therefore, 
\[n^2 u^2 \leq \min \left\{ 1, \frac{56 u^2 n^2}{3}\right\},\]
from which Eq.~(\ref{eq:C42}) follows easily.  Finally, from Eqs.~(\ref{eq:C7}) and (\ref{eq:C42}), we obtain
\begin{equation}
\label{eq:C43}
\gamma^\circ_n(u,u)\geq \frac{\alpha_n}{3}n^2 u^2\geq \frac{2n}{n+1}\frac{nu^2}{6\pi^2}\geq \frac{nu^2}{6\pi^2}. 
\end{equation}
for all $u< 1/n$. Combining the last inequality with Eq.~(\ref{eq:C40}), one deduces
\[
r_n(u)=\sqrt{\frac{\gamma_n(u,u)}{\gamma^\circ_n(u,u)}}\leq \sqrt{\frac{6\pi^2}{n u}}.
\]
Going back to Eq.~(\ref{eq:C39}) and using the inequalities $|g_n(u)|\leq 1$ and $ |g^\circ_n(u)|\leq 1$, one obtains
\begin{eqnarray}
\label{eq:C44}\nonumber 
\left|r_n(u)-1+\frac{1}{2}\left[g_n(u)-g^\circ_n(u)\right]\right|\leq |r_n(u)|+2 \\ \leq \sqrt{\frac{6\pi^2}{n u}}+\frac{2}{\sqrt{nu}} = \left(2+\sqrt{6\pi^2}\right)R(nu).
\end{eqnarray}
for all $u<1/n$. 

We now turn our attention to the case $u\geq 1/n$. From Eqs.~(\ref{eq:C9}), (\ref{eq:C10}), and (\ref{eq:C41}) we learn that 
\begin{equation}
\label{eq:C45}
|g_n(u)|\leq \frac{1}{n} \frac{1}{\sin(\pi u)} \leq  \frac{1}{2nu} 
\end{equation}
and 
\begin{equation}
\label{eq:C46}
|g^\circ_n(u)|\leq \frac{1}{3n} \frac{1}{\sin(\pi u)} \leq \frac{1}{6nu},
\end{equation}
respectively.
Moreover, since $u\geq 1/n$, we also have $|g_n(u)|\leq 1/2$ and $|g^\circ_n(u)|\leq 1/6$. Let 
\[h(x,y)= \sqrt{\frac{1-x}{1-y}}.\]
The function $h(x,y)$ has continuous partial derivatives of any order on the rectangle $D=[-1/2, 1/2]\times [-1/6,1/6]$. Performing a series expansion about origin to the first order and employing the Taylor formula with remainder, we learn that for each $(x,y)\in D$ there is $(x',y')\in D$ such that 
\begin{eqnarray*}
h(x,y)-1+\frac{1}{2}(x-y)= \frac{1}{2}\bigg[\frac{\partial^2h}{\partial x^2}(x',y')x^2\\+2\frac{\partial^2h}{\partial x \partial y}(x',y')xy+\frac{\partial^2h}{\partial y^2}(x',y')y^2\bigg].
\end{eqnarray*}
Thus, 
\begin{eqnarray}
\label{eq:C47}\nonumber
\left|h(x,y)-1+\frac{1}{2}(x-y)\right|\leq \frac{1}{2}\bigg[\left|\frac{\partial^2h}{\partial x^2}(x',y')\right|\\ +2\left|\frac{\partial^2h}{\partial x \partial y}(x',y')\right|+\left|\frac{\partial^2h}{\partial y^2}(x',y')\right|\bigg]\\ \times \max\{x^2, y^2\} \leq M_1 \max\{x^2, y^2\},\nonumber
\end{eqnarray}
where
\begin{eqnarray*}
M_1=\frac{1}{2}\sup_{(x',y')\in D} \bigg[\left|\frac{\partial^2h}{\partial x^2}(x',y')\right|+2\left|\frac{\partial^2h}{\partial x \partial y}(x',y')\right|\\+\left|\frac{\partial^2h}{\partial y^2}(x',y')\right|\bigg]
\end{eqnarray*}
is finite because the second order derivatives of $h(x,y)$ are continuous on the compact set $D$, thus bounded. 
 
Combining Eq.~(\ref{eq:C47}) with Eqs.~(\ref{eq:C45}) and (\ref{eq:C46}), we learn that 
\begin{eqnarray}
\label{eq:C48}&&\nonumber 
\left|r_n(u)-1+\frac{1}{2}\left[g_n(u)-g^\circ_n(u)\right]\right|\leq M_1 \\&& \times \max \left\{\frac{1}{(2nu)^2}, \frac{1}{(6nu)^2}\right\}= \frac{M_1}{4} R(nu)
\end{eqnarray}
for all $u\geq 1/n$. 
The statement of the proposition follows immediately from Eqs.~(\ref{eq:C44}) and (\ref{eq:C48}) by setting 
\[M_0 = \max \left\{2+\sqrt{6\pi^2}, {M_1}/{4}\right\}.\]
The proof is concluded. \hspace{\stretch{1}} $\Box$

\begin{C5}
\label{Pr:C5}
 Let $k_i\to \infty$ and $n_i\to \infty$ be two sequences such that $n_i/k_i \to a$. Let
 \begin{eqnarray*}&&
h_0(a,u)= \left[1-\int_1^\infty \frac{\cos(2 a u \nu)}{\nu^2} \ud \nu\right]^{1/2}\\&& \times  \left[ 1-\frac{1}{3}\int_1^4 {\cos(2 a u \nu)} \ud \nu\right]^{-1/2}-1 \\&&+\frac{1}{2}\left[\int_1^\infty \frac{\cos(2 a u \nu)}{\nu^2} \ud \nu-\frac{1}{3}\int_1^4 {\cos(2 a u \nu)} \ud \nu\right]
\end{eqnarray*}
and define
\begin{equation}
\label{eq:C49}
H_0(a)=\int_0^\infty h_0(a,u)\sin(u)\ud u.
\end{equation}
If $t\in (0,1/2]$,  then
\begin{equation}
\label{eq:C50}
\lim_{i \to \infty} \left[k_i\pi  \Psi_{n_i,k_i}(t)\right]= H_0(a).
\end{equation} 
Moreover,
\begin{equation}
\label{eq:C51}
\sup_{t \in (0,1/2]} \left|k\pi  \Psi_{n,k}(t)\right|\leq 12 \pi M_0 ,
\end{equation}
where $M_0$ is the constant appearing in Eq.~(\ref{eq:C39}).
\end{C5}

\emph{Proof.}
We begin by performing the transformation of coordinates
\begin{eqnarray*}
k_i\pi \int_{0}^{t} \psi_{n_i,k_i}(u)\ud u =  \int_{0}^{k_i \pi t} \psi_{n_i,k_i}\left(\frac{u}{\pi k_i}\right)\ud u \\ = \int_{0}^{\infty} I_{k_i \pi t}(u)\psi_{n_i,k_i}\left(\frac{u}{\pi k_i}\right)\ud u,
\end{eqnarray*}
where $I_{k_i \pi t}(u)$ is the indicator function of the interval $[0,{k_i \pi t}]$. 
As $k_i \to \infty$, we have $I_{k_i \pi t}(u)\to 1$ for all $t>0$.

Let us find the limit of the sequence 
\[\psi_{n_i,k_i}\left(\frac{u}{\pi k_i}\right).\]
We start with the observation that 
\begin{eqnarray}
\label{eq:C60a} \nonumber
\lim_{i \to \infty}g_n\left(\frac{u}{ \pi k_i} \right)= \lim_{i \to \infty} {n_i}\sum_{j=n_i+1}^\infty \frac{\cos(2ju/k_i)}{j^2}\\=\lim_{i \to \infty} \frac{n_i}{k_i}\sum_{\nu_{i,j}=(n_i+1)/k_i}^\infty \frac{\cos(2u\nu_{i,j})}{\nu_{i,j}^2} \frac{1}{k_i},
\end{eqnarray}
where $\nu_{i,j}= j/k_i$. 
Remembering that $n_i/k_i \to a$, we notice that the last sum is a Riemann sum approximation of the continuous function  $\cos(2u\nu)/\nu^2$, regarded as a function of $\nu$ over the interval $(a, \infty)$. For each $i$, the function $\cos(2u\nu)/\nu^2$ is evaluated on the grid made up of the equally spaced points  $\{\nu_{i,j}\}_{j> n}$. The mesh of the grid is $1/k_i$ and decays to zero as $i \to \infty$.  A little thought shows that the limit in Eq.~(\ref{eq:C60a}) is
\begin{equation}
\label{eq:C52}
\lim_{i \to \infty}g_n\left(\frac{u}{ \pi k_i} \right)= a \int_a^\infty \frac{\cos(2 u \nu)}{\nu^2} \ud \nu =\int_1^\infty \frac{\cos(2a u \nu)}{\nu^2} \ud \nu.
\end{equation}
Proceeding in a similar way, one shows
\begin{equation}
\label{eq:C53}
\lim_{i \to \infty}g^\circ_n\left(\frac{u}{ \pi k_i} \right)=\frac{1}{3}\int_1^4 {\cos(2 a u \nu)} \ud \nu.
\end{equation}
From Eqs.~(\ref{eq:C52}) and (\ref{eq:C53}) the reader easily concludes that 
\[
\lim_{i \to \infty}\psi_{n_i,k_i}\left(\frac{u}{\pi k_i}\right)=h_0(a, u) \sin(u) 
\]
and therefore,
\begin{equation}
\label{eq:C54}
\lim_{i \to \infty}I_{k_i \pi t}(u)\psi_{n_i,k_i}\left(\frac{u}{\pi k_i}\right)=h_0(a, u) \sin(u). 
\end{equation}

Let us notice that 
\begin{equation}
\label{eq:C55}
\left|I_{k_i \pi t}(u)\psi_{n_i,k_i}\left(\frac{u}{\pi k_i}\right)\right|\leq M_0 R\left(\frac{n_i u}{k_i \pi }\right).
\end{equation}
The inequality is trivial if $u> k_i \pi t$. Otherwise, we have $u/(k_i \pi)\leq t \leq 1/2$ and the inequality follows from Proposition~\ref{Pr:C4} and the fact that $|\sin(u)|\leq 1$. 
Now, the continuity of the function $R(u)$ as well as the fact that $n_i/k_i \to a$ imply
\[
\lim_{i \to \infty} R\left(\frac{n_i u}{k_i \pi }\right) = R\left(\frac{a u}{\pi }\right).
\]
Moreover, 
\begin{eqnarray*}\nonumber
\lim_{i \to \infty}\int_0^\infty R\left(\frac{n_i u}{k_i \pi }\right) \ud u =\lim_{i \to \infty}\frac{k_i \pi }{n_i}\int_0^\infty R\left(u\right) \ud u\\  = \frac{\pi}{a} \int_0^\infty R\left(u\right) \ud u= \int_0^\infty R\left(\frac{a u}{\pi}\right) \ud u.
\end{eqnarray*}
In these conditions, the dominated convergence theorem and Eq.~(\ref{eq:C54}) allow us to conclude that
\begin{equation*}
\lim_{i \to \infty}\int_0^\infty I_{k_i \pi t}(u)\psi_{n_i,k_i}\left(\frac{u}{\pi k_i}\right)\ud u =\int_0^\infty h_0(a, u) \sin(u) \ud u  
\end{equation*}
and so, the equality~(\ref{eq:C50}) is proved. 

From the inequality~(\ref{eq:C55}) we learn that 
\begin{eqnarray*}\limsup_{t \in (0,1/2]} \left|\int_0^\infty I_{k \pi t}(u)\psi_{n,k}\left(\frac{u}{\pi k}\right)\ud u\right| \leq  M_0  \int_0^\infty R\left(\frac{n u}{k \pi }\right) \ud u\\  =M_0 \frac{\pi k}{n}\int_0^\infty R\left(u\right) \ud u= M_0 \frac{3 \pi  k }{n}\leq 12 \pi M_0 \end{eqnarray*}
and the proposition is concluded. \hspace{\stretch{1}} $\Box$

\begin{C6}
\label{Pr:C6}
 Let $k_i\to \infty$ and $n_i\to \infty$ be two sequences such that $n_i/k_i \to a$ and the $k_i$'s are either odd or even. Let
\begin{eqnarray}\nonumber 
\label{eq:C56}&&
H(a)=H_0(a)-\frac{1}{2}\bigg\{1-a \ln\left(\left|\frac{1+2a}{1-2a}\right|\right)\quad\\&& + \frac{1}{12a} \left[\ln\left(\left|\frac{1+2a}{1-2a}\right|\right)-\ln\left(\left|\frac{1+8a}{1-8a}\right|\right)\right]\bigg\},\quad
\end{eqnarray} 
with $H_0(a)$ defined by Eq.~(\ref{eq:C49}).  Assume $a \neq 1/2$ and $t\in [0,1]$.  Then,
\begin{equation}
\label{eq:C57}
\lim_{i \to \infty} \left[k_i\pi  G_{n_i,k_i}(t)\right]=\left\{\begin{array}{ l l } 0, & t=0,\\ H(a), & 0<t<1,\\ \mbox{} [1-(-1)^{k_i}]H(a), & t=1. \end{array} \right. 
\end{equation} 
Moreover, there are positive constants $c''_1$, $c''_2$, $c''_3$ and $c''_4$ such that 
\begin{equation}
\label{eq:C58}
\sup_{t \in [0,1]} \left|k\pi  G_{n,k}(t)\right|\leq c''_1-c''_2\ln \left(\left|1-\frac{2n+1}{k}\right|\right), 
\end{equation}
if $k\neq 2n+1$ and
\begin{equation}
\label{eq:C58a}
\sup_{t \in [0,1]} \left|k\pi  G_{n,k}(t)\right|\leq c''_3+c''_4\ln \left(2n+1\right), 
\end{equation}
if $k=2n+1$.
\end{C6}

\emph{Proof.} First, let us notice that the theorem is true if $t\in [0,1/2]$. Eq.~(\ref{eq:C57}) is trivially true if $t=0$. If $t\in(0,1/2]$, then Eq.~(\ref{eq:C57}) follows from Eqs.~(\ref{eq:C14}), (\ref{eq:C26}), and (\ref{eq:C50}). Moreover, the existence of the constants $c''_1$, $c''_2$, $c''_3$, and $c''_4$ with the desired properties 
follows quickly from Eqs.~(\ref{eq:C14}), (\ref{eq:C27}),(\ref{eq:C27a}), and (\ref{eq:C51}), and from  the fact that $G_{n,k}(0)=0$. 

In the remainder of the proof, we use symmetry arguments to show that the theorem holds  on the whole interval $[0,1]$. 
Let $t>1/2$ and write
\begin{eqnarray*}
G_{n,k}(t)=G_{n,k}({1}/{2})+\int_{1/2}^1 [r_{n}(u)-1]\sin(k \pi u)\ud u\\ -\int_{t}^1  [r_{n}(u)-1]\sin(k \pi u)\ud u.
\end{eqnarray*}
Then the symmetry relation $ [r_{n}(u)-1]\sin(k\pi u)=(-1)^{k}[r_{n}(1-u)-1]\sin[k \pi (1-u)]$ implies
\[
\int_{1/2}^1 [r_{n}(u)-1]\sin(k \pi u)\ud u=(-1)^{k}G_{n,k}({1}/{2})
\]
and 
\[
\int_{t}^1 [r_{n}(u)-1]\sin(k\pi u)\ud u=(-1)^{k}G_{n,k}(1-t).
\]
Therefore, 
\begin{equation}
\label{eq:C59}
G_{n,k}(t)=[1+(-1)^{k}]G_{n,k}({1}/{2})-(-1)^{k}G_{n,k}(1-t).
\end{equation}
This last equation implies  the inequality
\[|G_{n,k}(t)|\leq 2|G_{n,k}({1}/{2})|+|G_{n,k}(1-t)|,\]
from which it follows that
\begin{eqnarray*}
 \sup_{t\in (1/2,1]}|G_{n,k}(t)|\leq 3 \sup_{t\in [0,1/2]}|G_{n,k}(t)|.  
\end{eqnarray*}
We then have 
\begin{eqnarray*}
 \sup_{t\in [0,1]}|G_{n,k}(t)|\leq \max\Big\{  \sup_{t\in [0,1/2]}|G_{n,k}(t)|, \\
 \sup_{t\in (1/2,1]}|G_{n,k}(t)|\Big\}\leq 3 \sup_{t\in [0,1/2]}|G_{n,k}(t)|,
\end{eqnarray*}
from which Eqs.~(\ref{eq:C58}) and (\ref{eq:C58a}) follow.

Now, if $t=1$, then  $G_{n_i,k_i}(1-t)=0$ and we have
\begin{eqnarray}
\label{eq:C60}\nonumber
\lim_{i \to \infty} \left[k_i\pi  G_{n_i,k_i}(t)\right]=[1+(-1)^{k_i}] \lim_{i \to \infty} \left[k_i\pi  G_{n_i,k_i}({1}/{2})\right]\\=[1+(-1)^{k_i}]H(a), \qquad
\end{eqnarray} 
where we also used the facts that the sequence $\{k_i\}_{i\geq0}$ is either even or odd  and that the theorem is true for $t=1/2$. 
If $1/2<t<1$, then $0<1-t<1/2$ and  
\begin{eqnarray*}
\lim_{i \to \infty} k_i\pi \left[ G_{n_i,k_i}(1/2)-  G_{n_i,k_i}(1-t)\right]=H(a)-H(a)=0.
\end{eqnarray*}
Using this last observation in Eq.~(\ref{eq:C59}), we obtain  
\begin{eqnarray}
\label{eq:C61}
\lim_{i \to \infty} \left[k_i\pi  G_{n_i,k_i}(t)\right]=\lim_{i \to \infty} \left[k_i\pi  G_{n_i,k_i}(1/2)\right]=H(a). \qquad
\end{eqnarray}
Eqs.~(\ref{eq:C60}) and (\ref{eq:C61}) demonstrate Eq.~(\ref{eq:C57}) and the proof of the proposition is concluded. \hspace{\stretch{1}} $\Box$

We are now in the position to state the first theorem of Appendix~A, which concerns the primitives of the functions $f_{n,k}(u)$. We define the functions
\begin{equation}
\label{eq:C62}
F_{n,k}(t)= -\cos(k\pi t)+k\pi \left[G_{n,k}(t)-G_{n,k}(1/2)\right].
\end{equation}
Eq.~(\ref{eq:C14a}) shows that the functions $F_{n,k}/(k\pi)$ are primitives of $f_{n,k}$. They satisfy the following theorem. 

\begin{C7}
\label{th:C7}
 Let $k_i\to \infty$ and $n_i\to \infty$ be two sequences such that $n_i/k_i \to a$ and the $k_i$'s are either odd or even.  Assume  $a \neq 1/2$.  Then,
\begin{equation}
\label{eq:C63}
\lim_{i \to \infty} F_{n_i,k_i}(0)= -[1+H(a)]. 
\end{equation} 
\begin{equation}
\label{eq:C64}
\lim_{i \to \infty} F_{n_i,k_i}(1)= -(-1)^{k_i}[1+H(a)]. 
\end{equation} 
If $\xi(t):[0,1] \to \mathbb{R}$ is  integrable, then 
\begin{equation}
\label{eq:C65}
\lim_{i \to \infty} \int_0^1 F_{n_i,k_i}(t)\xi(t)\ud t= 0 
\end{equation} 
and
\begin{equation}
\label{eq:C66}
\lim_{i \to \infty} \int_0^1 F_{n_i,k_i}(t)^2\xi(t)\ud t= \frac{1}{2} \int_0^1 \xi(t)\ud t. 
\end{equation} 
If $\xi'(u,\tau):[0,1]\times [0,1]\to \mathbb{R}$ is integrable, then 
\begin{equation}
\label{eq:C67}
\lim_{i \to \infty}\int_0^1\! \! \int_0^1 F_{n_i,k_i}(u)F_{n_i,k_i}(\tau) \xi'(u,\tau)  \ud u \ud \tau =0.
\end{equation}
Finally, there are positive constants $c_1$, $c_2$, $c_3$, and $c_4$ such that
\begin{equation}
\label{eq:C68}
\sup_{t \in [0,1]} \left| F_{n,k}(t)\right|\leq c_1-c_2\ln \left(\left|1-\frac{2n+1}{k}\right|\right), 
\end{equation}
if $k\neq 2n+1$ and
\begin{equation}
\label{eq:C69}
\sup_{t \in [0,1]} \left| F_{n,k}(t)\right|\leq c_3+c_4\ln \left(2n+1\right), 
\end{equation}
if $k=2n+1$.
\end{C7}

\emph{Proof.} 
Eqs.~(\ref{eq:C63}),(\ref{eq:C64}), (\ref{eq:C68}), and (\ref{eq:C69}) follow immediately from Eqs.~(\ref{eq:C57}), (\ref{eq:C58}), (\ref{eq:C58a}), and (\ref{eq:C62}). One also utilizes the inequality $|\cos(k\pi t)|\leq 1$.   

We turn our attention to Eq.~(\ref{eq:C65}). From the Riemann-Lebesgue lemma we learn that 
\begin{equation}
\label{eq:C79}
\lim_{i \to \infty} \int_0^1 \xi(t)\cos(k_i\pi t) \ud t =0. 
\end{equation}
For this proof, it is convenient to define the auxiliary function 
\begin{equation}
\label{eq:C80}
G^\circ_{n,k}(t)=k\pi  \left[ G_{n,k}(t)- G_{n,k}(1/2)\right],
\end{equation}
so that Eq.~(\ref{eq:C62}) becomes
\begin{equation}
\label{eq:C81}
F_{n,k}(t)= -\cos(k\pi t)+G^\circ_{n,k}(t).
\end{equation} 
Since $n_i/k_i \to a \neq 1/2$,  Eq.~(\ref{eq:C57}) implies 
\begin{equation}
\label{eq:C82}
\lim_{i \to \infty}G^\circ_{n_i,k_i}(t)=0\quad \forall t\in (0,1). 
\end{equation}
Also, from Eq.~(\ref{eq:C58}), we learn that 
\begin{equation}
\label{eq:C83}
\sup_{t \in [0,1]} \left| G^\circ_{n_i,k_i}(t)\right|  \leq 2c''_1-2c''_2\ln \left(\left|1-\frac{2n_i+1}{k_i}\right|\right). 
\end{equation}
Since $n_i/k_i \to a \neq 1/2$, the above equation shows that there are $i_0 \geq 1$ and $M>0$ such that $| G^\circ_{n_i,k_i}(t)|\leq M$ for all $i \geq i_0$. Then the dominated convergence theorem implies
\begin{equation}
\label{eq:C84}
\lim_{i \to \infty} \int_0^1 \xi(t)G^\circ_{n_i,k_i}(t) \ud t =0, 
\end{equation}
because the integrand converges to zero pointwise by Eq.~(\ref{eq:C82}) and is dominated by the integrable function $M|\xi(t)|$.
Eq.~(\ref{eq:C65}) follows from  Eqs.~(\ref{eq:C79}) and  (\ref{eq:C84}).
 
The same identities utilized to prove Eq.~(\ref{eq:C84}) and the same dominated convergence theorem also imply
\begin{equation}
\label{eq:C85}
\lim_{i \to \infty} \int_0^1 \xi(t)G^\circ_{n_i,k_i}(t)^2 \ud t =0, 
\end{equation}
\begin{equation}
\label{eq:C86}
\lim_{i \to \infty} \int_0^1 \xi(t)G^\circ_{n_i,k_i}(t)\cos(k_i\pi t) \ud t =0, 
\end{equation}
\begin{equation}
\label{eq:C87}
\lim_{i \to \infty} \int_0^1 \int_0^1 \xi'(u, \tau)G^\circ_{n_i,k_i}(u)G^\circ_{n_i,k_i}(t) \ud u \ud \tau =0, 
\end{equation}
and 
\begin{equation}
\label{eq:C88}
\lim_{i \to \infty} \int_0^1 \int_0^1 \xi'(u, \tau)G^\circ_{n_i,k_i}(u)\cos(k_i \pi \tau) \ud u \ud \tau =0. 
\end{equation}
In all cases, the integrands converge to zero pointwise by Eq.~(\ref{eq:C82}) and are dominated by $M^2 |\xi(t)|$, $M |\xi(t)|$, $M^2|\xi'(u,\tau)|$, and $M|\xi'(u,\tau)|$, respectively.
Now, the Riemann-Lebesgue lemma shows that 
\begin{eqnarray}
\label{eq:C89}\nonumber &&
\lim_{i \to \infty} \int_0^1 \xi(t) \cos(k_i \pi t)^2 \ud t= \frac{1}{2} \int_0^1 \xi (t) \ud t\\ &&+ \frac{1}{2}\lim_{i \to \infty} \int_0^1 \xi(t) \cos(2 k_i \pi t) \ud t = \frac{1}{2} \int_0^1 \xi (t) \ud t.\qquad
\end{eqnarray}
The Riemann-Lebesgue lemma also implies that 
\[
\lim_{i \to \infty} \cos(k_i \pi u) \int_0^1 \xi'(u,\tau)\cos(k_i \pi \tau)  \ud \tau= 0.
\]
for almost all $u \in [0,1]$ because $\xi'(u,\tau)$ as a function of $\tau$ is integrable for almost all $u\in [0,1]$ (this statement is part of the Fubini theorem). Moreover, the above integrand is dominated by the integrable function 
$
\int_0^1 |\xi'(u,\tau)|\ud \tau
$
and then the dominated convergence theorem and Fubini theorem imply 
\begin{equation}
\label{eq:C90}
\lim_{i \to \infty}\int_0^1\int_0^1  \xi'(u,\tau) \cos(k_i \pi u) \cos(k_i \pi \tau) \ud u \ud \tau = 0.
\end{equation}

Simple algebraic manipulations by means of Eq.~(\ref{eq:C81}) show that Eqs.~(\ref{eq:C85}), (\ref{eq:C86}), and (\ref{eq:C89}) imply Eq.~(\ref{eq:C66}), while Eqs.~(\ref{eq:C87}), (\ref{eq:C88}), and (\ref{eq:C90}) imply Eq.~(\ref{eq:C67}). The proof of the theorem is concluded. \hspace{\stretch{1}} $\Box$

We conclude the section by proving the following theorem, which  is crucially used in the main text. 
\begin{C8}
\label{th:C8}
Let $D\subset [0,1]\times [0,1]$ be the triangle $D=\{(u,\tau): 0\leq \tau \leq u,\; 0\leq u \leq 1\}$. Let $K(u,\tau):D \to \mathbb{R}$ be a continuous function with continuous first order derivatives such that $\frac{\partial^2}{\partial u \partial \tau} K(u,\tau)$ is also continuous. Then
\begin{eqnarray}
\label{eq:C91a}\nonumber &&
\lim_{n \to \infty} n \sum_{k=n+1}^{4n} \int_0^1 \ud u \int_0^u \ud \tau K(u,\tau) f_{n,k}(u)f_{n,k}(\tau) \\&& = \frac{1}{2\pi^2}\left\{\int_{1/4}^1 [1+H(a)]^2 \ud a\right\}\left[K(0,0)+K(1,1)\right]\quad \\&&-\frac{3}{16\pi^2}\left[K(1,1)-K(0,0)\right] \nonumber +\frac{3}{8\pi^2}\int_0^1 \frac{\partial}{\partial \tau}K(u,\tau)\Big|_{\tau=u} \ud u, 
\end{eqnarray}
where the function $H(a)$ is defined by Eq.~(\ref{eq:C56}). 
\end{C8}

\emph{Proof.} Successive integration by parts first against the variable $\tau$ and then against the variable $u$ produces the identity
\begin{eqnarray*}
\int_0^1 \ud u \int_0^u \ud \tau K(u,\tau) f_{n,k}(u)f_{n,k}(\tau)= \frac{A_{n,k}}{k^2 \pi^2},
\end{eqnarray*}
where 
\begin{widetext}
\begin{eqnarray*}
A_{n,k}&=& \frac{1}{2} \left[F_{n,k}(0)^2 K(0,0)-2F_{n,k}(0)F_{n,k}(1)K(1,0)+F_{n,k}(1)^2 K(1,1)\right] 
\\& 
-&\frac{1}{2} \int_0^1 F_{n,k}(u)^2 \frac{\partial}{\partial u} K(u,u)\ud u + F_{n,k}(0)\int_0^1 F_{n,k}(u) \frac{\partial}{\partial u} K(u,0) \ud u- F_{n,k}(1)\int_0^1 F_{n,k}(\tau) \frac{\partial}{\partial \tau} K(1,\tau)  \ud \tau
 \\ &
 + &\int_0^1 F_{n,k}(u)^2 \frac{\partial}{\partial u} K(u,\tau)\Big|_{\tau=u} \ud u + \int_0^1 \ud u \int_0^u \ud \tau  F_{n,k}(u)F_{n,k}(\tau)\frac{\partial^2}{\partial u \partial \tau } K(u,\tau).
\end{eqnarray*}
\end{widetext}

By Theorem~\ref{th:C7}, the numbers $A_{n,k}$ enjoy the following properties. Let 
\begin{eqnarray*}
M&=& \frac{1}{2} \left[|K(0,0)|+2|K(1,0)|+| K(1,1)|\right] \\ & +&\frac{1}{2}
\int_0^1\left| \frac{\partial}{\partial u} K(u,u)\right|\ud u + \int_0^1  \left|\frac{\partial}{\partial u} K(u,0)\right| \ud u \\ & +& \int_0^1 \left| \frac{\partial}{\partial \tau} K(1,\tau) \right| \ud \tau  + \int_0^1 \left|\frac{\partial}{\partial u} K(u,\tau)\Big|_{\tau=u}\right| \ud u \\ &+& \int_0^1 \ud u \int_0^u \ud \tau  \left|\frac{\partial^2}{\partial u \partial \tau } K(u,\tau)\right|. 
\end{eqnarray*}
Clearly,  $M<\infty$. Then from Eqs.~(\ref{eq:C68}) and (\ref{eq:C69}) we learn that 
\begin{equation}
\label{eq:C91}
|A_{n,k}|\leq M \left[ c_1-c_2\ln \left(\left|1-\frac{2n+1}{k}\right|\right)\right]^2,
\end{equation}
if $k \neq 2n+1$ and
\begin{equation}
\label{eq:C92}
|A_{n,k}|\leq M \left[ c_3+c_4\ln \left(2n+1\right)\right]^2,
\end{equation}
if $k = 2n+1$.
Moreover, if $k_i$ and $n_i$ are two sequences such that $\lim_{i \to \infty} n_i/k_i = a \neq 1/2$ and the $k_i$'s are either odd or even, then 
\begin{eqnarray}
\label{eq:C93}\nonumber
\lim_{i \to \infty} A_{n_i, k_i} = \frac{1}{2} \left[1+H(a)\right]^2\big[K(0,0)+K(1,1)\\ -2(-1)^{k_i}K(1,0)\big] -\frac{1}{4} \int_0^1 \frac{\partial}{\partial u} K(u,u)\ud u \\ \nonumber+\frac{1}{2}\int_0^1  \frac{\partial}{\partial u} K(u,\tau)\Big|_{\tau=u} \ud u . 
\end{eqnarray}
This last property also follows from Theorem~\ref{th:C7}. The straightforward argument is left for the reader. 

We now notice that the limit in Eq.~(\ref{eq:C91a}) does not change if the summation over $k$ is restricted to the set $\mathcal{I}_n=\{k: n+1 \leq k \leq 4n, k\neq 2n+1\}$. Indeed, the inequality
\[
 n \frac{|A_{n,2n+1}|}{\pi^2(2n+1)^2} \leq \frac{n M \left[ c_3+c_4\ln \left(2n+1\right)\right]^2 }{\pi^2(2n+1)^2}
\]
implies 
\[
\lim_{n \to \infty}  n \frac{A_{n,2n+1}}{\pi^2(2n+1)^2}=0,
\]
from which our claim follows. In these conditions, the limit in Eq.~(\ref{eq:C91a}) becomes
\begin{eqnarray}
\label{eq:C94}\nonumber
\lim_{n \to \infty} \frac{n}{\pi^2} \sum_{k \in \mathcal{I}_n} \frac{A_{n,k}}{k^2}=\lim_{n \to \infty} \frac{n}{\pi^2} \sum_{k \in \mathcal{I}_n; k \text{-even}} \frac{A_{n,k}}{k^2}\\ +\lim_{n \to \infty} \frac{n}{\pi^2} \sum_{k \in \mathcal{I}_n; k \text{-odd}} \frac{A_{n,k}}{k^2}.
\end{eqnarray}
We broke the sum into two parts over the even terms and the odd terms respectively, because these partial sums are easier to analyze. 

Let us begin with the sums over the even terms. We notice that
\begin{eqnarray*}
 \frac{n}{\pi^2} \sum_{k \in \mathcal{I}_n; k \text{-even}} \frac{A_{n,k}}{k^2}=  \frac{n}{\pi^2}  \sum_{k =\left[n/2\right]+1}^{2n}  \frac{A_{n,2k}}{4k^2}.
\end{eqnarray*}
Since  $2n/(2n+1) \to 1$ as $n \to \infty$, we have
\begin{eqnarray*}&&
\lim_{n \to \infty} \frac{n}{\pi^2} \sum_{k \in I_n; k \text{-even}} \frac{A_{n,k}}{k^2}\\&&=  \frac{1}{2\pi^2}\lim_{n \to \infty}  \frac{1}{2n+1} \sum_{k =\left[n/2\right]+1}^{2n}  A_{n,2k}\frac{(2n+1)^2}{4k^2}.
\end{eqnarray*}
Now, for $\left[n/2\right]+1\leq k \leq 2n$, let us define the intervals 
\[
I_{n,k}= \left(\frac{n+k}{2n+1}, \frac{n+k+1}{2n+1}\right].
\] 
By construction, the intervals are disjoint and their reunion is the semi-open interval
\begin{equation}
\label{eq:C95}
\left(\frac{n+[n/2]+1}{2n+1}, \frac{3n+1}{2n+1}\right] \subset \left[{1}/{2}, {3}/{2}\right] .
\end{equation}
Now, on the interval $[1/2, 3/2]$, we define the sequence of step functions
\[
\zeta_n(x)=\left\{\begin{array}{l l} A_{n, 2k} \frac{(2n+1)^2}{4k^2}, & \text{if} \ x \in I_{n,k}, \\ 0, & \text{otherwise}.\end{array} \right.
\]
By the construction of the functions $\zeta_n(x)$,
\begin{eqnarray*} \frac{1}{2n+1} \sum_{k =\left[n/2\right]+1}^{2n}  A_{n,2k}\frac{(2n+1)^2}{4k^2}= \int_{1/2}^{3/2} \zeta_n(x) \ud x.
\end{eqnarray*}

In this paragraph, we use Eq.~(\ref{eq:C93}) to show that the sequence of functions $\zeta_n(x)$ is almost surely convergent on the interval $[1/2, 3/2]$.  
Since the function $\zeta_n(x)$ vanishes outside the interval given by Eq.~(\ref{eq:C95}), it follows that 
$\lim_{n \to \infty} \zeta_n(x)= 0$ for all $x< 3/4$. Now, let us pick an arbitrary $x$ such that $3/4 < x < 1$ or $1<x < 3/2$. For each large enough $n$, there is a unique $k_n$ such that $x \in I_{n,k_n}$. Clearly, 
\[
\lim_{n \to \infty} \frac{n+k_n}{2n+1} =x,
\]
from which it follows that $k_n/n \to 2x-1$. With the help of Eq.~(\ref{eq:C93}), we obtain
\begin{eqnarray*}&&
\lim_{n \to \infty} \zeta_n(x)= \lim_{n \to \infty} A_{n,2k_n}\frac{(2n+1)^2}{4k_n^2}=  \frac{1}{2(2x-1)^2}\\&& \times \left[1+H\left(\frac{1}{4x-2}\right)\right]^2\big[K(0,0)+K(1,1) -2K(1,0)\big] \\&&-\frac{K(1,1)-K(0,0)}{4(2x-1)^2} +\frac{1}{2(2x-1)^2}\int_0^1  \frac{\partial}{\partial u} K(u,\tau)\Big|_{\tau=u} \ud u . 
\end{eqnarray*}

Next, we use Eq.~(\ref{eq:C91}) to show that the sequence of functions $\zeta_n(x)$ is dominated by some integrable function. 
Let us consider the function \[ \eta(x)= M \left[ c_1-c_2\ln \left(\left|1-\frac{1}{x}\right|\right)\right]^2. \]
We leave it for the reader to show that $\eta(x)$ is increasing on the interval $[1/2,1)$ and decreasing on the interval  $(1, 3/2]$ (this property is not dependent on the values of the positive constants $c_1$ and $c_2$). Moreover, $\eta(x)$ is integrable on the interval $[1/2, 3/2]$ because the singularity at the point $x=1$ is of logarithmic type.  Now, if $x \in I_{n,k}$ with $k \geq n+1$, then $x>1$ and by Eq.~(\ref{eq:C91}) we have
\begin{eqnarray*}&&
|\zeta_n(x)|= \left| A_{n,2k}\frac{(2n+1)^2}{4k^2}\right|\leq 4\left| A_{n,2k} \right| \\ && \leq 4 \eta\left(\frac{2k}{2n+1}\right)  \leq 4 \eta\left(\frac{n+k+1}{2n+1}\right) \leq 4 \eta (x). \end{eqnarray*}
If $x \in I_{n,k}$ with $k \leq n$, then $x\leq 1$ and again
\begin{eqnarray*}
|\zeta_n(x)| \leq 4 \eta\left(\frac{2k}{2n+1}\right)  \leq 4 \eta\left(\frac{n+k}{2n+1}\right) \leq 4 \eta (x).  
\end{eqnarray*}
If $x$ is a point lying outside any $I_{n,k}$, then $\zeta_n(x)= 0$ and the inequality $\zeta_n(x) \leq 4 \eta (x)$ is trivially true.

From the facts presented in the last two paragraphs, we conclude that the sequence $\zeta_n(x)$ is dominated by the integrable function $4\eta(x)$ and is pointwise convergent almost everywhere to $0$ if $x\in [1/2, 3/4]$ and to 
\begin{eqnarray*}
\frac{1}{2(2x-1)^2} \left[1+H\left(\frac{1}{4x-2}\right)\right]^2\big[K(0,0)+K(1,1)\\ -2K(1,0)\big] -\frac{1}{4(2x-1)^2}[K(1,1)-K(0,0)] \\ +\frac{1}{2(2x-1)^2}\int_0^1  \frac{\partial}{\partial u} K(u,\tau)\Big|_{\tau=u} \ud u 
\end{eqnarray*}
otherwise. An application of  the dominated convergence theorem produces
\begin{eqnarray*}\lim_{n \to \infty} \frac{1}{2n+1} \sum_{k =\left[n/2\right]+1}^{2n}  A_{n,2k}\frac{(2n+1)^2}{4k^2} = \lim_{n \to \infty} \int_{1/2}^{3/2} \zeta_n(x) \ud x\\ = \left\{\int_{3/4}^{3/2} \frac{1}{2(2x-1)^2} \left[1+H\left(\frac{1}{4x-2}\right)\right]^2 \ud x\right\} \\ \times \big[K(0,0)+K(1,1) -2K(1,0)\big] +\left\{\int_{3/4}^{3/2}\frac{1}{2(2x-1)^2} \ud x\right\} \\ \times \left[-\frac{K(1,1)-K(0,0)}{2}+\int_0^1  \frac{\partial}{\partial u} K(u,\tau)\Big|_{\tau=u} \ud u \right].
\end{eqnarray*}
Performing the substitution $a= 1/(4x-2)$, we compute
\begin{eqnarray}
\label{eq:C96}\nonumber &&
\lim_{n \to \infty} \frac{n}{\pi^2} \sum_{k \in \mathcal{I}_n; k \text{-even}} \frac{A_{n,k}}{k^2}  = \frac{1}{4\pi^2} \left\{\int_{1/4}^{1}  \left[1+H\left(a\right)\right]^2 \ud a\right\}\\ && \times \big[K(0,0)+K(1,1) -2K(1,0)\big]-\frac{3}{32\pi^2}\qquad \\ && \times [K(1,1)-K(0,0)] +\frac{3}{16\pi^2}   \int_0^1  \frac{\partial}{\partial u} K(u,\tau)\Big|_{\tau=u} \ud u. \nonumber
\end{eqnarray}

The computation of the limit 
\begin{eqnarray*}
\lim_{n \to \infty} \frac{n}{\pi^2} \sum_{k \in \mathcal{I}_n; k \text{-odd}} \frac{A_{n,k}}{k^2} =\frac{1}{2\pi^2} \lim_{n \to \infty} \frac{1}{2n+1} \\ \times \Bigg[ \sum_{k=[\frac{n+1}{2}]+1}^n A_{n, 2k-1} \frac{(2n+1)^2}{(2k-1)^2} \\+ \sum_{k=n+1}^{2n-1} A_{n, 2k+1} \frac{(2n+1)^2}{(2k+1)^2}   \Bigg]
\end{eqnarray*}
proceeds in a similar way. For $\left[(n+1)/2\right] +1 \leq k \leq 2n-1$, we define the intervals 
\[
I_{n,k}= \left(\frac{n+k}{2n+1}, \frac{n+k+1}{2n+1}\right].
\] 
By construction, the intervals are disjoint and their reunion is the semi-open interval
\[
\left(\frac{n+[(n+1)/2]+1}{2n+1}, \frac{3n}{2n+1}\right] \subset \left[{1}/{2}, {3}/{2}\right].
\]
Now, on the interval $[1/2, 3/2]$, we define the sequence of step functions
\[
\zeta_n(x)=\left\{\begin{array}{l l} A_{n, 2k+1} \frac{(2n+1)^2}{(2k+1)^2}, & \text{if} \ x \in I_{n,k}\ \text{and}\ k\geq n+1, \\  A_{n, 2k-1} \frac{(2n+1)^2}{(2k-1)^2}, & \text{if} \ x \in I_{n,k}\ \text{and}\ k\leq n,\\ 0, & \text{otherwise}.\end{array} \right.
\]
By the construction of the functions $\zeta_n(x)$,
\begin{eqnarray*}&&
 \frac{1}{2n+1}\Bigg[ \sum_{k=[\frac{n+1}{2}]+1}^n A_{n, 2k-1} \frac{(2n+1)^2}{(2k-1)^2} \\ &&+ \sum_{k=n+1}^{2n-1} A_{n, 2k+1} \frac{(2n+1)^2}{(2k+1)^2}   \Bigg] =\int_{1/2}^{3/2} \zeta_n(x) \ud x.
\end{eqnarray*}

I leave it for the reader to use arguments similar to those employed for the sums over the even coefficients and prove that the sequence $\zeta_n(x)$ is dominated by the integrable function $4\eta(x)$ and is pointwise convergent almost everywhere to $0$ if $x\in [1/2, 3/4]$ and to 
\begin{eqnarray*}
\frac{1}{2(2x-1)^2} \left[1+H\left(\frac{1}{4x-2}\right)\right]^2\big[K(0,0)+K(1,1)\\ +2K(1,0)\big] -\frac{1}{4(2x-1)^2}[K(1,1)-K(0,0)] \\ +\frac{1}{2(2x-1)^2}\int_0^1  \frac{\partial}{\partial u} K(u,\tau)\Big|_{\tau=u} \ud u 
\end{eqnarray*}
otherwise. Then, with the help of the dominated convergence theorem and of the transformation of coordinates $a= 1/(4x-2)$, one concludes
\begin{eqnarray}
\label{eq:C97}\nonumber &&
\lim_{n \to \infty} \frac{n}{\pi^2} \sum_{k \in \mathcal{I}_n; k \text{-odd}} \frac{A_{n,k}}{k^2}  = \frac{1}{4\pi^2} \left\{\int_{1/4}^{1}  \left[1+H\left(a\right)\right]^2 \ud a\right\}\\ && \times \big[K(0,0)+K(1,1) +2K(1,0)\big]-\frac{3}{32\pi^2}  \\ && \times [K(1,1)-K(0,0)] +\frac{3}{16\pi^2}   \int_0^1  \frac{\partial}{\partial u} K(u,\tau)\Big|_{\tau=u} \ud u. \nonumber
\end{eqnarray}

The statement of the theorem follows easily from Eqs.~(\ref{eq:C94}), (\ref{eq:C96}), and (\ref{eq:C97}). \hspace{\stretch{1}} $\Box$

\section{}

In this appendix, we shall prove that
\begin{equation}
\label{eq:D1}
\lim_{n \to \infty} n^3 \tilde{\gamma}_n(u,\tau)^2 = \frac{1}{3\pi^4}\delta(u-\tau)
\end{equation}
and 
\begin{equation}
\label{eq:D2}
\lim_{n \to \infty} n^3 \tilde{\gamma}_n(u,\tau)\gamma_n(u,\tau) = \frac{1}{4\pi^4}\delta(u-\tau),
\end{equation}
in the sense of distributions. More precisely, we shall prove the following theorem. 
\begin{D1}
\label{th:D1}
If $h:[0,1]\times [0,1] \to \mathbb{R}$ is a continuous function, then
\begin{equation}
\label{eq:D3}
\lim_{n \to \infty} n^3 \int_0^1 \int_0^1 \tilde{\gamma}_n(u,\tau)^2 h(u,\tau) \ud u \ud \tau = \frac{1}{3\pi^4}\int_0^1 h(u,u) \ud u
\end{equation}
and
\begin{eqnarray}
\label{eq:D4} \nonumber
\lim_{n \to \infty} n^3 \int_0^1 \int_0^1 \tilde{\gamma}_n(u,\tau) \gamma_n(u,\tau) h(u,\tau) \ud u \ud \tau \\ = \frac{1}{4\pi^4}\int_0^1 h(u,u) \ud u.
\end{eqnarray}
\end{D1}

Before giving the proof of the theorem, we demonstrate the following proposition, which will eventually allow us to replace $\tilde{\gamma}_n(u,\tau)$ by $\gamma^\circ_n(u,\tau)$ in Eqs.~(\ref{eq:D3}) and (\ref{eq:D4}) without changing the limits. 
\begin{D2}
\label{Pr:D2}
For any continuous function $h(u,\tau)$, we have
\begin{equation}
\label{eq:D5}
\lim_{n \to \infty} n^3 \int_0^1 \int_0^1 \left[\tilde{\gamma}_n(u,\tau)- \gamma^\circ_n(u,\tau) \right]^2 |h(u,\tau)| \ud u \ud \tau =0.
\end{equation}
\end{D2}

\emph{Proof.} Since the function $h(u,\tau)$ is continuous on the compact set $[0,1]\times[0,1]$, it is bounded  by some positive constant $M$. Use of the inequality $|h(u,\tau)|\leq M$ in Eq.~(\ref{eq:D5}) shows that it is enough to prove 
\begin{equation}
\label{eq:D6}
\lim_{n \to \infty} n^3 \int_0^1 \int_0^1 \left[\tilde{\gamma}_n(u,\tau)- \gamma^\circ_n(u,\tau) \right]^2 \ud u \ud \tau =0.
\end{equation}

To continue, we need to establish a series of useful identities. 
An application of the Cauchy-Schwartz inequality in Eq.~(\ref{eq:43}) produces the relation
\begin{equation}
\label{eq:D7}
\gamma^\circ_n(u,\tau)^2 \leq \gamma^\circ_n(u,u) \gamma^\circ_n(\tau, \tau). 
\end{equation}
The same Cauchy-Schwartz inequality implies
\begin{equation}
\label{eq:D8}
(a+b+c)^2= (1\cdot a+ 1\cdot b+1\cdot c)^2 \leq 3(a^2+b^2+c^2).
\end{equation}
On the other hand, by the orthogonality of the functions $\{ \sin(k\pi \tau),\; k\geq 1\}$, 
\begin{equation}
\label{eq:D9}
\int_0^1 \gamma^\circ_n(u,\tau)^2 \ud \tau = \frac{2\alpha_n^2}{9n^2}\sum_{k=n+1}\sin(k \pi u)^2= \frac{\alpha_n}{3n} \gamma^\circ_n(u,u).
\end{equation}
With the help of Eq.~(\ref{eq:D8}), one proves
\begin{eqnarray*}\nonumber &&
[r_n(u)r_n(\tau)-1]^2=\big\{[r_n(u)-1][r_n(\tau)-1] \\&& +[r_n(u)-1] +[r_n(\tau)-1]\big\}^2 \leq 3[r_n(u)-1]^2 \qquad \\&& \times [r_n(\tau)-1]^2 +3[r_n(u)-1]^2+3[r_n(\tau)-1]^2, \nonumber
\end{eqnarray*}
from which it follows that 
\begin{eqnarray*}
\left[\tilde{\gamma}_n(u,\tau)- \gamma^\circ_n(u,\tau) \right]^2= [r_n(u)r_n(\tau)-1]^2 \gamma^\circ_n(u,\tau)^2 \\ \leq  3[r_n(u)-1]^2 [r_n(\tau)-1]^2 \gamma^\circ_n(u,\tau)^2 \\ +3\left\{[r_n(u)-1]^2+[r_n(\tau)-1]^2\right\} \gamma^\circ_n(u,\tau)^2.
\end{eqnarray*}
This last inequality together with Eqs.~(\ref{eq:D7}) and (\ref{eq:D9}) implies 
\begin{eqnarray*}
\int_0^1 \int_0^1 \left[\tilde{\gamma}_n(u,\tau)- \gamma^\circ_n(u,\tau) \right]^2 \ud u \ud \tau \\ \leq  3 \left\{\int_0^1  [r_n(u)-1]^2 \gamma^\circ_n(u,u)\ud u \right\}^2 \\ +\frac{2\alpha_n}{n} \int_0^1  [r_n(u)-1]^2 \gamma^\circ_n(u,u)\ud u.
\end{eqnarray*}
Setting 
\[
I_n=n^{3/2}\int_0^1  [r_n(u)-1]^2 \gamma^\circ_n(u,u)\ud u 
\]
and employing the inequality $\alpha_n \leq 1/(\pi^2 n)$ [see Eq.~(\ref{eq:C7})], we obtain
\begin{eqnarray*}
n^3\int_0^1 \int_0^1 \left[\tilde{\gamma}_n(u,\tau)- \gamma^\circ_n(u,\tau) \right]^2 \ud u \ud \tau  \leq  3 I_n^2  +\frac{2}{\pi^2n^{1/2}} I_n.
\end{eqnarray*}
This inequality shows that in order to prove the statement of Eq.~(\ref{eq:D6}) and therefore the proposition, it is enough to show that $I_n \to 0$ as $n \to \infty$. 

Remembering the definition of the functions $r_n(u)$, we compute 
\begin{eqnarray*}
\nonumber &&
I_n=n^{3/2}\int_0^1 \left[\sqrt{ \gamma_n(u,u)}-\sqrt{ \gamma^\circ_n(u,u)}\right]^2\ud u \\&& \nonumber = 2 n^{3/2}\left[ \alpha_n - \int_0^1 \sqrt{ \gamma_n(u,u)}\sqrt{ \gamma^\circ_n(u,u)} \ud u\right]\\&& =2 \alpha_n n^{3/2}\left[ 1 - \int_0^1 \sqrt{ 1-g_n(u)}\sqrt{1- g^\circ_n(u)} \ud u\right] \\&& \leq \frac{2n^{1/2}}{\pi^2}\int_0^1\left[ 1 -  \sqrt{ 1-g_n(u)}\sqrt{1- g^\circ_n(u)} \right]\ud u, \nonumber
\end{eqnarray*}
where we also employed Eqs.~(\ref{eq:C5a}) and (\ref{eq:C5b}). The functions $g_n(u)$ and $g^\circ_n(u)$ satisfy the inequalities $|g_n(u)|\leq 1$ and $|g^\circ_n(u)|\leq 1$ as well as 
\[|g_n(u)|\leq \frac{1}{n}\frac{1}{\sin(\pi u)}\quad \text{and} \quad |g^\circ_n(u)|\leq \frac{1}{3n}\frac{1}{\sin(\pi u)} \]
[for the last two inequalities, see Eqs.~(\ref{eq:C9}) and (\ref{eq:C10})]. 

I leave it for the reader to prove the inequality
\[
\sqrt{1-x}\geq \sqrt{1-|x|}\geq 1-|x|^a 
\]
for all $x\in [-1,1]$ and  $a\in [0,1]$. Then, clearly
\[\sqrt{1-x}\sqrt{1-y}\geq \left(1-|x|^a\right)\left(1-|y|^a\right)\geq 1-|x|^a-|y|^a\]
and so
\[1-\sqrt{1-x}\sqrt{1-y}\leq |x|^a+|y|^a.\]
It follows  that 
\begin{eqnarray*}
I_n \leq \frac{2n^{1/2}}{\pi^2}\int_0^1\left[ |g_n(u)|^a +|g^\circ_n(u)|^a \right]\ud u \\ \leq  \frac{2n^{1/2-a}}{\pi^2}\left(1+3^{-a}\right)\int_0^1\frac{1}{\sin(\pi u)^a} \ud u.
\end{eqnarray*}
Notice that the integral 
\[
\int_0^1\frac{1}{\sin(\pi u)^a} \ud u
\]
is finite for all $a< 1$. Choosing $a=3/4$, we obtain
\begin{eqnarray*}
I_n \leq n^{-1/4}  \frac{2}{\pi^2}\left(1+3^{-3/4}\right)\int_0^1\frac{1}{\sin(\pi u)^{3/4}} \ud u,
\end{eqnarray*}
from which it follows that $I_n \to 0$ as desired. The proof of the proposition is concluded. \hspace{\stretch{1}} $\Box$

\emph{Proof of the Theorem.} By decomposing the function $h(u,\tau)$ in its positive and negative parts, which are also continuous functions, we may assume that $h(u,\tau)$ is positive. The proof is organized in two steps. The second step is left for the reader. 

\emph{Step~1.} In this step, we use Proposition~\ref{Pr:D2} to show that we can replace the functions $\tilde{\gamma}_n(u,\tau)$ in Eqs.~(\ref{eq:D3}) and (\ref{eq:D4})  with $\gamma^\circ_n(u,\tau)$ without changing the respective limits. Because $h(u,\tau)$ is positive, it turns out that it is convenient to consider the $L^2$ scalar product
\[
\left\langle f | g \right \rangle =\int_0^1 \int_0^1 f(u,\tau)g(u,\tau) h(u,\tau) \ud u \ud \tau
\]
and the associated norm $\|f\|= \langle f|f \rangle^{1/2} $. With the new notations, the claims of this step are
\begin{equation}
\label{eq:D10}
\lim_{n \to \infty} n^3 \|\tilde{\gamma}_n\|^2 = \lim_{n \to \infty} n^3 \|{\gamma}^\circ_n\|^2 \end{equation}
and 
\begin{equation}
\label{eq:D11}
\lim_{n \to \infty} n^3 \langle \tilde{\gamma}_n| \gamma_n \rangle = \lim_{n \to \infty} n^3 \langle {\gamma}^\circ_n| \gamma_n \rangle,
 \end{equation}
while the statement of Proposition~\ref{Pr:D2} is 
\begin{equation}
\label{eq:D12}
\lim_{n \to \infty} n^3  \|\tilde{\gamma}_n-{\gamma}^\circ_n\|^2 =0 \quad \text{or} \quad \lim_{n \to \infty} n^{3/2}  \|\tilde{\gamma}_n-{\gamma}^\circ_n\| =0.
\end{equation}

Using Eq.~(\ref{eq:D12}) and the standard inequality
\[
| \| \tilde{\gamma}_n\|- \| {\gamma}^\circ_n\|\, | \leq \|\tilde{\gamma}_n-{\gamma}^\circ_n\|,
\]
one easily argue that
\[
\lim_{n \to \infty} n^{3/2} \| \tilde{\gamma}_n\| = \lim_{n \to \infty} n^{3/2} \| {\gamma}^\circ_n\|,
\]
from which Eq.~(\ref{eq:D10}) follows. On the other hand, by Cauchy-Schwartz inequality, 
\begin{eqnarray*}
\lim_{n \to \infty} n^3 | \langle \tilde{\gamma}_n- \gamma^\circ_n | \gamma_n \rangle \,| \leq 
\left(\lim_{n \to \infty} n^{3/2} \| \tilde{\gamma}_n- \gamma^\circ_n \|\right) \\ \left(\lim_{n \to \infty} n^{3/2} \|\gamma_n \|\right).
\end{eqnarray*}
Eq.~(B5) of  Ref.~(\onlinecite{Pre02a}) says that
\[
\lim_{n \to \infty} n^{3/2} \|\gamma_n \|= \frac{1}{(3\pi^4)^{1/2}} \left[ \int_0^1 h(u,u) \ud u\right]^{1/2} < \infty. 
\]
Then Eq.~(\ref{eq:D12}) implies 
\begin{eqnarray*}
\lim_{n \to \infty} n^3 | \langle \tilde{\gamma}_n- \gamma^\circ_n | \gamma_n \rangle \,| =0,
\end{eqnarray*}
from which Eq.~(\ref{eq:D11}) follows readily.

\emph{Step~2.} By the results of the first step, it is enough to prove
\begin{equation}
\label{eq:D13}
\lim_{n \to \infty} n^3 \int_0^1 \int_0^1 {\gamma}_n^\circ(u,\tau)^2 h(u,\tau) \ud u \ud \tau = \frac{1}{3\pi^4}\int_0^1 h(u,u) \ud u
\end{equation}
and
\begin{eqnarray}
\label{eq:D14} \nonumber
\lim_{n \to \infty} n^3 \int_0^1 \int_0^1 {\gamma}_n^\circ(u,\tau) \gamma_n(u,\tau) h(u,\tau) \ud u \ud \tau \\ = \frac{1}{4\pi^4}\int_0^1 h(u,u) \ud u,
\end{eqnarray}
respectively. I leave it for the reader to adapt the arguments utilized in Appendix~B of Ref.~(\onlinecite{Pre02a}) and prove these last two statements. In the cited reference, the authors proved 
\[
\lim_{n \to \infty} n^3 \int_0^1 \int_0^1 {\gamma}_n(u,\tau)^2 h(u,\tau) \ud u \ud \tau = \frac{1}{3\pi^4}\int_0^1 h(u,u) \ud u
\]
and the same technique can be used for Eqs.~(\ref{eq:D13}) and (\ref{eq:D14}).  \hspace{\stretch{1}} $\Box$

\section{Harmonic oscillator}

The $n$-th order RW-WFPI approximation of the partition 
function for an harmonic oscillator centered at the origin has the
expression 
\[
Z^{\text{RW}}_{n}(\beta)=\int_{\mathbb{R}} \ud a_0 \int_{\mathbb{R}}\ud a_1 \ldots \int_{\mathbb{R}} \ud a_{4n} \rho^{\text{RW}}_{n}(a_0, \ldots, a_{4n};\beta),
\]
where
\begin{eqnarray*} 
\rho^{\text{RW}}_{n}(a_0,\ldots, a_{4n};\beta)=\frac{1}{\sqrt{2\pi \sigma^2}} (2\pi)^{-2n}\exp\left(-\sum_{k=1}^{4n} a_k^2/2\right)\\ \exp\left\{-\beta\frac{m_0\omega^2} {2} \int_{0}^{1} \left[ \sum_{k=0}^{4n} a_k \Omega_{n,k}(u)\right]^2\ud u\right\}
\end{eqnarray*}
with
\[
\Omega_{n,k}(u)=\left\{\begin{array}{l l} 
1, & k=0, \\
\sigma \Lambda_k(u), & 1\leq k\leq n, \\ 
\sigma \tilde{\Lambda}_{n,k}(u), & n< k \leq 4n.
\end{array} \right. 
\]
The functions $\Lambda_k(u)$ and $\tilde{\Lambda}_{n,k}(u)$ are defined at the beginning of Section~IV. Notice that $a_0$ stands for the physical coordinate $x$. 

Let $\delta_{i,j}$ denote the Kronecker symbol 
\[
\delta_{i,j}=\left\{ \begin{array}{l l} 1, & i=j, \\ 0,& i\neq j \end{array}\right.
\]
and define the positive definite symmetric matrix of entries
\[
A_{i,j}^{(n)}=\left\{\begin{array}{l l}
 \beta m_0 \omega^2, & i=j=0,\\
\delta_{i,j}+\beta m_0 \omega^2 \int_0^1 \Omega_{n,i}(u)\Omega_{n,j}(u)\ud u, &
 \text{otherwise,}
 \end{array} \right.
\]
with $0\leq i,j\leq 4n$. 
Then, 
\begin{eqnarray*} 
\rho^{\text{RW}}_{n}(a_0,\ldots, a_{4n};\beta)=\frac{1}{\sqrt{2\pi \sigma^2}} (2\pi)^{-2n}
\\ \times \exp\left(-\frac{1}{2}\sum_{i,j}A^{(n)}_{i,j} a_i a_j \right)
\end{eqnarray*}
and
\[
Z^{\text{RW}}_n(\beta)=1 \left/ \left[\sigma^2 \det(A^{(n)})\right]^{1/2}. \right.
\]
The computation of the entries of the matrix $A^{(n)}$ has been performed numerically by Gauss-Legendre quadrature in $8n$ points.


\begin{thebibliography}{99}
\bibitem{Footnote}{The reader is remainded that a sequence $x_n$ has $o(1/n^s)$ asymptotic convergence if $\lim_{n \to \infty} n^s x_n=0$ and it has  $O(1/n^s)$ asymptotic convergence if there is a constant $M>0$ such that $ n^s |x_n|\leq M$ for all $n \geq 1$. In the latter case, if it exists and is not zero, the limit $\lim_{n \to \infty} n^s x_n$ is called the convergence constant of the sequence $x_n$. }
\end{thebibliography}
\end{document}